%% file: main.tex
\documentclass[sigconf]{acmart}
\copyrightyear{2026}
\acmYear{2026}
\setcopyright{cc}
\setcctype{by}
\acmConference[CCS '26]{Proceedings of the 2026 ACM SIGSAC Conference on Computer and Communications Security}{November 15--19, 2026}{The Hague, Netherlands}
\acmBooktitle{Proceedings of the 2026 ACM SIGSAC Conference on Computer and Communications Security (CCS '26), November 15--19, 2026, The Hague, Netherlands}
\acmDOI{10.1145/3830454.3832587}
\acmISBN{979-8-4007-2871-6/2026/11}

\AtBeginDocument{%
  }

\usepackage[english]{babel}
\usepackage{bold-extra}
\usepackage{booktabs}

\usepackage{amsmath}
\DeclareMathAlphabet{\mathcal}{OMS}{cmsy}{m}{n}
\usepackage{amsfonts}
\usepackage{mathtools}

\usepackage[group-separator={,},group-minimum-digits=3]{siunitx}
\usepackage{numprint}
\usepackage{calculator}
\usepackage{fp}

\usepackage{pifont} % for \ding

\usepackage{graphicx}
\usepackage{color}
\usepackage{colortbl}
\usepackage{tikz}
\usepackage{subcaption}
\usepackage{graphicx}

\usepackage{footnote}
\makesavenoteenv{tabular}
\makesavenoteenv{table}

\usepackage{paralist}
\usepackage{multirow}
\usepackage{array}
\usepackage{adjustbox}
\usepackage{tabularx}
\usepackage{xspace}
\usepackage{xhfill}
\usepackage{xstring}
\usepackage{makecell}
\usepackage{collcell}
\usepackage{lipsum}
\usepackage{blindtext}
\usepackage{wasysym}
\usepackage{tabularx}
\usepackage{xcolor}
\usepackage{pgf}
\usepackage[binary-units=true, group-separator={,}, group-minimum-digits={3}]{siunitx}
\usepackage{hyperref}

\definecolor{mycolor1}{rgb}{1, 1, 1} % White
\definecolor{mycolor2}{rgb}{0.941, 0.939, 0.969} % Lightened purple
\definecolor{mycolor3}{rgb}{0.882, 0.876, 0.937} % Medium lightened purple
\definecolor{mycolor4}{rgb}{0.780, 0.783, 0.890} % Darker lightened purple

\definecolor{USENIXPurpleTrans}{cmyk}{0.44, 0.92,0,0.01}
\newlength\maxlen

\def\header{Rating}
\newlength\maxlenweb

\def\headerweb{MonthlyVisitIsAB}
\ccsdesc[500]{Security and privacy~Embedded systems security}
\ccsdesc[500]{Security and privacy~Mobile and wireless security}
\ccsdesc[500]{Human-centered computing~Virtual reality}
\ccsdesc[500]{Human-centered computing~Mixed / augmented reality}

\keywords{Virtual reality; traffic analysis; wireless power; eavesdropping}

\begin{document}
\input{macro}

\title[Long-Range and Through-Wall AR/VR Privacy Eavesdropping via Wireless Power Side Channel]{Penetrating the Reality Wall: Long-Range and Through-Wall AR/VR Privacy Eavesdropping via Wireless Power Side Channel}

% \affiliation{%

% \affiliation{%

% \affiliation{%

% \affiliation{%

\author{Tao Ni}
\authornote{Tao Ni and Zehua Sun are co-first authors.}
\affiliation{%
  \institution{King Abdullah University of Science and Technology}
  % \city{Thuwal}
  % \state{Beijing Shi}
  \country{Saudi Arabia}
  }
\email{tao.ni@kaust.edu.sa}

\author{Zehua Sun}
\authornotemark[1]
\affiliation{%
  \institution{National University of Singapore}
  % \city{San Antonio}
  % \state{Texas}
  \country{Singapore}}
\email{zehua.sun@nus.edu.sg}

\author{Yihe Zhou}
\affiliation{%
  \institution{City University of Hong Kong}
  % \city{Kowloon Tong}
  % \country{Hong Kong}}
  % \city{Hong Kong}
  \country{Hong Kong}}
\email{yihezhou2@cityu.edu.hk}

\author{Yuefeng Du}
\affiliation{%
  \institution{City University of Hong Kong}
  % \city{Kowloon Tong}
  % \country{Hong Kong}}
  % \city{Hong Kong}
  \country{Hong Kong}}
\email{yf.du@cityu.edu.hk}

% \author{Yihe Zhou}
% \affiliation{%
%   \institution{City University of Hong Kong}
%   % \city{Hong Kong}
%   \country{Hong Kong SAR}}
% \email{yihezhou2@cityu.edu.hk}

\author{Qingchuan Zhao}
\authornote{Qingchuan Zhao and Cong Wang are co-corresponding authors.}
\affiliation{%
  \institution{City University of Hong Kong}
  % \city{Kowloon Tong}
  % \country{Hong Kong}}
  % \city{Hong Kong}
  \country{Hong Kong}}
\email{qizhao@cityu.edu.hk}

\author{Cong Wang}
\authornotemark[2]
\affiliation{%
  \institution{City University of Hong Kong}
  % \city{Kowloon Tong}
  % \country{Hong Kong}}
  % \city{Hong Kong}
  \country{Hong Kong}}
\email{congwang@cityu.edu.hk}

\renewcommand{\shortauthors}{Tao Ni et al.}
%% No italics, no superscripts, not anonymous
%% Use footnote or author note to identify equal contribution, shared contribution, and/or contact author info

% \renewcommand{\shortauthors}{T. Ni, Z. Sun, Y. Du, Y. Zhou, Q. Zhao, C. Wang}

% \begin{abstract}
%   A clear and well-documented \LaTeX\ document is presented as an
%   article formatted for publication by ACM in a conference proceedings
%   or journal publication. Based on the ``acmart'' document class, this
%   article presents and explains many of the common variations, as well
%   as many of the formatting elements an author may use in the
%   preparation of the documentation of their work.
% \end{abstract}

\input{sections_major/00_abstract}

\maketitle

\input{sections_major/01_introduction}
\input{sections_major/02_background}
\input{sections_major/03_threat_model}
\input{sections_major/04_attack_design}
\input{sections_major/05_evaluation}
\input{sections_major/06_discussion}
\input{sections_major/07_related_work}

\input{sections_major/08_conclusion}
\bibliographystyle{ACM-Reference-Format}
\balance
\bibliography{paper}

\appendix

\section{Ethical Considerations}
We propose \sysname as a proof-of-concept attack framework to validate the wireless power side-channel leakage in AR/VR devices.
To this end, our study involves collecting Wi-Fi power traces when users perform different activities \major{while} wearing VR devices.
Recognizing the potential misuse of \sysname and the impact on human participants, we consider the ethical implications of our work and implement mitigation strategies throughout the research process.
\looseness=-1

\noindent\textbf{Stakeholders.}
Our study affects two key stakeholders, including AR/VR users and corresponding manufacturers.
Specifically, AR/VR users may face safety risks when the Metaverse systems expose side-channel vulnerabilities that may leak private information (\eg, UI-level activities, identification, and keystrokes) to unauthorized adversaries.
In addition, AR/VR manufacturers may experience user privacy breaches and loss of user trust if the wireless power side-channel attacks are not \major{addressed} in their products.
Therefore, we hope that our study can raise public awareness about these attacks.
\noindent\textbf{Ethics in Physical Experiments.}
As our experiments involve $20$ human participants (\ie, $10$ males and $10$ females), we obtained IRB approval from our institution to implement strict procedures to ensure informed consent and protect the privacy of all participants.
Specifically, all participants are informed about the experimental details, such as wearing VR headsets and \major{performing} different activities.
Then, all participants need to sign the consent form to allow the usage of their data (\eg, activities in VR) collected in different conditions (\eg, distances, VR devices) for experiment evaluations.
\looseness=-1

\noindent\textbf{Mitigating Potential Harm.}
To mitigate potential misuse of the \sysname, we only used our own accounts on the Meta Oculus Quest and PICO platforms to browse websites, run apps, play VR games, and type keystrokes. \sysname and our attack devices and inference models have never been released to any other parties.
Furthermore, data collection from volunteer participants \major{was} conducted \major{under} the supervision of \major{the} IRB committee.
We asked the participants to type on the virtual keyboards in VR devices for 30 minutes to get familiar with VR typing before the official data collection.
Considering the dizziness and motion sickness associated with VR usage, we require participants to rest for $1$--$10$ minutes to report their conditions after collecting data with different time durations in each trial.
In addition, the captured data from all participants were stored on an encrypted server to prevent any form of data breaches.
\looseness=-1

\noindent\textbf{\major{Broader Societal Implications.}}
\major{The emergence of long-range, through-wall side-channel attacks on AR/VR devices raises broader societal concerns beyond individual privacy. As AR/VR platforms increasingly integrate into sensitive domains such as healthcare consultations, financial transactions, remote education, and confidential workplace collaborations, the unintended leakage of user behaviors, identities, and keystrokes could enable surveillance, identity theft, corporate espionage, or targeted harassment without the victim's awareness. Unlike conventional surveillance that requires LoS or physical access, the passive and non-intrusive features of \sysname makes detection difficult, which could disproportionately impact vulnerable populations (\eg, journalists, activists) who rely on the perceived privacy of indoor environments.}

\noindent\textbf{\major{Responsible Disclosure.}}
\major{Following responsible disclosure practices, we have notified leading AR/VR manufacturers, including Meta and PICO, of the vulnerabilities identified in this work prior to publication. We provided detailed technical descriptions, threat scenarios, and recommended countermeasures (\eg, signal obfuscation, dynamic transmission power, traffic padding) discussed in \autoref{subsec:countermeasures}, and we are actively coordinating with their security teams to support remediation efforts. We will continue to assist manufacturers in evaluating and deploying defenses, and we encourage further dialogue between academia and industry to ensure timely mitigation of side-channel risks in next-generation AR/VR products.}

% \vspace{-0.05in}
\section{Open Science}
We fully support the open science policy and believe it will greatly benefit the
advancement of the entire community. In line with the open science policy of ACM CCS, we provide our artifact, including additional experimental details (\eg, time-domain and frequency-domain features), complete lists of websites and apps for evaluations, source code for data processing and model training, datasets, and supporting instructions needed for artifact evaluation in the following permanent repository: \url{https://doi.org/10.5281/zenodo.21312850}. In addition, we also provide the design and implementation of our customized wireless power sniffer, which can be easily imported and produced on platforms like EasyEDA to accelerate the reproduction of our hardware platforms.
\looseness=-1

% \appendix

% \input{tables/appendix_app_list_oculus}

% \input{tables/appendix_app_list_pico}

% \input{tables/appendix_website_list}









\end{document}

%% file: macro.tex
\newcommand{\bheading}[1]{{\vspace{0.02in}\noindent{\textbf{#1}}}}
\newcommand{\iheading}[1]{{\vspace{0.01in}\noindent{\textit{#1}}}}
% \newcolumntype{?}{!{\vrule width 1pt}}

\newcommand{\draft}[1]{\textcolor{blue}{#1}}
%terminology and formatting
\newcounter{note}[section]
\renewcommand{\thenote}{\thesection.\arabic{note}}
\colorlet{Mycolor1}{green!10!orange!90!}
\newcommand{\xz}[1]{\refstepcounter{note}{\bf\textcolor{purple}{$\ll$Xiaokuan~\thenote: {\small\sf #1}$\gg$}}}
\newcommand{\tao}[1]{\refstepcounter{note}{\bf\textcolor{purple}{$\ll$Tao: {\small\sf #1}$\gg$}}}

\renewcommand{\sectionautorefname}{\S{}\hspace{-0.2em}} % PS
\renewcommand{\subsectionautorefname}{\S{}\hspace{-0.2em}} 
\renewcommand{\subsubsectionautorefname}{\S{}\hspace{-0.2em}}
\renewcommand{\appendixautorefname}{}

\newcommand{\figurewidth}{\columnwidth}
\newcommand{\secref}[1]{\mbox{Sec.~\ref{#1}}\xspace}
\newcommand{\secrefs}[2]{\mbox{Sec.~\ref{#1}--\ref{#2}}\xspace}
\newcommand{\figref}[1]{\mbox{Fig.~\ref{#1}}}
\newcommand{\tabref}[1]{\mbox{Table~\ref{#1}}}
\newcommand{\appref}[1]{\mbox{Appendix~\ref{#1}}}

\newcommand{\appname}[1]{\textsf{\small #1}}

% writing styles
\newcommand{\etc}{\textit{etc.}\xspace}
\newcommand{\eg}{\textit{e.g.}\xspace}
\newcommand{\ie}{\textit{i.e.}\xspace}

\newcommand{\cf}{\textit{cf.}\xspace}
\newcommand{\aka}{\textit{a.k.a.}\xspace}
\newcommand{\etal}{\textit{et al.}\xspace}

\newcommand{\ios}{\textsc{iOS}\xspace}
\newcommand{\android}{\textsc{Android}\xspace}
\newcommand{\paypal}{\textsc{PayPal}\xspace}
\newcommand{\youtube}{\textsc{YouTube}\xspace}
\newcommand{\netflix}{\textsc{Netflix}\xspace}
\newcommand{\facebook}{\textsc{Facebook}\xspace}
\newcommand{\twitter}{\textsc{Twitter}\xspace}
\newcommand{\amperes}{\textsc{Amperes 4}\xspace}
\newcommand{\bnnook}{\textsc{B\&N NOOK}\xspace}
\newcommand{\duolingo}{\textsc{Duolingo}\xspace}
\newcommand{\visionpro}{\textsc{Vision Pro}\xspace}
\newcommand{\visionos}{\textsc{visionOS}\xspace}
\newcommand{\populationone}{\textsc{Population: ONE}\xspace}
\newcommand{\twitchtv}{\textsc{twitch.tv}\xspace}
\newcommand{\ticktockcom}{\textsc{tiktok.com}\xspace}
\newcommand{\cnncom}{\textsc{cnn.com}\xspace}
\newcommand{\youtubecom}{\textsc{youtube.com}\xspace}
\newcommand{\googlecom}{\textsc{google.com}\xspace}
\newcommand{\baiducom}{\textsc{baidu.com}\xspace}
\newcommand{\amazoncom}{\textsc{amazon.com}\xspace}
\newcommand{\yahoocom}{\textsc{yahoo.com}\xspace}
\newcommand{\gunraider}{\textsc{Gun Raider}\xspace}
\newcommand{\metaquestbrowser}{\textsc{Meta Quest Browser}\xspace}
\newcommand{\oculusquest}{\textsc{Oculus Quest}\xspace}
\newcommand{\youtubevr}{\textsc{YouTube VR}\xspace}
\newcommand{\messenger}{\textsc{Messenger}\xspace}
\newcommand{\meetinvr}{\textsc{MeetinVR}\xspace}
\newcommand{\virtualspeech}{\textsc{VirtualSpeech}\xspace}
\newcommand{\plutotv}{\textsc{Pluto TV}\xspace}
\newcommand{\cardstankards}{\textsc{Cards and Tankards}\xspace}
\newcommand{\multiverse}{\textsc{Multiverse}\xspace}
\newcommand{\maloka}{\textsc{Maloka}\xspace}
\newcommand{\spotify}{\textsc{Spotify}\xspace}

% units
\newcommand{\gbytes}{\ensuremath{\mathrm{GB}}\xspace}
\newcommand{\mbytes}{\ensuremath{\mathrm{MB}}\xspace}
\newcommand{\kbytes}{\ensuremath{\mathrm{KB}}\xspace}
\newcommand{\bytes}{\ensuremath{\mathrm{B}}\xspace}

\newcommand{\ghertz}{\ensuremath{\mathrm{GHz}}\xspace}
\newcommand{\msecs}{\ensuremath{\mathrm{ms}}\xspace}
\newcommand{\usecs}{\ensuremath{\mathrm{\mu{}s}}\xspace}
\newcommand{\nsecs}{\ensuremath{\mathrm{ns}}\xspace}
\newcommand{\secs}{\ensuremath{\mathrm{s}}\xspace}
\newcommand{\gbits}{\ensuremath{\mathrm{Gb}}\xspace}

\newcommand{\cmark}{\ding{51}}%
\newcommand{\xmark}{\ding{53}}%
\newcommand{\tabincell}[2]{\begin{tabular}{@{}#1@{}}#2\end{tabular}}
\newcommand{\whiteding}[1]{\ding{\numexpr171+#1\relax}}
\newcommand{\blackding}[1]{\ding{\numexpr181+#1\relax}}

\newcounter{packednmbr}
\newenvironment{packedenumerate}{
\begin{list}{\thepackednmbr.}{\usecounter{packednmbr}
\setlength{\itemsep}{0pt}
\addtolength{\labelwidth}{4pt}
\setlength{\leftmargin}{12pt}
\setlength{\listparindent}{\parindent}
\setlength{\parsep}{3pt}
\setlength{\topsep}{3pt}}}{\end{list}}

\newenvironment{packeditemize}{
\begin{list}{$\bullet$}{
\setlength{\labelwidth}{0pt}
\setlength{\itemsep}{2pt}
\setlength{\leftmargin}{\labelwidth}
\addtolength{\leftmargin}{\labelsep}
\setlength{\parindent}{0pt}
\setlength{\listparindent}{\parindent}
\setlength{\parsep}{1pt}
\setlength{\topsep}{1pt}}}{\end{list}}

\newcommand{\e}[1]{{\mathbb E}\left[ #1 \right]}

\newcommand*\mycircTiny[1]{%
\tiny
  \begin{tikzpicture}[baseline=(C.base)]
    \node[draw,circle,inner sep=.3pt](C) {\scriptsize{#1}};
  \end{tikzpicture}}  
\newcommand*\myYescirc{\CIRCLE\xspace}
\newcommand*\myHalfcirc{\LEFTcircle\xspace}
\newcommand*\myNocirc{\Circle\xspace}

\newcommand{\major}[1]{\textcolor{black}{#1}}
\newcommand{\empirical}[1]{\textcolor{red}{\uwave{#1}}}
\newcommand{\shep}[1]{\textcolor{red}{#1}}
% Change \paragraph spacing
% \renewcommand{\paragraph}{\vspace{0.01in}\noindent\textbf}
% \renewcommand{\paragraph}[1]{\vspace{0.02in}\noindent{\bf{#1}}}
\addtolength{\abovedisplayskip}{-6pt}
\addtolength{\belowdisplayskip}{-6pt}

\newcommand{\sysname}{\textnormal{\textsf{MetaSpy}}\xspace}

% Author notes

%\newcommand{\authnote}[2]{{\bf[\textcolor{red}{#1}: \textcolor{blue}{#2}]}}
%\newcommand{\QC}[1]{{\authnote{Zhao}{#1}}}
%\newcommand{\XK}[1]{{\authnote{Zhang}{#1}}}
%\newcommand{\CS}[1]{{\authnote{Zuo}{#1}}}
%\newcommand{\XP}[1]{{\authnote{Luo}{#1}}}
%\newcommand{\JF}[1]{{\authnote{Li}{#1}}}

% Ignore content
%\newcommand{\ignore}[1]{}

% Tool Components
% \newcommand{\sysname}{\textsc{WiSERListener}\xspace}
%\newcommand{\sysname}{\textsc{WISERS}\xspace}
% \newcommand{\firstcompo}{\textsf{Wireless Charger Status Inference}\xspace}
%\newcommand{\firstcompo}{\textsf{Attack Preparation}\xspace}
%\newcommand{\secondcompo}{\textsf{Switch Inference}\xspace}
%\newcommand{\thirdcompo}{\textsf{Activity Uncovering}\xspace}
%\newcommand{\forthcompo}{\textsf{Privacy Revelation}\xspace}
% \newcommand{\nfirstcompo}{Wireless Charger Status Inference\xspace}
%\newcommand{\nfirstcompo}{Attack Preparation\xspace}
%\newcommand{\nsecondcompo}{Switch Inference\xspace}
%\newcommand{\nthirdcompo}{Activity Uncovering\xspace}
%\newcommand{\nforthcompo}{Privacy Revelation\xspace}

%% Appendix
% \makeatletter
% \renewcommand{\appendix}{%
%   \par
%   \setcounter{section}{0}%
%   \renewcommand\thesection{\appendixname~\Alph{section}}%
%   \renewcommand{\sectionautorefname}{\appendixname} % Redefine the autoref name
% }
% \makeatother

%% file: sections_major/00_abstract.tex
\begin{abstract}
Over the past decade, Metaverse technology (AR/VR) has seen widespread adoption in numerous areas, including gaming, social networking, and online collaboration.
Although recent studies have exposed vulnerabilities in VR systems, they often apply methods (\eg, pre-installed malicious apps or scripts) from conventional mobile platforms directly, demonstrating impractical attacks that fail to investigate unique privacy leakages specific to VR platforms.
In this paper, we present \major{an orthogonal} attempt to analyze the attack surface of the power side channel on AR/VR's network connection by capturing the \textit{through-wall} wireless power in the network transmission medium (\eg, Wi-Fi signals) non-intrusively in a \major{comparatively} long distance (\eg, \SI{5.0}{\meter}--\SI{8.0}{\meter}) and further reconstructing multi-level privacy-aware user behaviors.
\looseness=-1

We propose \sysname to validate the feasibility and practicality of our disclosed
\major{wireless power-based traffic analysis} attack.
It first captures through-wall wireless power transmitted from the Wi-Fi router of an adjacent room, identifies the device type, then infers \textit{(i)} user interactions such as browsing websites, running AR/VR apps, and performing various in-app activities, \textit{(ii)} unseen VR-specific activities in VR games (\eg, \populationone),
and \textit{(iii)} avatar-based user re-identification and virtual keystroke inference corresponding to head movements.
We evaluated the effectiveness of \sysname in four commercial-off-the-shelf (COTS) VR headsets, and the empirical results demonstrate its competitive accuracy in unveiling multi-level user privacy within virtual scenes in through-wall settings at a longer distance ($1.8\times$–$80\times$) than most of the recent studies, and it shows resilience to a series of external interference and impact factors.
In addition, we also propose and evaluate effective countermeasures to protect data confidentiality in AR/VR wireless transmission.
\looseness=-1

\end{abstract}

%% file: sections_major/01_introduction.tex
\section{Introduction}
\label{sec:introduction}

The Metaverse technologies, such as augmented reality (AR) and virtual reality (VR), have experienced rapid growth over the past decade due to their unique capacity to deliver immersive experiences without physical constraints. 
A recent survey reports that the global AR/VR market will exceed $\$46.6$ billion by 2025 and is expected to reach approximately $\$62.0$ billion by 2029~\cite{arvrmarket2025}.
Despite the prosperity of AR/VR in a variety of fields, \ie, healthcare, education, finance, and online collaborations, they also expose various security and privacy issues that hinder trustworthiness between the public and this emerging paradigm of human-computer interaction.
\looseness=-1

\begin{figure}[t]
    \centering
    \includegraphics[width=\linewidth]{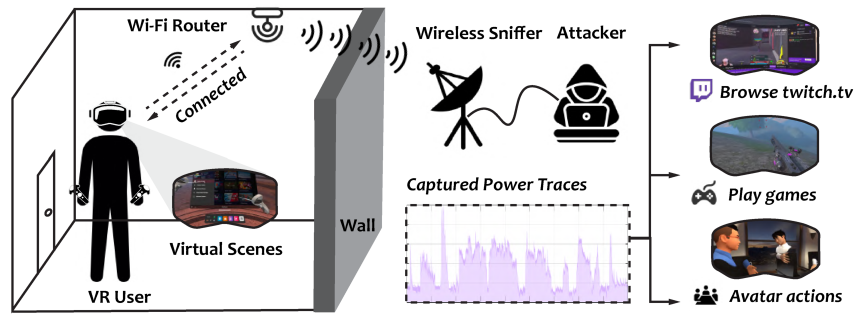}
    \vspace{-0.2in}
    \caption{An illustration of \sysname attack scenario. The victim wears the VR headset that is connected to the Wi-Fi router, then browses the website \twitchtv to watch the video streaming, plays the VR game \populationone, and interact with others through his/her avatar in \meetinvr. An attacker from an adjacent room exploits a wireless power sniffer to capture the through-wall Wi-Fi signals to infer fine-grained user privacy (\eg, in-app actions, virtual keystrokes).}
    \vspace{-0.15in}
    \label{fig:motivating_example}
\end{figure}

Recent literature has revealed that attackers can leverage user movements (\eg, hand and body movements) and variations in wireless sensing to infer user privacy, \ie, sensitive keystroke information within virtual scenes.
Specifically, it is a unique characteristic of VR platforms that \major{leverage} body movements for interaction, and numerous attack surfaces have been studied to exploit VR users' motion data for user privacy inference.
Some attacks extract user movements from the built-in motion sensors of VR headsets through pre-installed malware~\cite{wu2023privacy, slocum2023going, luo2022holologger, ling2019know, zhang2023s, ye2025bpsniff, zhang2023facereader, zhang2023passive, cayir2025speak, choi2023revisiting, son2025side, nair2023unique, nair2022going}, while others record videos or audio using nearby devices such as hand trackers~\cite{meteriz2022keylogging}, smartphones~\cite{nguyen2024penetration, gopal2023hidden, luo2024eavesdropping, khalili2024virtual} or radars~\cite{mei2024mmspyvr}.
On the other hand, AR/VR systems are suspected to suffer from the threat of traffic-based side-channel attacks that analyze packet characteristics (\eg, size, speed, and energy), as shown in recent studies~\cite{van2020flowprint, li2022foap, lakshmanan2022privacy, ozmen2024physical, ni2023eavesdropping, yamansavascilar2017application} on mobile devices because they share a similar interaction medium with other mobile devices by communicating with the server through downlink/uplink network transmission \major{via} the connected Wi-Fi router.
Research has shown that attackers could exploit wireless transmission variations when users are stationary between transceivers~\cite{al2021vr} or lure them into unauthorized online meetings to share their avatars, \major{which are} digital representations of users in virtual scenes~\cite{su2024remote, wang2024gazeploit}. These methods can extract user head and hand activities in real time via unencrypted network packets or virtual camera feeds.
\looseness=-1

However, these studies may only scratch the tip of the iceberg due to their various limitations that leave the security of VR largely unexploited.
\ding{182} With respect to \textit{motion data-based attacks}, the pre-installed malware in previous studies~\cite{wu2023privacy, slocum2023going, luo2022holologger, ling2019know, zhang2023s, cayir2025speak, ye2025bpsniff, zhang2023facereader, zhang2023passive, choi2023revisiting} is not \major{easy to deploy} on AR/VR devices.
Unlike most Android smartphones, which allow users to download apps from websites or USB-connected laptops, AR/VR platforms typically restrict app downloads to their official stores (\eg, Meta Quest Store~\cite{oculusapps}, PICO VR Store~\cite{picoapps}), significantly \major{limiting} an adversary's ability to trick users into installing malware on VR headsets.
Another limitation is the scalability of these attacks, as they often target specific open-source or less-restricted VR SDKs/APIs (\eg, OpenVR~\cite{openvr}, WebXR~\cite{webxr, mukherjee2025shadowed}). These methods may not be effective on closed-source platforms, such as Apple's visionOS~\cite{visionos}, making the attacks non-universal and difficult to execute across different VR ecosystems.
Moreover, the illegal deployment of smartphones to record videos/audios is usually unrealistic, \major{as it} could raise suspicions and be ineffective in low-light  or noisy environments~\cite{gopal2023hidden, nguyen2024penetration, meteriz2022keylogging, luo2024eavesdropping}.
\ding{183} Similarly, recent \textit{wireless metadata-based attacks} rely mainly on unrealistic assumptions that limit their practicality in real-world scenarios. For example, imposing strict physical restrictions on VR users is impractical, as they usually move within the play zone during interactions~\cite{al2021vr, mei2024mmspyvr, li2025synergistic}. 
Moreover, extracting hand movements from unencrypted network packet metadata is effective only for a limited set of protocols in specific multi-user VR apps (\eg, Photon protocol in \textsc{RecRoom}~\cite{su2024remote}).
\ding{184} Additionally, existing attacks lack stealthiness and scalability, and nearly all focus on user privacy at the user interface (UI) layer (\eg, running apps, virtual keystrokes), leaving in-depth VR-specific and systematic vulnerabilities largely unexplored.
\looseness=-1

\input{tables/comparison_prior_works_v6}

In this work, we disclose a through-wall eavesdropping attack in wireless transmission between the VR headset and the connected Wi-Fi router that not only can reveal fine-grained user privacy, but also can be launched at a long distance (\eg, \SI{5.0}{\meter}--\SI{8.0}{\meter}) in more stealthy, practical, and scalable real-world scenarios.
The key insight is that various VR interactions involve different loads of network traffic that lead to variations in network transmission, \ie, the power of Wi-Fi signals.
Therefore, we propose \sysname to study the feasibility of the newly identified \major{wireless power-based traffic analysis} attack, which leverages the captured wireless power traces to infer multi-level user privacy within virtual scenes, ranging from \textit{seen} UI-level activities to \textit{unseen} in-game and avatar-based user behaviors.
In \autoref{tab:comparison_prior_works_v4}, we present a thorough comparison between \sysname and relevant VR attacks (\eg, ~\cite{wu2023privacy, slocum2023going, zhang2023s, luo2022holologger, al2021vr, gopal2023hidden, nguyen2024penetration, meteriz2022keylogging, luo2024eavesdropping, su2024remote, wang2024gazeploit, mei2024mmspyvr, ni2025non, shi2021face, zhang2023facereader, li2024dangers, huang2025siren, wei2025vreaves, cayir2025speak, son2025side}) to highlight \major{their} distinct advantages.
First, instead of relying on invasive malware installation \major{like} some established methods~\cite{wu2023privacy, slocum2023going, luo2022holologger, zhang2023s, shi2021face, zhang2023facereader, cayir2025speak}, \sysname utilizes the power side channel inherent to Wi-Fi energy to launch attacks non-intrusively.
Second,  \sysname operates with minimal prerequisites, which neither demand prior knowledge of the VR device or the joining of shared virtual spaces, nor
apply physical constraints such as specific fixed positions (\eg, \cite{gopal2023hidden, al2021vr, luo2024eavesdropping, meteriz2022keylogging, su2024remote, wang2024gazeploit, mei2024mmspyvr, ni2025non}).
Third, \sysname stands out in identifying not only UI-related user behaviors, \eg, generic website browsing, but also unseen VR-specific activities such as in-game activities in VR games, avatar-based user re-identification, \major{and} keystrokes on virtual keyboards.
Last, \sysname achieves a longer distance than previous studies~\cite{al2021vr, huang2025siren, wei2025vreaves, meteriz2022keylogging, gopal2023hidden, nguyen2024penetration, luo2024eavesdropping, ni2025non}, showing nearly $1.8\times$–$80\times$ enhancements.
\looseness=-1

\noindent\textbf{Contributions.} We make the following contributions:
\begin{packeditemize}
    \item We present an exploration of a long-range, through-wall eavesdropping attack that utilizes the wireless power \major{in the network traffic} between the Wi-Fi router and the VR headset to unveil user privacy, providing \major{an orthogonal} attack vector.
    \item We propose the attack, \sysname, by designing a customized wireless power sniffer and applying open-world reconstruction models to infer fine-grained privacy-aware user behaviors, which is \major{a step-forward} study to reveal unseen privacy in VR.
    \item We conducted a series of evaluations on four COTS VR headsets to show \sysname's effectiveness in uncovering multi-level user privacy, including UI-level website/app/in-app activity fingerprinting to unseen VR-specific privacy like in-game actions in \populationone, as well as user's identity and virtual keystrokes from its avatar activities. 
\end{packeditemize}

%% file: tables/comparison_prior_works_v6.tex
\begin{table}[t]
\centering
\scriptsize
\setlength{\tabcolsep}{2.5pt}
\renewcommand{\arraystretch}{1.0}
\caption{Comparing \sysname with existing VR attacks.}
\vspace{-0.1in}
\begin{tabular}{c|c|c|c|c|c}
\toprule
\textbf{AR/VR Attack} & \textbf{Non-intrusive} & \textbf{\begin{tabular}[c]{@{}c@{}}No Prior\\  Knowledge\end{tabular}} & \textbf{\begin{tabular}[c]{@{}c@{}}No Fixed\\ Position\end{tabular}} & \textbf{\begin{tabular}[c]{@{}c@{}}Multi-Level \\ Inference\end{tabular}} & \textbf{\begin{tabular}[c]{@{}c@{}}Attack \\ Distance\end{tabular}} \\ \hline\hline
\multicolumn{1}{c|}{\begin{tabular}[c]{@{}c@{}}Wu \etal~\cite{wu2023privacy}\end{tabular}} & \multicolumn{1}{c|}{\myNocirc} & \multicolumn{1}{c|}{\myNocirc} & \multicolumn{1}{c|}{\myYescirc} & \myNocirc & -- \\ \hline
\multicolumn{1}{c|}{\begin{tabular}[c]{@{}c@{}}TyPose~\cite{slocum2023going}\end{tabular}} & \multicolumn{1}{c|}{\myNocirc} & \multicolumn{1}{c|}{\myNocirc} & \multicolumn{1}{c|}{\myYescirc} & \myNocirc & -- \\ \hline
\multicolumn{1}{c|}{\begin{tabular}[c]{@{}c@{}}HoloLogger~\cite{luo2022holologger}\end{tabular}} & \multicolumn{1}{c|}{\myNocirc} & \multicolumn{1}{c|}{\myNocirc} & \multicolumn{1}{c|}{\myYescirc} & \myNocirc & -- \\ \hline
\multicolumn{1}{c|}{\begin{tabular}[c]{@{}c@{}}LineTalker~\cite{li2024dangers}\end{tabular}} & \multicolumn{1}{c|}{\myNocirc} & \multicolumn{1}{c|}{\myNocirc} & \multicolumn{1}{c|}{\myYescirc} & \myNocirc & -- \\ \hline
\multicolumn{1}{c|}{\begin{tabular}[c]{@{}c@{}}OVRWatcher~\cite{son2025side}\end{tabular}} & \multicolumn{1}{c|}{\myNocirc} & \multicolumn{1}{c|}{\myNocirc} & \multicolumn{1}{c|}{\myYescirc} & \myNocirc & -- \\ \hline
\multicolumn{1}{c|}{\begin{tabular}[c]{@{}c@{}}ImmerSpy~\cite{cayir2025speak}\end{tabular}} & \multicolumn{1}{c|}{\myNocirc} & \multicolumn{1}{c|}{\myNocirc} & \multicolumn{1}{c|}{\myYescirc} & \myNocirc & -- \\ \hline
\multicolumn{1}{c|}{\begin{tabular}[c]{@{}c@{}}BPSniff~\cite{ye2025bpsniff}\end{tabular}} & \multicolumn{1}{c|}{\myNocirc} & \multicolumn{1}{c|}{\myNocirc} & \multicolumn{1}{c|}{\myYescirc} & \myNocirc & -- \\ \hline
\multicolumn{1}{c|}{\begin{tabular}[c]{@{}c@{}}GAZEploit~\cite{wang2024gazeploit}\end{tabular}} & \multicolumn{1}{c|}{\myNocirc} & \multicolumn{1}{c|}{\myYescirc} & \multicolumn{1}{c|}{\myYescirc} & \myNocirc & -- \\ \hline
\multicolumn{1}{c|}{\begin{tabular}[c]{@{}c@{}}Face-Mic~\cite{shi2021face}\end{tabular}} & \multicolumn{1}{c|}{\myNocirc} & \multicolumn{1}{c|}{\myNocirc} & \multicolumn{1}{c|}{\myYescirc} & \myYescirc & -- \\ \hline
\multicolumn{1}{c|}{\begin{tabular}[c]{@{}c@{}}Zhang \etal~\cite{zhang2023s}\end{tabular}} & \multicolumn{1}{c|}{\myNocirc} & \multicolumn{1}{c|}{\myNocirc} & \multicolumn{1}{c|}{\myYescirc} & \myYescirc & -- \\ \hline
\multicolumn{1}{c|}{\begin{tabular}[c]{@{}c@{}}FaceReader~\cite{zhang2023facereader}\end{tabular}} & \multicolumn{1}{c|}{\myNocirc} & \multicolumn{1}{c|}{\myNocirc} & \multicolumn{1}{c|}{\myYescirc} & \myYescirc & -- \\ \hline
\multicolumn{1}{c|}{\begin{tabular}[c]{@{}c@{}}VR-Spy~\cite{al2021vr}\end{tabular}} & \multicolumn{1}{c|}{\myYescirc} & \multicolumn{1}{c|}{\myNocirc} & \multicolumn{1}{c|}{\myNocirc} & \myNocirc & \SI{1.27}{\meter} \\ \hline
\multicolumn{1}{c|}{\begin{tabular}[c]{@{}c@{}}Huang~\etal~\cite{huang2025siren}\end{tabular}} & \multicolumn{1}{c|}{\myYescirc} & \multicolumn{1}{c|}{\myNocirc} & \multicolumn{1}{c|}{\myYescirc} & \myNocirc & $<$~\SI{0.1}{\meter} \\ \hline
\multicolumn{1}{c|}{\begin{tabular}[c]{@{}c@{}}VReaves~\cite{wei2025vreaves}\end{tabular}} & \multicolumn{1}{c|}{\myYescirc} & \multicolumn{1}{c|}{\myNocirc} & \multicolumn{1}{c|}{\myYescirc} & \myNocirc & $<$~\SI{2.0}{\meter} \\ \hline
\multicolumn{1}{c|}{\begin{tabular}[c]{@{}c@{}}Meteriz~\etal~\cite{meteriz2022keylogging}\end{tabular}} & \multicolumn{1}{c|}{\myYescirc} & \multicolumn{1}{c|}{\myNocirc} & \multicolumn{1}{c|}{\myYescirc} & \myNocirc & $<$~\SI{0.8}{\meter}\\ \hline
\multicolumn{1}{c|}{\begin{tabular}[c]{@{}c@{}}Gopal \etal~\cite{gopal2023hidden}\end{tabular}} & \multicolumn{1}{c|}{\myYescirc} & \multicolumn{1}{c|}{\myNocirc} & \multicolumn{1}{c|}{\myYescirc} & \myNocirc & $<$~\SI{4.0}{\meter} \\ \hline
\multicolumn{1}{c|}{\begin{tabular}[c]{@{}c@{}}Intrude~\cite{nguyen2024penetration}\end{tabular}} & \multicolumn{1}{c|}{\myYescirc} & \multicolumn{1}{c|}{\myNocirc} & \multicolumn{1}{c|}{\myYescirc} & \myNocirc & $<$~\SI{4.5}{\meter} \\ \hline
\multicolumn{1}{c|}{\begin{tabular}[c]{@{}c@{}}Heimdall~\cite{luo2024eavesdropping}\end{tabular}} & \multicolumn{1}{c|}{\myYescirc} & \multicolumn{1}{c|}{\myNocirc} & \multicolumn{1}{c|}{\myYescirc} & \myNocirc & $<$~\SI{2.2}{\meter} \\ \hline
\multicolumn{1}{c|}{\begin{tabular}[c]{@{}c@{}}Su \etal~\cite{su2024remote}\end{tabular}} & \multicolumn{1}{c|}{\myYescirc} & \multicolumn{1}{c|}{\myYescirc} & \multicolumn{1}{c|}{\myYescirc} & \myNocirc & -- \\ \hline 
\multicolumn{1}{c|}{\begin{tabular}[c]{@{}c@{}}VRecKey~\cite{ni2025non}\end{tabular}} & \multicolumn{1}{c|}{\myYescirc} & \multicolumn{1}{c|}{\myYescirc} & \multicolumn{1}{c|}{\myYescirc} & \myNocirc & $<$~\SI{4.0}{\meter} \\ \hline
\multicolumn{1}{c|}{\begin{tabular}[c]{@{}c@{}}mmSpyVR~\cite{mei2024mmspyvr}\end{tabular}} & \multicolumn{1}{c|}{\myYescirc} & \multicolumn{1}{c|}{\myYescirc} & \multicolumn{1}{c|}{\myYescirc} & \myNocirc & \SI{2.0}{\meter}--\SI{8.0}{\meter}\\ \hline\hline
\multicolumn{1}{c|}{\begin{tabular}[c]{@{}c@{}}\sysname \end{tabular}} & \multicolumn{1}{c|}{\myYescirc} & \multicolumn{1}{c|}{\myYescirc} & \multicolumn{1}{c|}{\myYescirc} & \myYescirc & \SI{5.0}{\meter}--\SI{8.0}{\meter} \\ \bottomrule
\end{tabular}%
%\vspace{0.05in}
\vspace{-0.25in}
\label{tab:comparison_prior_works_v4}
\end{table}

%% file: sections_major/02_background.tex
\section{Background}
\label{sec:background}

\subsection{Primer on Virtual Reality (VR)}
\label{subsec:primer_on_vr_devices}

A typical VR device (\eg, Meta Oculus Quest series, PICO 4 All-in-One) usually consists of \textit{(i)} a headset that employs a head-mounted display (HMD) with various components to build virtual scenes, \ie, front/corner cameras, embedded sensors, and network communication modules (\eg, Wi-Fi \major{antennas,} and BLE). \textit{(ii)} two handheld controllers with buttons that contain position-tracking and built-in motion sensors (\eg, accelerometer, gyroscope) to achieve various user \major{interactions,} such as pressing icons to run apps and typing \major{on} virtual keyboards to enter accounts or passwords in VR. 
Due to its network-intensive characteristic, a VR system is usually
connected to the Wi-Fi router to access the network to render virtual scenes and provide immersive services and experiences (\eg, app launching, video streaming, and gaming) to the VR user, while the Wi-Fi router transmits heavy downlink/uplink network packets over the air in this process, resulting in apparent changes in network traffic.
\looseness=-1

\begin{figure}[t]
    \centering
	\begin{subfigure}[b]{.49\linewidth}
         \centering
         \includegraphics[width=\linewidth]{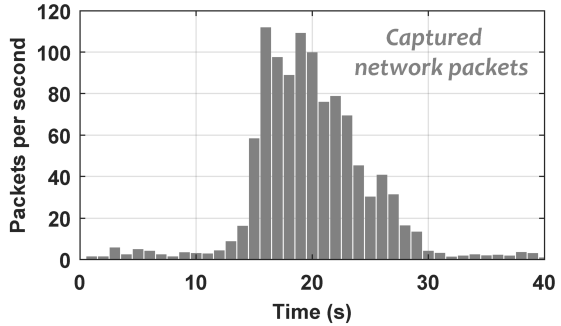}
         \vspace{-0.2in}
         \caption{Captured packets.}
         \label{fig:captured_network_packets}
    \end{subfigure}
    \centering
	\begin{subfigure}[b]{.49\linewidth}
         \centering
         \includegraphics[width=\linewidth]{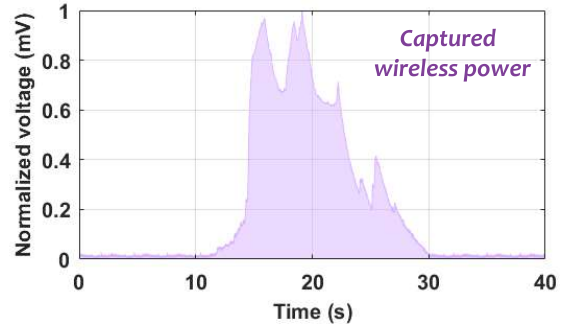}
         \vspace{-0.2in}
         \caption{Captured Wi-Fi energy.}
         \label{fig:harvested_rf_energy_populationone}
    \end{subfigure}
    \vspace{-0.1in}
    \caption{Captured network packets and Wi-Fi power when launching the VR game \populationone.}
    \vspace{-0.2in}
    \label{fig:harvested_rf_energy_network_packets}
\end{figure}

\subsection{Signal Power in Wi-Fi Network Traffic}
\label{subsec:rf_energy_network_traffic}

When connected to the VR headset, the Wi-Fi router radiates Wi-Fi signals into the ambient wireless channels~\cite{kim2014ambient} to transmit data for network services.
In particular, the emitted Wi-Fi signal is a \major{type} of wireless signal with specific frequencies (\eg, \SI{2.4}{\giga\hertz}/\SI{5}{\giga\hertz}), which also carries signal power, denoted as wireless power, during wireless transmission over the air.
Intuitively, as a VR user engages in varied activities within the virtual scenes, such as running apps or gaming, the number of packets in data transmission varies. Consequently, the emitted RF signals have different energy strengths \major{due to} the density of network packets, making wireless power an indicator traffic load changes relevant to corresponding user behaviors.
\looseness=-1

Theoretically, 
when transmitting a set of network packets $\mathcal{S}_{P}$,
assuming the transmitting power of a Wi-Fi router is $P_{t}$, according to Friis' transmission equation~\cite{wadhwa2009impact}, the wireless power $P_{r}$ captured at a distance $d$ can be expressed as:
\looseness=-1
\begin{equation}
\label{eq:friis_transmission_equation}
\small
    \mathcal{S}_{P}\propto P_{r} = P_{t}G_{t}G_{r}(\frac{\lambda}{4\pi})^{2}(\frac{1}{d})^{2} - PL,
\end{equation}

\noindent where $\lambda$ is the wavelength of the RF signal, $G_{t}$ and $G_{r}$ are the \major{gains} of the transmitting antenna and the receiving antenna, respectively. In practice, there are many partitions and obstacles (\eg, walls) that can block the LoS wireless power transmission and result in path losses $PL$, which can be modeled at the distance $d$ as follows:
\looseness=-1
\begin{equation}
\small
\begin{split}
    PL(d)\left[dB\right]&=PL(d_{0})+10nlog(\frac{d}{d_{0}}) +FAF\left[dB\right]\\
    &+p\ast AF_{partition}\left[dB\right]+q\ast AF_{wall}\left[dB\right],
\label{eq:pathlossmodel}
\end{split}
% \smallskip
\end{equation}

\noindent where $PL(d_{0})$ is the path loss at a reference distance (\ie, typically $d_{0}=$\SI{1}{\meter}), $FAF$ is the floor attenuation, $p$ and $q$ are the \major{numbers} of soft partitions (\eg, furniture \major{such as sofas}, and \major{cushions}) and the walls, and $AF_{partition}$ and $AF_{wall}$ are the corresponding attenuation factors, respectively.

When the VR user performs network-related interactions inside the virtual scenes, the extra packets $\Delta \mathcal{S}_{P}$ in the wireless transmission lead to power changes $\Delta P_{t}$ on the side of the Wi-Fi router, which further induces changes in the transmitted wireless power $\Delta P_{r}$ as shown in \autoref{eq:changes_rf_energy}:
\begin{equation}
\label{eq:changes_rf_energy}
% \small
    (\Delta \mathcal{S}_{P} \propto  \Delta P_{t}) \cap (\Delta P_{t} \propto \Delta P_{r}) \Rightarrow \Delta \mathcal{S}_{P} \propto \Delta P_{r}.
% \smallskip
\end{equation}

Therefore, changes in wireless power could reflect the extra network packets induced by VR user activities. \autoref{fig:captured_network_packets} and \autoref{fig:harvested_rf_energy_populationone} separately show the captured network packets using the ALFA AWUS036ACH Wi-Fi sniffer~\cite{alfawifisniffer} and the energy captured by our wireless power sniffer (\autoref{subsec:rf_energy_harvester}) when the VR headset is launching the VR game in \populationone, where the captured wireless energy is proportional to the traffic load of the Wi-Fi network, \ie, data packets per second. Specifically, when the game app is buffering content, \eg, playing animations, or loading game interfaces, we see sharp increases in both network traffic and wireless power. Therefore, since wireless signals are spherically emitted \major{into} the ambient environment, the Wi-Fi power traces can be detected \major{in} any distance $d$ and direction to monitor \major{their} changes \major{and} infer changes in network traffic and the corresponding user privacy.
\looseness=-1

%% file: sections_major/03_threat_model.tex
\section{Threat Model}
\label{sec:threat_model}

\begin{figure}[t]
\centering
    \begin{subfigure}[b]{.49\linewidth}
         \centering
         \includegraphics[width=\linewidth]{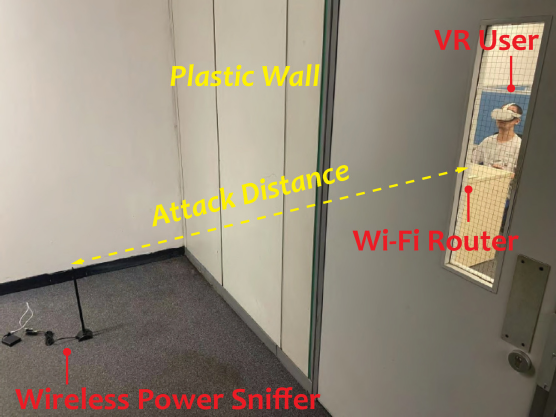}
         \vspace{-0.2in}
         \caption{Scenario 1: Plastic wall}
         \label{fig:scenario_plastic}
    \end{subfigure}
    \centering
	\begin{subfigure}[b]{.49\linewidth}
         \centering
         \includegraphics[width=\linewidth]{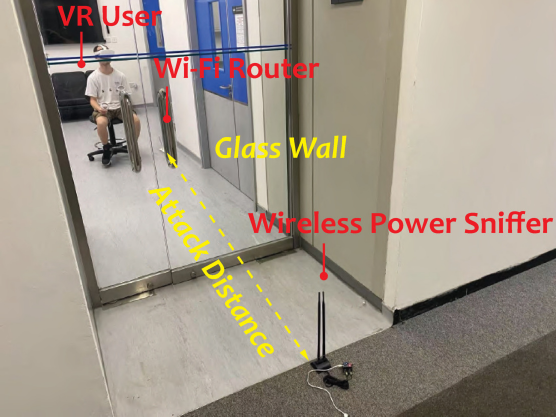}
         \vspace{-0.2in}
         \caption{Scenario 2: Glass wall}
         \label{fig:scenario_glass}
    \end{subfigure}
    \begin{subfigure}[b]{.49\linewidth}
         \centering
         \includegraphics[width=\linewidth]{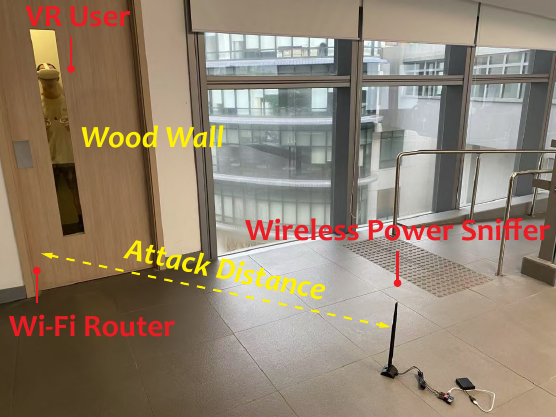}
         \vspace{-0.2in}
         \caption{Scenario 3: Wood wall}
         \label{fig:scenario_wood}
    \end{subfigure}
    \begin{subfigure}[b]{.49\linewidth}
         \centering
         \includegraphics[width=\linewidth]{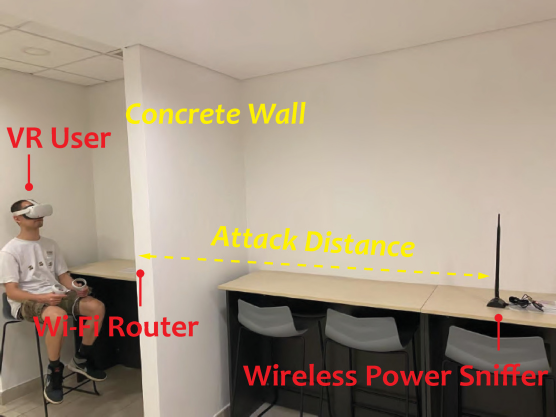}
         \vspace{-0.2in}
         \caption{Scenario 4: Concrete wall}
         \label{fig:scenario_concrete}
    \end{subfigure}
    \vspace{-0.15in}
    \caption{Through-wall attack scenarios.}
    \vspace{-0.2in}
    \label{fig:experiment_setup}
\end{figure}

\begin{figure*}[t]
    \centering
    \includegraphics[width=\linewidth]{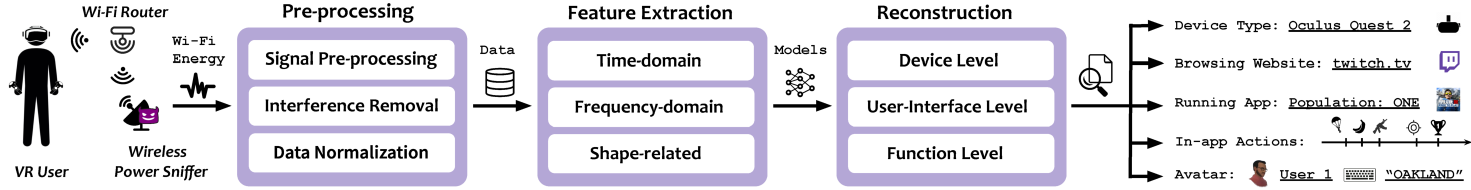}
    \vspace{-0.25in}
    \caption{Overview of \sysname.}
    \vspace{-0.15in}
    \label{fig:system_overview}
\end{figure*}

\noindent\textbf{Attack Scenario.}
As shown in \autoref{fig:motivating_example}, we present
an illustration of long-range and through-wall eavesdropping
attacks to infer fine-grained VR user activities from the captured wireless power traces.
The VR user first connects the VR headset to the Wi-Fi network, browses the websites \twitchtv to watch video streaming, taps the app icon to open the popular VR game \populationone,
plays the game with online players, joins the online meeting in \meetinvr, and interacts with other participants through the avatar's movements (\eg, typing on the virtual keyboard).
These activities change the load (\eg, the number of packets) in the transmission of network traffic.
As mentioned in \autoref{subsec:rf_energy_network_traffic}, changes in network traffic could also lead to changes in the power of transmitted wireless signals, which present detectable features that can be leveraged to recognize corresponding VR user activities.
\looseness=-1

% \begin{figure*}[t]
%     \centering
%     \includegraphics[width=\linewidth]{figures/overview_vrsnoop_v5.pdf}
%     \vspace{-0.25in}
%     \caption{Overview of \sysname.}
%     \vspace{-0.1in}
%     \label{fig:system_overview}
% \end{figure*}

In practice, as shown in \autoref{fig:experiment_setup}, the attacker can place a wireless power sniffer through the wall (\eg, plastic board, wood door, glass wall, and concrete wall) in an adjacent room at a \major{comparatively} longer distance (\eg, \SI{5.0}{\meter}--\SI{8.0}{\meter}) and remotely analyze the captured Wi-Fi power traces to recognize fine-grained VR user activities and unveil multi-level privacy in virtual scenes.
This scenario is prevalent in daily life in various indoor spaces such as homes and offices. It is rational because \textit{(i)} the vast majority of VR apps, multi-player VR games, and avatar-based activities require network connections with the server during interactions to transmit packets containing rendering animations and user movements, and \textit{(ii)} a wireless power sniffer located in the \major{neighboring} room behind the wall can stealthily capture the wireless power radiated from the target Wi-Fi router that \major{connects} to the VR headset.
\looseness=-1

\begin{figure}[t]
    \centering
    \includegraphics[width=.9\linewidth]{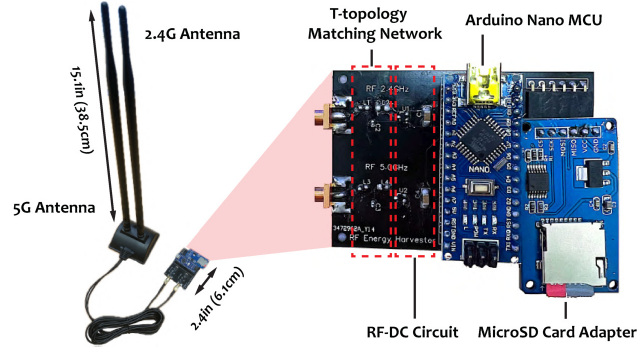}
    \vspace{-0.1in}
    \caption{Prototype of the custom-built wireless power sniffer.  It can capture Wi-Fi power at \SI{2.4}{\giga\hertz} and \SI{5}{\giga\hertz}.}
    \vspace{-0.2in}
    \label{fig:prototype_rf_energy_harvester}
\end{figure}

\noindent\textbf{Attacker's Capabilities.}
Unlike previous AR/VR attacks, we do not assume that the attacker can compromise the VR headset by installing malware to obtain motion sensor data or monitor changes in system workload~\cite{slocum2023going, zhang2023s, wu2023privacy, luo2022holologger, ling2019know, zhang2023facereader, zhang2023passive, cayir2025speak, ye2025bpsniff}. Next, the attacker has no Line-of-Sight (LoS) view of VR users and cannot place cameras or hand trackers to record videos of the victim's head movements and hand gestures directly~\cite{gopal2023hidden, nguyen2024penetration, meteriz2022keylogging,khalili2024virtual}, nor can a suspicious device be placed in close proximity (\eg, less than \SI{2.0}{\meter}) to record button-clicking \major{sounds} or emitted infrared signals from VR controllers~\cite{luo2024eavesdropping, ni2025non}. Moreover, we do not assume that the victim would join unauthorized shared virtual spaces~\cite{meng2023anonymization, wang2024gazeploit} or multi-user VR apps~\cite{su2024remote} to share avatar data publicly (\eg, hand movements, gaze information). Furthermore, the victim can \major{use} the VR device while sitting on the stationary boundary or moving in the playing zone, without the physical restrictions of a fixed position required in other side-channel attacks~\cite{al2021vr, mei2024mmspyvr}. Note that we only assume that the attacker can place the wireless power sniffer in the adjacent room to capture the Wi-Fi signals from the target router and use the measurable wireless power patterns to infer multi-level and fine-grained privacy from the AR/VR users. \major{Following the research line of prior wireless side-channel studies~\cite{al2021vr, mei2024mmspyvr, ni2023eavesdropping}, our threat model inherently requires the attacker to be physically present in a space adjacent to the victim within a limited yet practical proximity (\eg, \SI{5.0}{\meter}--\SI{8.0}{\meter}). This scenario is naturally satisfied in apartments, hotels, and shared offices without requiring same-room device placement or malware installation, although the attack's effectiveness depends on the propagation characteristics of the wireless medium and is influenced by intervening obstacles and co-channel devices in the environment.}
\looseness=-1

%% file: sections_major/04_attack_design.tex
\section{Attack Design}
\label{sec:attack_design}

\begin{figure}[t]
\centering
    \begin{subfigure}[b]{.49\linewidth}
         \centering
         \includegraphics[width=\linewidth]{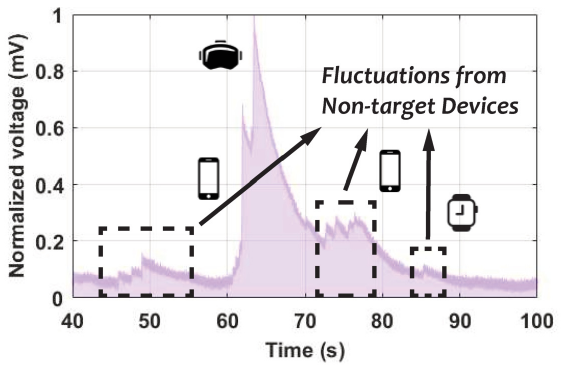}
         \vspace{-0.2in}
         \caption{Before interference removal.}
         \label{fig:before_noise_mitigation}
    \end{subfigure}
    \begin{subfigure}[b]{.49\linewidth}
         \centering
         \includegraphics[width=\linewidth]{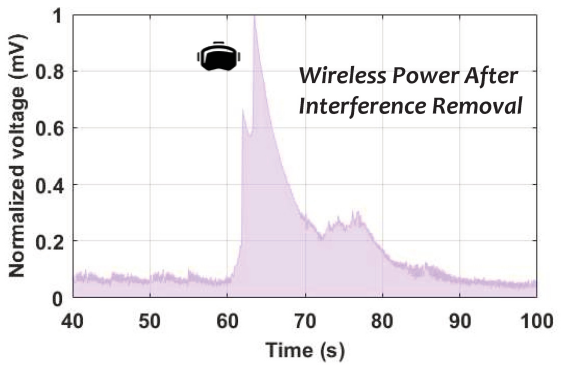}
         \vspace{-0.2in}
         \caption{After interference removal}
         \label{fig:after_noise_mitigation}
    \end{subfigure}
    \vspace{-0.1in}
    \caption{Wireless power before and after removing interference caused by other non-target connected devices.}
    \vspace{-0.25in}
    \label{fig:interference_removal}
\end{figure}

\autoref{fig:system_overview} presents the overview of \sysname. An attacker first places a wireless power sniffer (\autoref{subsec:rf_energy_harvester}) behind the wall of the target Wi-Fi router, which is connected to the VR headset to provide various online services \major{through} network transmission.
When the user wears the VR headset and plays in the virtual scene, the wireless power sniffer captures the transmitted RF signal and converts it \major{into} a measurable voltage signal (\autoref{subsec:rf_energy_harvester}).
Then, the recorded power traces are filtered, segmented, and normalized to obtain the informative signal segments (\autoref{subsec:preprocessing_harvested_voltage}), where \sysname extracts different types of energy-based features for pre-trained open-world inference models to recognize user activities in different virtual scenes (\autoref{subsec:energy_based_feature_extraction}).
Finally, the attacker can reconstruct multi-level, fine-grained, and privacy-related user activities in AR/VR (\autoref{subsec:vr_activity_reconstruction} and \autoref{subsec:vr_user_privacy_inference}),
such as \major{launching} different websites and VR apps,  VR-specific activities that describe unseen scenarios and in-game actions inside the virtual game scenes, as well as \major{the} recognition of user identities and \major{the} uncovering \major{of} virtual keystrokes from users' \major{avatars} (\autoref{subsec:vr_activity_reconstruction} and \autoref{subsec:vr_user_privacy_inference}).
\looseness=-1

% \begin{figure}[t]
%     \centering
%     \includegraphics[width=\linewidth]{figures/prototype_wifi_sniffer_embed.pdf}
%     \vspace{-0.25in}
%     \caption{Prototype of the custom-built wireless power sniffer.  It can capture Wi-Fi power at \SI{2.4}{\giga\hertz} and \SI{5}{\giga\hertz}.}
%     \vspace{-0.25in}
%     \label{fig:prototype_rf_energy_harvester}
% \end{figure}

% \begin{figure}[t]
%     \centering
%     \includegraphics[width=.9\linewidth]{figures/prototype_wifi_sniffer_embed.pdf}
%     \vspace{-0.15in}
%     \caption{Prototype of the custom-built wireless power sniffer.  It can capture Wi-Fi power at \SI{2.4}{\giga\hertz} and \SI{5}{\giga\hertz}.}
%     \vspace{-0.22in}
%     \label{fig:prototype_rf_energy_harvester}
% \end{figure}

\subsection{Custom-built Wireless Power Sniffer}
\label{subsec:rf_energy_harvester}

% Our custom-built wireless power sniffer (\autoref{fig:prototype_rf_energy_harvester}) consists of
% the following items on a customized $2.4$in$\times 1.8$in (\SI{6.1}{\centi\meter}$\times$\SI{4.6}{\centi\meter}) PCB board.
% To achieve longer attack distances compared \major{to} previous studies (\eg, \cite{al2021vr, ni2023eavesdropping, ni2025non, luo2024eavesdropping, mei2024mmspyvr}) and enable through-wall eavesdropping, we select two high-gain dipole antennas with 
% \SI{36}{\deci\bel}i~\cite{antenna24g} and \major{an} impedance of \SI{50}{\ohm} 
% to capture the \SI{2.4}{\giga\hertz} and \SI{5}{\giga\hertz} Wi-Fi \major{signals}, respectively.
Our custom-built wireless power sniffer (\autoref{fig:prototype_rf_energy_harvester}) consists of the following items on a $2.4$in$\times 1.8$in (\SI{6.1}{\centi\meter}$\times$\SI{4.6}{\centi\meter}) PCB board.
To achieve longer attack distances compared \major{to} previous studies (\eg, \cite{al2021vr, ni2023eavesdropping, ni2025non, luo2024eavesdropping, mei2024mmspyvr}) and enable through-wall eavesdropping, we select two high-gain dipole antennas with \SI{36}{\deci\bel}i~\cite{antenna24g} and \major{an} impedance of \SI{50}{\ohm} to capture the \SI{2.4}{\giga\hertz} and \SI{5}{\giga\hertz} Wi-Fi \major{signals}, respectively. \major{Although the dipole antennas measure approximately \SI{15}{\centi\meter} in length each, the entire wireless sniffer assembly fits within a compact footprint of approximately \SI{15}{\centi\meter}$\times$\SI{6}{\centi\meter}$\times$\SI{5}{\centi\meter}, comparable in size to a paperback book. The sniffer operates passively without any signal emission and requires no network connection, which allows it to be concealed inside a bag, drawer, or behind furniture in the adjacent room without detectable electromagnetic or visual signatures, making \sysname significantly more concealable than prior attacks requiring conspicuous cameras~\cite{gopal2023hidden, meteriz2022keylogging}, smartphones~\cite{khalili2024virtual, nguyen2024penetration}, or mmWave radars~\cite{mei2024mmspyvr}.}
In the T-topology matching networks, we choose to configure the inductors $L_{1}=$ \SI{100}{\nano\henry}, $L_{3}$= \SI{50}{\nano\henry}, and $L_{2}=L_{4}=$ \SI{470}{\nano\henry}, and the capacitors $C_{2}=C_{3}=$ \SI{0.01}{\pico\farad}. \major{With} respect \major{to} the on-chip RF-DC converter, we choose the Powercast PCC110~\cite{pcc110} due to its high efficiency in harvesting RF signals at a wide frequency range (\ie, \SI{10}{\mega\hertz} to \SI{6}{\giga\hertz}). The output voltage of the capacitors $C_{1}$ and $C_{4}$ is used as the final output for wireless power at the two mentioned frequencies, where we set the capacitance $C_{1}=C_{4}=$ \SI{1}{\mu\farad} to ensure a noticeable change in the converted voltage signal. Finally, the Arduino Nano MCU takes the voltage signals and stores the data in a \SI{32}{\giga\byte} microSD card.
\looseness=-1

% \begin{figure}[t]
%     \centering
%     \includegraphics[width=\linewidth]{figures/prototype_wifi_sniffer_embed.pdf}
%     \vspace{-0.25in}
%     \caption{Prototype of the custom-built wireless power sniffer.  It can capture Wi-Fi power at \SI{2.4}{\giga\hertz} and \SI{5}{\giga\hertz}.}
%     \vspace{-0.25in}
%     \label{fig:prototype_rf_energy_harvester}
% \end{figure}

\subsection{Pre-processing of Captured Wi-Fi Signal}
\label{subsec:preprocessing_harvested_voltage}

\noindent\textbf{Pre-processing.} After capturing the raw voltage signal from the wireless power sniffer, we apply a low pass Savitzky-Golay (S-G) filter~\cite{chen2004simple} to remove high-frequency noise and smooth the signal while maintaining the original time and frequency domain patterns. We then \major{use} the average values of the first one-second data as the DC offset and \major{deduct} this offset value from the following voltage signals. Since \major{the} converted voltage signals contain both inactive and active voltage changes, we utilize a moving-variance window with \major{an} empirical threshold (\eg, $0.05$) to find the start and end indices of the voltage patterns induced by the burst of the transmitted Wi-Fi energy in network traffic, and then obtain the signal segments that reflect the specific VR activities
of the VR user (\eg, browsing websites, running apps, and playing games).
\looseness=-1

% \begin{figure}[t]
% \centering
%     \begin{subfigure}[b]{.49\linewidth}
%          \centering
%          \includegraphics[width=\linewidth]{figures/before_mitigation_v2.pdf}
%          \vspace{-0.2in}
%          \caption{Before interference removal.}
%          \label{fig:before_noise_mitigation}
%     \end{subfigure}
%     \begin{subfigure}[b]{.49\linewidth}
%          \centering
%          \includegraphics[width=\linewidth]{figures/after_mitigation_v3.pdf}
%          \vspace{-0.2in}
%          \caption{After interference removal}
%          \label{fig:after_noise_mitigation}
%     \end{subfigure}
%     \vspace{-0.1in}
%     \caption{Wireless power before and after removing interference caused by other non-target connected devices.}
%     \vspace{-0.2in}
%     \label{fig:interference_removal}
% \end{figure}

\noindent\textbf{Interference Removal.} In a real-life scenario, a Wi-Fi router usually connects to multiple mobile devices (\eg, smartphones, tablets, and laptops). Hence, these non-target devices could result in extra traffic loads in the wireless network, and the various and unpredictable packet \major{energies} can be picked up by the wireless power sniffer.
To mitigate such interference, we propose and design a generalized interference removal approach based on short-time energy computation with adaptive filtering~\cite{zhang2023facereader}. For each segment of the captured wireless power, our approach continuously calculates the total energy within a designated time window. This energy computation is then compared against a pre-defined power threshold, established during a baseline measurement when only the VR headset is connected to the Wi-Fi router, to identify non-target devices in the wireless traffic. Upon detection of non-target devices, we utilize an adaptive-filter-based method to reduce the interference. 
Specifically, given the wireless power signal $S_{P}(t)$, we implement the adaptive filter by addressing the following optimization problem to determine the adaptive weight vector $\mathcal{V}$:
\looseness=-1
\begin{equation}
\small
\begin{aligned}
& \underset{\mathcal{V}}{\text{argmin}}
& & \sum_{t\in \mathcal{T}}\mathcal{E}(t) = \sum_{t \in \mathcal{T}} \mathcal{R}(t) \log \left( \mathcal{R}(t)/\mathcal{S}^{'}_{P}(t) \right) \\
& \text{subject to}
& & \sum_{t \in \mathcal{T}} \mathcal{V}(t) = \alpha \sum_{t \in \mathcal{T}} \mathcal{V}(t) + \beta \sum_{t \in \mathcal{T}} \mathcal{E}(t) \cdot \mathcal{S}_{P}(t), \\
& & & \sum_{t \in \mathcal{T}} \mathcal{S}^{'}_{P}(t) = \sum_{t \in \mathcal{T}} \mathcal{V}(t) \cdot \mathcal{S}_{P}(t), \sum_{t \in \mathcal{T}} \mathcal{S}_{P}(t)^2 \leq \mathcal{P}.
\end{aligned}
\end{equation}

% \begin{figure}[t]
% \centering
%     \begin{subfigure}[b]{.49\linewidth}
%          \centering
%          \includegraphics[width=\linewidth]{figures/before_mitigation_v2.pdf}
%          \vspace{-0.2in}
%          \caption{Before interference removal.}
%          \label{fig:before_noise_mitigation}
%     \end{subfigure}
%     \begin{subfigure}[b]{.49\linewidth}
%          \centering
%          \includegraphics[width=\linewidth]{figures/after_mitigation_v3.pdf}
%          \vspace{-0.2in}
%          \caption{After interference removal}
%          \label{fig:after_noise_mitigation}
%     \end{subfigure}
%     \vspace{-0.15in}
%     \caption{Wireless power before and after removing interference caused by other non-target connected devices.}
%     \vspace{-0.25in}
%     \label{fig:interference_removal}
% \end{figure}

We choose Kullback-Leibler Divergence ($\mathcal{D}_{KL}$)~\cite{jerfel2021variational} to quantify the difference in the distribution of the signals before and after mitigation and optimize the weights of the adaptive filter. $S^{'}_{P}(t)$ represents the Wi-Fi energy after interference mitigation. Furthermore, $\mathcal{E}$, $\mathcal{R}$, and $\mathcal{T}$ are the error function, the reference signal, and the signal index, and $\alpha$, $\beta$ are the optimization hyperparameters, respectively.
In addition, $\mathcal{P}$ represents the threshold for determining whether the signal segments need to be adaptively filtered.
\autoref{fig:before_noise_mitigation} and \autoref{fig:after_noise_mitigation} demonstrate wireless power before and after interference removal.
Specifically, we connect the Meta Oculus Quest 2 VR headset to the Wi-Fi router with two non-target devices (\eg, an iPhone 13 Pro and an Apple Watch 3) running in default settings, and we can observe that interference from an additional smartphone can be significantly mitigated, showing the effectiveness of the proposed interference removal method.
\looseness=-1

% \noindent\textbf{Data Normalization.} To eliminate the impact of distance and the varied transmitting power of different Wi-Fi routers, we use data normalization methods on segmented voltage signals. Since different user \major{activities} usually \major{present} different lengths of time in every attempt (\eg, loading a website, launching an app), we use min-max scaling to normalize the amplitude of the voltage signals to the range from $0$ to $1$ and utilize up-sampling algorithms such as interpolations~\cite{Olivier_2012} and down-sampling \major{algorithms} (\eg, decimation factor~\cite{kreindler2006effects}) to reshape the voltage signals \major{into} fixed-length segments (\eg, $128\times 1$), and then \major{extract} energy-related features from these normalized segments to train deep learning neural networks for VR user privacy inference.
% \looseness=-1

\noindent\textbf{Data Normalization.} To eliminate the impact of distance and the varied transmitting power of different Wi-Fi routers, we use data normalization methods on segmented voltage signals. Since different user \major{activities} usually \major{present} different lengths of time in every attempt (\eg, loading a website, launching an app), we use min-max scaling to normalize the amplitude of the voltage signals to the range from $0$ to $1$ and utilize up-sampling algorithms such as interpolations~\cite{Olivier_2012} and down-sampling \major{algorithms} (\eg, decimation factor~\cite{kreindler2006effects}) to reshape the voltage signals \major{into} fixed-length segments (\eg, $128\times 1$), and then \major{extract} energy-related features from these normalized segments to train deep learning neural networks for VR user privacy inference. \major{This normalization also makes \sysname inherently robust to varying \textit{Transmit Power Control} (TPC) configurations across router brands, as it strips away absolute power levels and preserves only the relative temporal patterns (\eg, burst shapes, durations, and amplitude ratios) that are driven by application-specific traffic. Furthermore, since TPC typically adjusts on slow timescales (\ie, seconds to minutes) based on link quality, while VR activity bursts occur on millisecond timescales, the two are clearly separable in the normalized signal.}
\looseness=-1

\subsection{Energy-based Feature Extraction}
\label{subsec:energy_based_feature_extraction}

As the processed voltage signals are one-dimensional time series, we extract a total of $15$ time-domain features and $2$ frequency-domain features from the segments of voltage signals.
The time-domain features include the mean, variance, median, maximum, minimum, range, root mean square, mean absolute deviation, coefficient of variation, interquartile range, skewness, and kurtosis. 
These statistical features encode the temporal (\eg, amplitude, rising/falling time) and spatial (\eg, position, peaks/troughs) features to describe the burst of Wi-Fi energy in the transmitted network traffic, which can reflect distinctive patterns of VR user activity in the virtual scenes.
For instance, when the VR user runs different apps, the launching animations take different time durations, and the network traffic loads, while the captured wireless power presents high \major{amplitudes} and drastic fluctuations.
Furthermore, to reduce the interference from other Wi-Fi-connected devices (\eg, smartphones, tablets), we extract another three features to describe the outline shape of the voltage signal in the time domain, including \textit{(i)} shape factor, \textit{(ii)} impulsive factor, and \textit{(iii)} crest factor, as shown in \autoref{eq:shape_factors}:
\looseness=-1
\begin{equation}
\label{eq:shape_factors}
\small
\underbrace{\frac{s_{rms}}{\frac{1}{N}\sum_{1}^{N}{\left |s_{i}  \right |}}}_{\mathclap{\text{Shape Factor}}}
\hphantom{\underbrace{\frac{s_{peak}}{\frac{1}{N}\sum_{1}^{N}{\left |s_{i}  \right |}}}_{\mathclap{\text{Impulsive Factor}}}}
\hfill
\underbrace{\frac{s_{peak}}{\frac{1}{N}\sum_{1}^{N}{\left |s_{i}  \right |}}}_{\mathclap{\text{Impulsive Factor}}}
\hphantom{\underbrace{\frac{s_{peak}}{\sqrt{\frac{1}{N}\sum_{1}^{N}{s^{2}_{i}}}}}_{\mathclap{\text{Crest Factor}}}}
\hfill
\underbrace{\frac{s_{peak}}{\sqrt{\frac{1}{N}\sum_{1}^{N}{s^{2}_{i}}}}}_{\mathclap{\text{Crest Factor}}},
\end{equation}
where $s_{rms}$ and $s_{peak}$ represent the root mean square and the peak value of segment $s$, and $s_{i}$ is the $i$th feature point.
Note that these three shape-related features maintain the informative patterns of the voltage signals to mitigate signal distortions resulting from extra wireless power induced by other devices that connect to the same Wi-Fi router.
In addition, we apply the \textit{Fast Fourier Transform} (FFT) to the signal segments to extract frequency-domain features. These features encompass the mean and entropy of wireless power~\cite{shi2021face}, effectively capturing the periodic characteristics of network traffic behaviors induced by various activities of the VR user.
In the end, the extracted energy-based features are fed into a pre-trained classifier for virtual activity reconstruction.
\looseness=-1

\subsection{Privacy Reconstruction Model}
\label{subsec:vr_activity_reconstruction}

\noindent\textbf{Model Architecture and Training.} The primary aim of \sysname's model training process is to interpret the learned time-series features and identify actions that could potentially compromise VR users' privacy. In particular, a spike in the captured wireless power traces corresponds to an immediate increase in network traffic. This observation indicates that both preceding and succeeding segments around a given time window offer valuable \major{contextual} information. Considering the value of this bidirectional context, we utilize the Bidirectional Long Short-Term Memory (Bi-LSTM) neural network as the backbone for our reconstruction model. This method is proficient in processing time-series signals~\cite{maia2021can} and transforming extracted feature vectors into identifiable VR information.
% \looseness=-1

\begin{figure*}[t]
%\centering
    \minipage{0.62\linewidth}
    \centering
    \begin{subfigure}[b]{.48\linewidth}
         \centering
         \includegraphics[width=\linewidth]{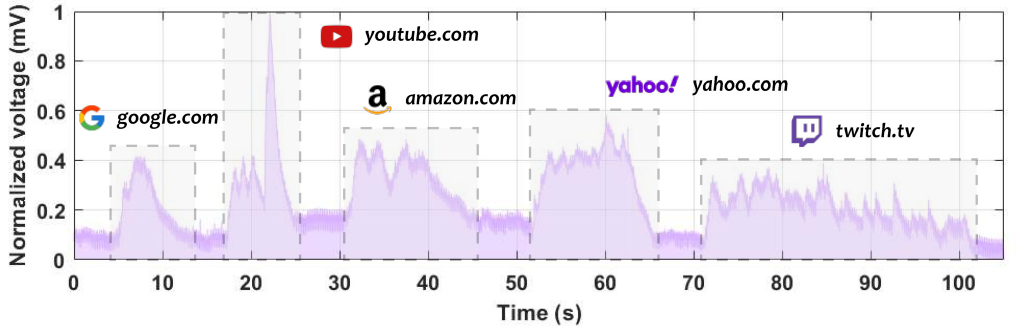}
         \vspace{-0.1in}
         \caption{Browsing different websites.}
         \vspace{0.08in}
         \label{fig:pi_website_fingerprinting}
    \end{subfigure}
    \centering
	\begin{subfigure}[b]{.48\linewidth}
         \centering
         \includegraphics[width=\linewidth]{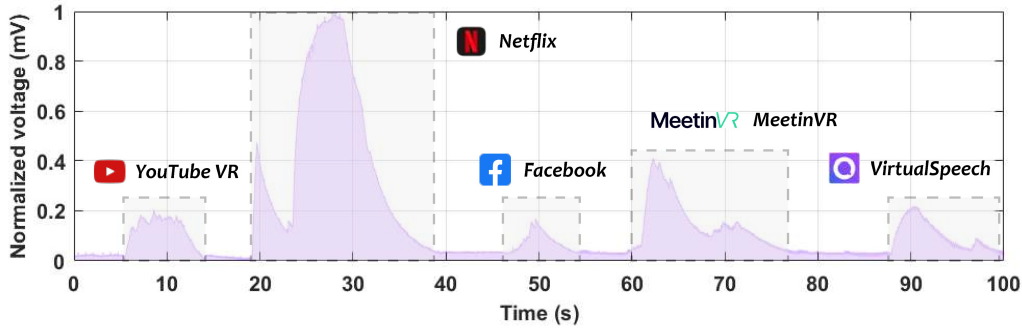}
         \vspace{-0.1in}
         \caption{Launching different VR apps.}
         \vspace{0.08in}
         \label{fig:pi_app_fingerprinting}
    \end{subfigure}
    \begin{subfigure}[b]{.48\linewidth}
         \centering
         \includegraphics[width=\linewidth]{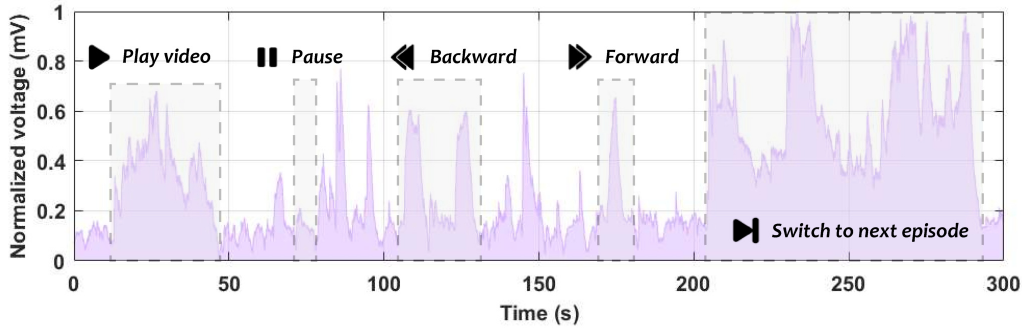}
         \vspace{-0.1in}
         \caption{Performing activities in \netflix.}
         \label{fig:pi_inapp_activity_recognition}
    \end{subfigure}
    \begin{subfigure}[b]{.48\linewidth}
         \centering
         \includegraphics[width=\linewidth]{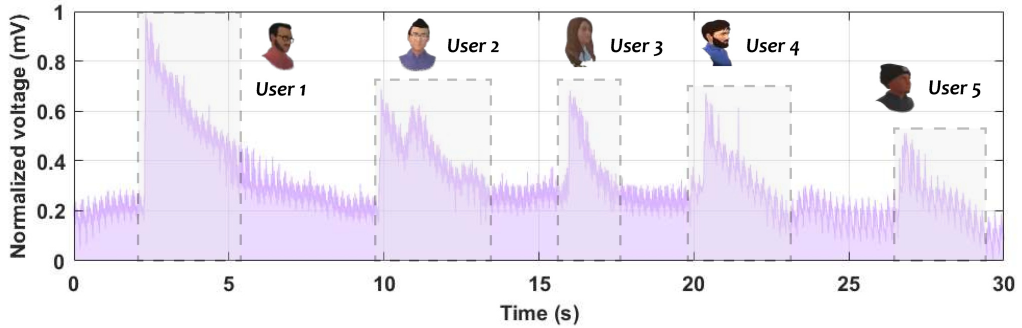}
         \vspace{-0.1in}
         \caption{Generating avatars in \meetinvr.}
         \label{fig:pi_user_identification}
    \end{subfigure}
    \vspace{-0.1in}
    \caption{Captured Wi-Fi power when performing different activities in VR.}
    % \vspace{-0.1in}
    \label{fig:pi_different_activities}
    \endminipage
    \hfill
    \minipage{0.37\linewidth}%
    \centering
    \includegraphics[width=\linewidth]{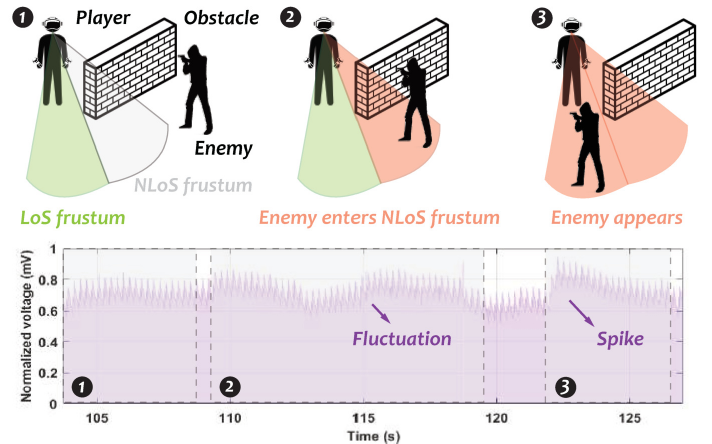}
    \vspace{-0.2in}
    \caption{Unseen VR-specific activity (\eg, ``enemy hide'' $\rightarrow$  ``enemy appear'') in the VR game \populationone.}
    \vspace{-0.15in}
    \label{fig:unseen_vr_specific_activity}
    \endminipage
    % \vspace{-0.1in}
\end{figure*}
\begin{figure*}
%\centering
  %   \minipage{0.245\textwidth}
  %   \centering
  %     \includegraphics[width=\linewidth]{figures/privacy_inference_af.pdf}
  %     \vspace{-0.2in}
  %     \caption{Wireless power when launching different apps.}
  %     \vspace{0.05in}
  %     \label{fig:pi_app_fingerprinting}
  %   \endminipage\hfill
  %   \minipage{0.245\textwidth}
  %   \centering
  %     \includegraphics[width=\linewidth]{figures/privacy_inference_iaf.pdf}
  %     \vspace{-0.2in}
  %     \caption{Wireless power when performing activities in \netflix.}
  %     \vspace{0.05in}
  % \label{fig:pi_inapp_activity_recognition}
  %   \endminipage\hfill
    \minipage{.58\textwidth}
    \centering
      \includegraphics[width=\linewidth]{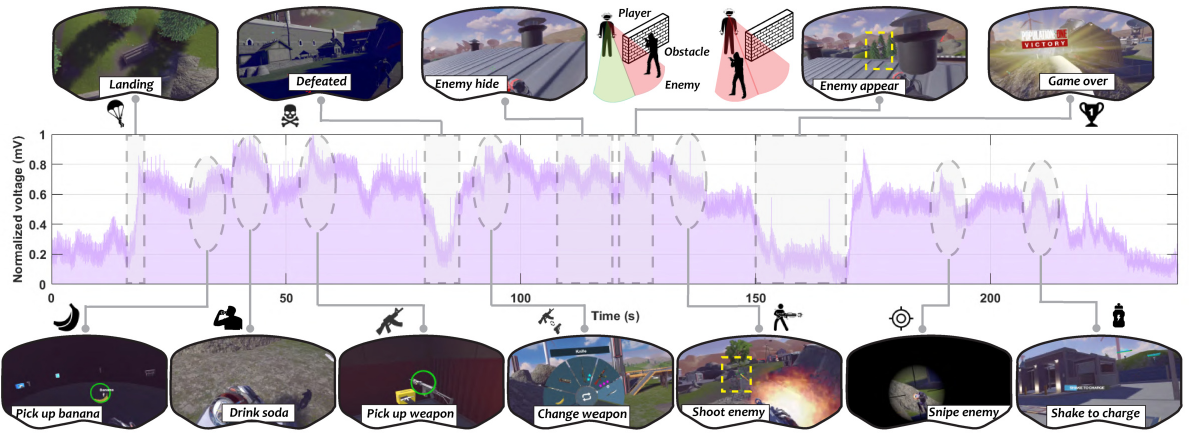}
      \vspace{-0.25in}
      \caption{Wireless power when performing various VR-specific activities in the VR game \populationone, involving five different game scenarios (\ie, landing, defeated, game over, enemy hide,\etc) and seven different in-game actions (\ie, pick up a banana, pick up a weapon, change weapon, drink soda, shoot enemy, snipe enemy, \etc).}
      \vspace{-0.15in}
  \label{fig:pi_vr_specific_activity_recognition}
    \endminipage\hfill
    \minipage{0.41\textwidth}
    \centering
    \includegraphics[width=\linewidth]{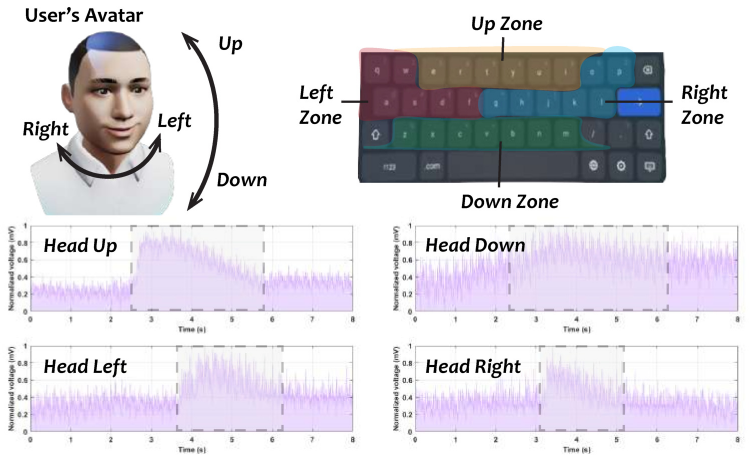}
    \vspace{-0.25in}
    \caption{Patterns of captured wireless power in the user avatar's head movements, which correspond to typing zones in the virtual keyboard.}
    \vspace{-0.1in}
    \label{fig:pi_keystroke_inference}
    \endminipage\hfill
    % \vspace{-0.1in}
\end{figure*}

% In the Bi-LSTM neural network, 
% an embedding layer takes in the input sequences and generates a numerical vector.
% Then, we utilize two layers of bidirectional LSTM with the size of $128$ to process the input data, and a dropout layer with a dropout rate of $50\%$ to regularize the network and prevent overfitting, where the input data is a sliding window of the feature vector extracted from the recorded voltage signal from the time series.
% Then, it \major{is followed by} a fully-connected layer to classify the feature vectors \major{into} the predicted class with the highest probability.
% For \major{the} training processes, we tailored the \major{constraints} of loss functions to suit different configurations of the user privacy inference task. 
% In practice, we utilize $80\%$ data to train the Bi-LSTM model and evaluate \sysname with the remaining $20\%$ data, \major{ensuring that} there is no overlap between the training and testing datasets.
% Finally, we transfer the trained Bi-LSTM model to an embedding model by removing the layers after the embedding layer and \major{use} the embedding model with a cosine distance to classify captured wireless power into different patterns, each of which is \major{associated} with a unique activity in the virtual scenes that contain different levels of user privacy.
% \looseness=-1
In the Bi-LSTM neural network,
an embedding layer takes in the input sequences and generates a numerical vector.
Then, we utilize two layers of bidirectional LSTM with the size of $128$ to process the input data, and a dropout layer with a dropout rate of $50\%$ to regularize the network and prevent overfitting, where the input data is a sliding window of the feature vector extracted from the recorded voltage signal from the time series.
Then, it \major{is followed by} a fully-connected layer to classify the feature vectors \major{into} the predicted class with the highest probability.
For \major{the} training processes, we tailored the \major{constraints} of loss functions to suit different configurations of the user privacy inference task.
\major{In practice, for each privacy inference task, we randomly partition the collected dataset into a training set ($80\%$) and a testing set ($20\%$) per task, and we ensure there is no sample overlap between the two sets. The randomized split is performed independently for each of the different privacy inference tasks to avoid any data leakage during evaluation.}
Finally, we transfer the trained Bi-LSTM model to an embedding model by removing the layers after the embedding layer and \major{use} the embedding model with a cosine distance to classify captured wireless power into different patterns, each of which is \major{associated} with a unique activity in the virtual scenes that contain different levels of user privacy.
\major{While FFT and Bi-LSTM are well-established techniques, \sysname's contribution lies in their end-to-end integration with custom dual-band hardware, adaptive interference removal, and an open-world embedding model to enable long-range, through-wall, multi-level VR privacy inference.}
\looseness=-1

\noindent\textbf{Open-world Privacy Inference.} Since our algorithm requires a threshold on the cosine distance to classify an item, we propose an adaptive threshold mechanism to ensure its scalability.
Specifically, we let $A_{T}=\{item^{i}_{T}\}^{m_{T}}_{i=1}$ (resp. $A_{I}=\{item^{i}_{I}\}^{m_{I}}_{i=1}$) be the set consisting of all activities in the training stage (resp. the identification stage). 
In particular, $A_{I}$ is the subset of $A_{T}$ ($A_{I}\subseteq A_{T}$) in the closed-world setting, while it could contain items that are unmonitored in the training stage ($A_{I}\nsubseteq  A_{T}$) in the open-world setting. Next, we choose the threshold based on the training set property.
Specifically, we first produce the threshold set $T = \{threshold^{i}_{T}\}^{m_{T}}_{i=1}$ for each closed-world $item^{i}_{T}$ class in the training stage. 
Next, we select the maximum threshold value $T_{max} = maximum\{T\}$ as the approximate open-world threshold. 
Accordingly, if the cosine distance of a new item exceeds $T_{max}$, the embedding model will classify it as an unmonitored item; otherwise, it will be classified as a monitored closed-world class $item^{i}_{T}$ if their distance is shorter than $T_{i}$.
In privacy inference, the embedding model generates the embedding $e_{t}$ of a new sequence of the wireless power signal $s_{t}$, and calculates its cosine distance to each embedded sequence $s_{i}^{j}$ of a class $C_{i}$. Hence, $s_{t}$ will be classified into class $C_{i}$ if one of the cosine \major{distances} between $s_{t}$ and $s_{i}^{j}$ is lower than the threshold.
\major{This embedding-based design also scales to significantly larger sets of applications or websites without full retraining, as new classes can be incrementally added by computing their embeddings from a few samples and updating $T$.}
\looseness=-1

\subsection{Multi-Level Privacy Inference}
\label{subsec:vr_user_privacy_inference}

We specifically focus on the scope of traffic-related privacy inference and
systematically understand \sysname's attack capabilities
in uncovering textbook-defined virtual user privacy~\cite{pal2020iot}.
\major{Note that the wireless power patterns illustrated in \autoref{fig:pi_different_activities}~(a)--(d) are representative single-user examples collected from one participant using the Meta Oculus Quest 2, intended to visually demonstrate the distinctive patterns associated with different VR activities.}
\looseness=-1

\noindent\textbf{(I) Device Level: Device Identification (DI).} Prior VR attacks~\cite{wu2023privacy, slocum2023going, zhang2023s, luo2022holologger, al2021vr, gopal2023hidden, nguyen2024penetration, meteriz2022keylogging, luo2024eavesdropping} usually assume the attacker knows the type of the victim's VR device to pre-install malware through \major{the} corresponding SDKs/APIs, whereas it is impractical and increases the difficulties of launching attacks in a real-life scenario.
Based on \autoref{eq:changes_rf_energy}, we know that when a VR headset connects to the Wi-Fi network, changes in network traffic will induce a burst of transmitted wireless power.
Therefore, it is reasonable to exploit the wireless power to identify different VR headsets and other \major{connected} mobile devices since they require a variety of traffic loads after the Wi-Fi connection.
\looseness=-1

\noindent\textbf{(II-1) UI Level: Website Fingerprinting (WF).} After detecting the connection of the VR headset, \sysname can then recognize various activities of VR users in the virtual scenes, \ie, opening the browser to browse websites.
The reason is that different websites have versatile rendering layouts and resource requests~\cite{sirinam2018deep} from the servers, resulting in different network traffic loads, as well as distinguishable patterns of captured wireless power, as shown in \autoref{fig:pi_website_fingerprinting}.
We noticed that the burst of captured wireless power is slight when opening the website \googlecom compared to opening the other three websites \youtubecom, \amazoncom, and \yahoocom, which load many figures and links to render the whole webpage.
When browsing the video streaming website \twitchtv, it automatically loads the video streaming, leading to continuous and drastic changes. 
% in the Wi-Fi energy.
\looseness=-1

\noindent\textbf{(II-2) UI Level: App Fingerprinting (AF).} 
Most COTS VR headsets (\eg, Meta Quest series, PICO 4) have developed their own app \major{stores} to provide VR users with thousands of VR apps from different categories, \major{most} of which require \major{a} Wi-Fi connection to support versatile functionalities and immersive experiences.
Hence, when running VR apps, the VR headset requests additional network traffic to transmit more packets~\cite{li2022foap}, which also changes the transmitted Wi-Fi energy.
\autoref{fig:pi_app_fingerprinting} shows the captured wireless power when launching five VR apps from different categories, where we find the power \major{is} highest when launching \netflix because it not only presents high-resolution image-enhanced video streaming interfaces but also renders the surroundings in cinema mode, leading to intensive traffic transmission and high Wi-Fi power.
In contrast, apps such as \facebook, load fewer images and more text information, presenting mild changes in Wi-Fi power.
\major{Therefore, The leakage of website browsing histories and installed/used VR apps is directly actionable, as it reveals users' interests, lifestyle, and consumption patterns (\eg, fitness routines, dating, religious or political content), which can be exploited for behavioral profiling, targeted advertising, or social engineering attacks.}
\looseness=-1

\noindent\textbf{(III-1) Function Level: Common In-app Activity (IAR).} We further investigate the patterns of captured wireless power when performing different in-app activities in a specific app. For example, \autoref{fig:pi_inapp_activity_recognition} shows the recorded voltage signal when we performed five common in-app activities inside the VR app \netflix. Among the five activities, the captured wireless power drastically changes when playing the video and switching to the next episode, \major{as} these in-app activities require more multimedia content delivery from the content/service provider to the VR headset.
In particular, VR apps like \netflix exploit Adaptive BitRate (ABR) streaming~\cite{stockhammer2011dynamic, sodagar2011mpeg} for content delivery, employing concurrent TCP or QUIC/UDP flows to deliver multiple data chunks simultaneously. Meanwhile, the VR headset uses buffers to store received chunks~\cite{gutterman2019requet}, and \major{requests} data chunks as fast as the Wi-Fi network can deliver to increase the buffer level, resulting in \major{an} intensive traffic load and high RF energy transmitted from the Wi-Fi router. 
Thus, it is feasible for \sysname to leverage the captured wireless power to uncover in-app activities.
\looseness=-1

\noindent\textbf{(III-2) Function Level: VR-specific Activity in VR Games (VAR).}
\autoref{fig:pi_vr_specific_activity_recognition} presents the captured wireless power when we play a popular battle royale game \major{on the} Oculus Quest 2, \populationone, which has over $10$k ratings.
Obviously, changes in the wireless power patterns could reflect the corresponding VR-specific activities, including five different game scenarios (\eg, landing, defeated, game over) and seven in-game actions (\eg, picking up a weapon, shooting \major{an} enemy).
Apart from recognizing VR-specific activities in the user's LoS view,
\sysname can also recognize unseen VR-specific activities that \major{occur} in the Non-Line-of-Sight (NLoS) view at the enemy side of other online players.
\autoref{fig:unseen_vr_specific_activity} shows an example of unseen VR-specific activities in the \populationone, which consists of three scenarios: \ding{182} an enemy hides behind an obstacle (\eg, wall) and is out of the NLoS frustum; \ding{183} the enemy enters the NLoS frustum (\textit{red area}) while still \major{remaining} unseen to the VR user (\textit{green area}), but the VR headset has requested more data transmission from the Wi-Fi network to pre-render the game scenes, resulting in fluctuations in the captured wireless power; and \ding{184} the enemy appears and enters the user's LoS view, leading to a spike in the patterns.
\looseness=-1

% \noindent\textbf{(III-3) Function Level: Avatar-based User Identification.} Apart from various VR games, VR also provides shared virtual spaces for real-time online meetings.
% In particular, the participants will be transformed into their avatars, digital representations of the users in the virtual scenes, to achieve engagement and \major{interaction} with others.
% Nevertheless, the rendering process of a user's avatar requires extra traffic loads, which inevitably \major{cause} fluctuations in the Wi-Fi energy in the wireless network.
% \autoref{fig:pi_user_identification} depicts the wireless power captured when five users enter the shared meeting room in \meetinvr, where we observe that different avatars present different patterns in wireless energy.
% Hence, an adversary could use Wi-Fi power as an indicator to distinguish different users and achieve re-identification attacks without compromising the VR headset to obtain virtual camera data or other sensor data in previous work~\cite{meng2023anonymization, yang2024can, tricomi2023you, sabra2024anonymizing}.
% \looseness=-1

\noindent\textbf{(III-3) Function Level: Avatar-based User Identification.} Apart from various VR games, VR also provides shared virtual spaces for real-time online meetings.
In particular, the participants will be transformed into their avatars, digital representations of the users in the virtual scenes, to achieve engagement and \major{interaction} with others.
Nevertheless, the rendering process of a user's avatar requires extra traffic loads, which inevitably \major{cause} fluctuations in the Wi-Fi energy in the wireless network.
\autoref{fig:pi_user_identification} depicts the wireless power captured when five users enter the shared meeting room in \meetinvr, where we observe that different avatars present different patterns in wireless energy.
Hence, an adversary could use Wi-Fi power as an indicator to distinguish different users and achieve re-identification attacks without compromising the VR headset to obtain virtual camera data or other sensor data in previous work~\cite{meng2023anonymization, yang2024can, tricomi2023you, sabra2024anonymizing}, \major{which is directly actionable as it enables linking real identities across sessions, tracking individuals' presence and social interactions in supposedly anonymous virtual spaces, and facilitating stalking or de-anonymization of sensitive users.}
\looseness=-1

\noindent\textbf{(III-4) Function Level: Avatar-based Keystroke Inference.} In practice, when rendering the user's avatar \major{in} the virtual environment, the user's hand movement data collected from the VR headset's motion sensors are also transmitted, thereby inducing extra changes in the captured wireless power.
However, such posture information reflected on the avatars could be maliciously exploited as a \major{new} attack vector to infer the user's head movement \major{and} uncover user privacy, \ie, keystrokes on virtual keyboards, as illustrated by previous orientation-based privacy inference works~\cite{slocum2023going}.
\autoref{fig:pi_keystroke_inference} shows the patterns of captured wireless power when the avatar's head rotates \major{in} four different directions inside the \meetinvr scenario (\eg: head up, head down, head left, and head right), where an adversary could exploit this power side channel to infer head movements, as well as further keystroke inference on the virtual keyboard. Specifically, we can separate the whole keyboard into four zones corresponding to different head movements and generate \major{candidates} for keystrokes.
\major{In particular, AKI poses the most severe and directly actionable threat: inferred virtual keystrokes can expose passwords, banking credentials, and private messages.
% , which enables severe consequence like financial fraud beyond the UI-level inferences in prior work~\cite{ling2019know, mei2024mmspyvr, slocum2023going}.
}
\looseness=-1

%% file: sections_major/05_evaluation.tex
\section{Evaluation}
\label{sec:evaluation}

\subsection{Experimental Setup}
\label{subsec:experimental_setup}

For the setup of the experiment, we considered four common indoor scenarios: home, office, corridor, and cafe. For the home and corridor, we assume that there is only one Wi-Fi router to which the VR headset is connected.
For the office and cafe, we assume that there are multiple Wi-Fi routers and that the VR user connects the VR headset (Meta Oculus Quest 2) to the target Wi-Fi router, while other routers act as interference sources.
Specifically, we placed the custom-built wireless power sniffer behind a plastic wall with \SI{6.7}{\centi\meter} thickness and \major{captured} wireless power from the Wi-Fi signals emitted from the target router (Huawei Pro 2) at a primary distance of \SI{5.0}{\meter} and collected the  recorded voltage signals.
\major{Note that all evaluations are conducted under Wi-Fi 5 (IEEE 802.11ac), the dominant protocol supported by all tested VR headsets. The underlying principle of \sysname remains valid under Wi-Fi 6/7 because the aggregate RF power emitted by the access point still scales with total downlink traffic volume regardless of sub-channel allocation.}
The sampling rate of the Arduino Nano MCU is set to \SI{20}{\kilo\hertz}, and the collected \major{wireless} power signals are stored \major{on} a \SI{32}{\giga\byte} SD card. Moreover, all data processing is performed on a desktop running Windows 10 with 32GB memory, an Intel \major{i7-9700K} CPU, and an NVIDIA GeForce RTX 2080Ti GPU.
\looseness=-1

\subsection{Data Collection}
\label{subsec:data_collection}

% This study has obtained IRB approval from our institution for data collection. We recruited $20$ participants ($10$ males, $10$ females) and built seven datasets
% in different experimental scenarios for each participant to demonstrate \sysname's effectiveness in \autoref{subsec:overall_effectiveness}.
This study has obtained IRB approval from our institution for data collection. We recruited 20 participants ($10$ males, $10$ females) \major{primarily from the university community, including students and staff with diverse backgrounds. 
% As shown in \autoref{tab:participant_demographics}, 
Specifically, participants ranged in age from $20$ to $53$ years and represented a variety of ethnicities (\ie, Asian, European, Latin American, Middle Eastern, and Australian) and prior VR experience levels, \ie, from no prior exposure to over six years of regular use. This diversity helps ensure that \sysname's effectiveness is evaluated across users with varying physical characteristics and interaction styles.} 
\major{Next, we} built seven datasets in different experimental scenarios for each participant to demonstrate \sysname's effectiveness in \autoref{subsec:overall_effectiveness}.
We asked participants to play with the VR devices for $30$ minutes to familiarize themselves with \major{VR typing} before the official data collection, and all participants were informed that only the Wi-Fi power signals in the VR wireless transmission would be recorded to infer their activities.
The datasets for each participant are constructed from four COTS VR devices ($D_{1}$--$D_{4}$: Meta Oculus Quest 2, PICO 4 All-in-One, Meta Quest 3, and Meta Quest Pro) and six other mobile devices ($D_{5}$--$D_{10}$: iPhone 13 Pro, iPhone 6S, OnePlus 10 Pro, iPad Pro, Apple Watch 3, and HomePod mini), to train embedding models in \sysname.
% and evaluate their performance.
\looseness=-1

\begin{packeditemize}
    \item $\mathcal{D}_{DI}$: We collected data samples from four VR devices and six other mobile devices. We collect samples of the recorded voltage signals by connecting each device to the target Wi-Fi router. This procedure is repeated $500$ times, and then the sliding window in \autoref{subsec:preprocessing_harvested_voltage} is applied to obtain the segmented samples ($10\times 500$ traces).
    \item $\mathcal{D}_{WF}$: The website fingerprinting dataset is also collected from the above four VR devices, where the VR user utilizes the default web browsers (\eg, \metaquestbrowser) of the tested VR devices to open $50$ most-visited websites~\cite{mostvisitedwebsites} 
    ($2\times 50\times 500$ traces).
    \item $\mathcal{D}_{AF}$: The app fingerprint dataset is collected from a total of $50$ apps from the official Oculus and PICO stores, which contain Wi-Fi-connected VR apps of varying categories, playing mode, user requirements, and ratings, and we collect samples of captured Wi-Fi power traces during the first five seconds of launching each app and repeat it $100$ times ($50\times 100\times 2$ traces).
    \item $\mathcal{D}_{IAR}$: We collected in-app activities from five VR apps (\youtubevr, \netflix, \facebook, \meetinvr, and \virtualspeech) in different categories. We perform five common in-app activities in each app (\eg, play videos in \youtubevr) for $100$ times to collect samples for in-app activity recognition ($5\times 5\times 100$ traces).
    \item $\mathcal{D}_{VAR}$: We collected data samples when we play the VR game \populationone in Oculus Quest 2. Specifically, we focus on five different scenarios (\eg, landing, enemy hide) and seven VR user activities (\eg, picking up weapons, shooting enemy, shaking to charge) in one round of the immersive VR game. We conduct $50$ rounds of the game and collect recorded voltage signals, where we select $10$ segments of the recorded voltage signals corresponding to the specific scenarios and in-game actions inside each round game ($5\times 50\times 10$ scenario recognition, $7\times 50\times 10$ in-game activities).
    \item $\mathcal{D}_{AUD}$: We collected data samples when different users enter into \meetinvr online meeting. Specifically, we customized $50$ avatars with different outlooks and collected recorded voltage signals that contain the user identity information, and each process was repeated for $100$ rounds ($50\times 100$ traces).
    \item $\mathcal{D}_{AKI}$: We build the avatar-based keystroke inference dataset by asking participants to enter the $26$ alphabetic keys on the default virtual keyboard inside \meetinvr. During this process, we collect the recorded voltage signals that represent the head movements of the user's avatar, which also corresponds to the specific key zones. To address natural variability in postures when typing at the same zone, we repeat it $100$ times ($26\times 100$ traces).
\end{packeditemize}

% \input{tables/user_demographics}

% \begin{figure*}[t]
%     \centering
% 	\begin{subfigure}[b]{.3\linewidth}
%          \centering
%          \includegraphics[width=\linewidth]{figures/overall_effectiveness_user_identification_2025.pdf}
%          \vspace{-0.2in}
%          \caption{Avatar-based user identification.}
%          \label{fig:overall_user_deanonymization_results}
%     \end{subfigure}
%     \centering
% 	\begin{subfigure}[b]{.35\linewidth}
%          \centering
%          \includegraphics[width=\linewidth]{figures/overall_effectiveness_keystroke_inference_2025.pdf}
%          \vspace{-0.2in}
%          \caption{Avatar-based keystroke inference.}
%          \label{fig:overall_keystroke_inference_results}
%     \end{subfigure}
%     \centering
% 	\begin{subfigure}[b]{.3\linewidth}
%          \centering
%          \includegraphics[width=\linewidth]{figures/topk_results_2025.pdf}
%          \vspace{-0.2in}
%          \caption{Top-k AKI.}
%          \label{fig:overall_keystroke_inference_top_results}
%     \end{subfigure}
%     \vspace{-0.1in}
%     \caption{Avatar-based user privacy inference results.}
%     \vspace{-0.1in}
%     \label{fig:avatar_level_inference}
% \end{figure*}

\subsection{Overall Effectiveness of \sysname}
\label{subsec:overall_effectiveness}

\begin{figure}[t]
    \centering
	\begin{subfigure}[b]{.3\linewidth}
         \centering
         \includegraphics[width=\linewidth]{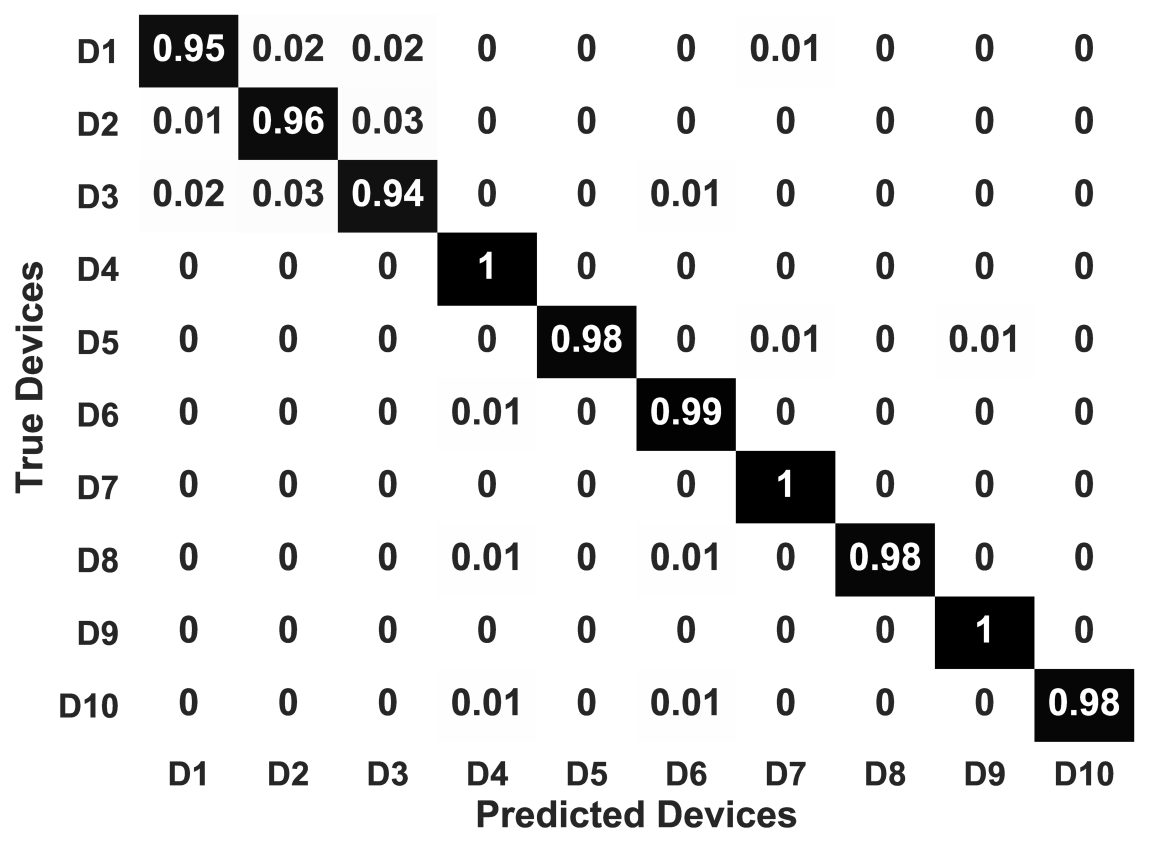}
         \vspace{-0.2in}
         \caption{DI.}
         \label{fig:overall_device_fingerprinting_results}
    \end{subfigure}
    \centering
	\begin{subfigure}[b]{.3\linewidth}
         \centering
         \includegraphics[width=\linewidth]{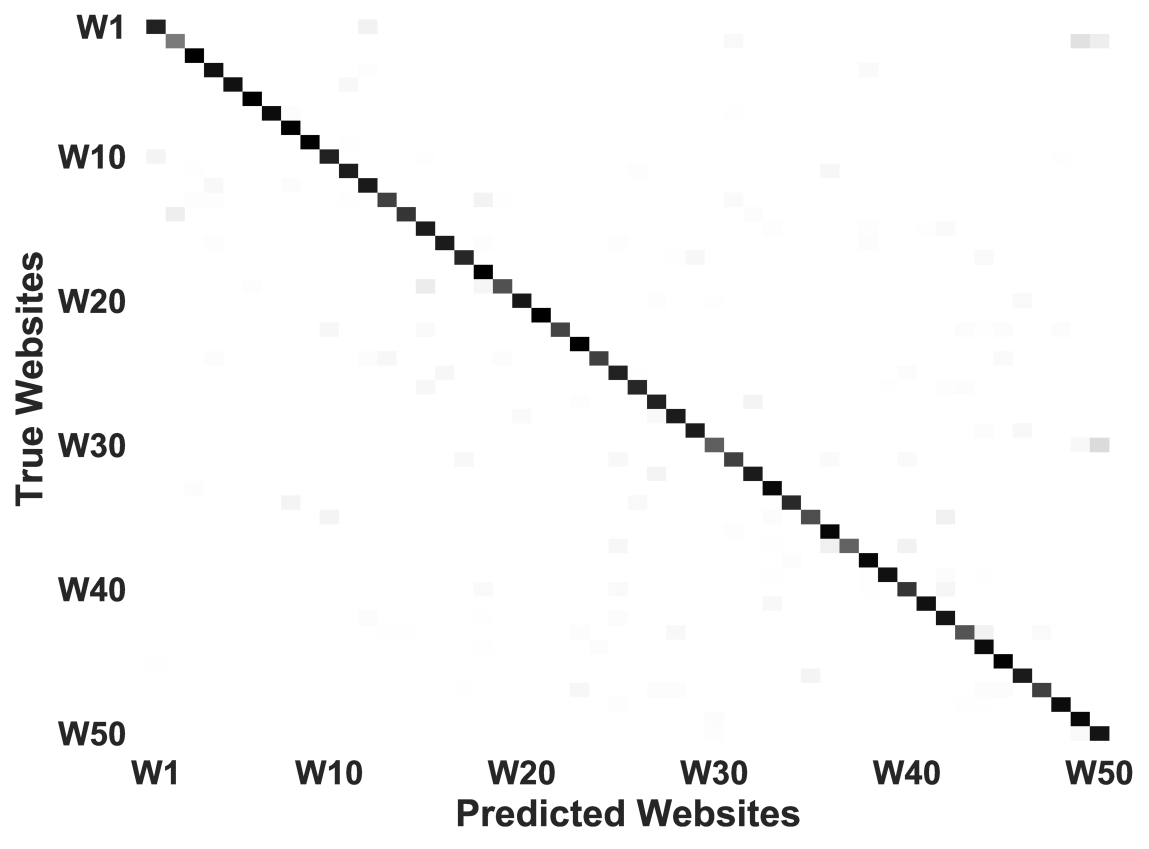}
         \vspace{-0.2in}
         \caption{WF.}
         \label{fig:overall_website_fingerprinting_results}
    \end{subfigure}
    \centering
	\begin{subfigure}[b]{.35\linewidth}
         \centering
         \includegraphics[width=\linewidth]{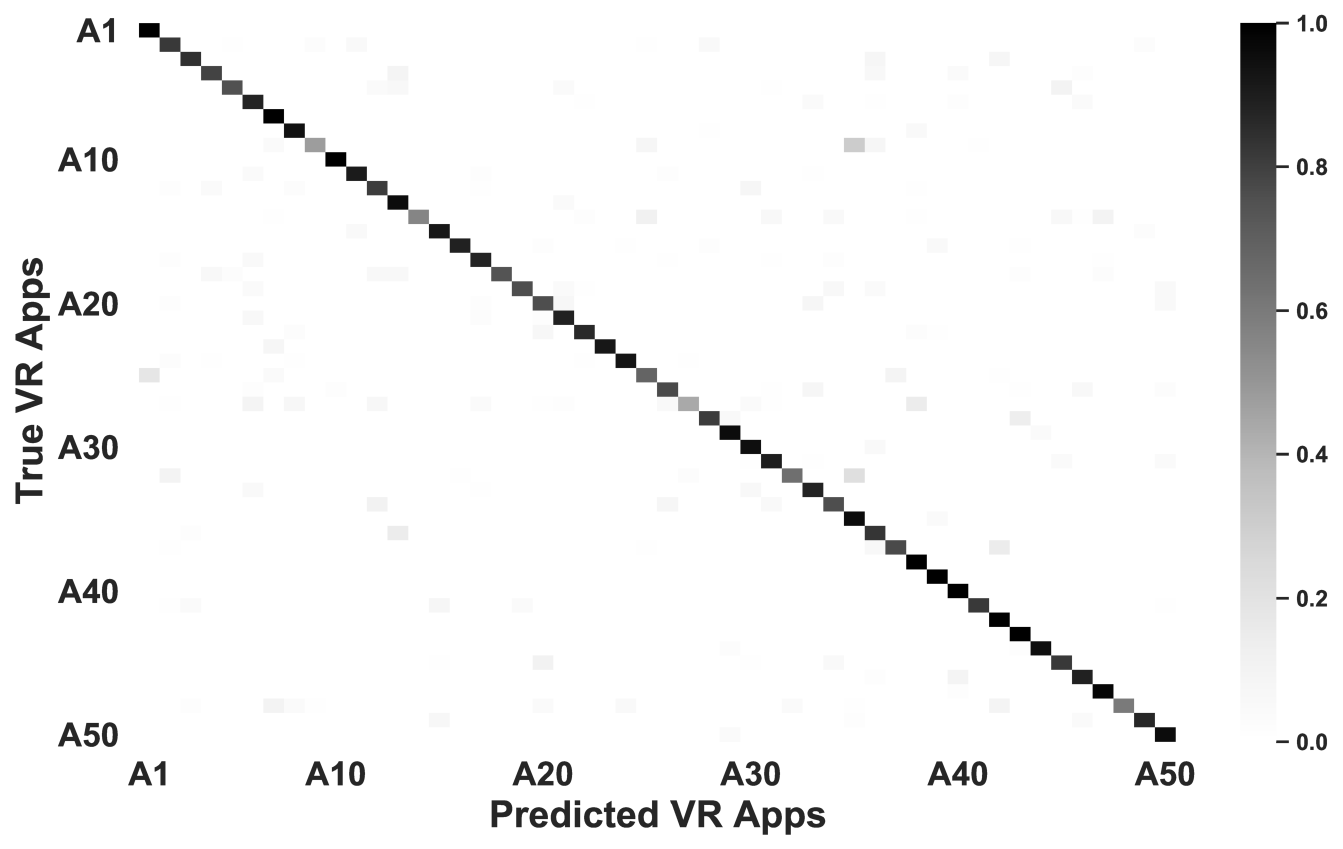}
         \vspace{-0.2in}
         \caption{AF.}
         \label{fig:overall_app_fingerprinting_results}
    \end{subfigure}
    \vspace{-0.1in}
    \caption{Device-level and UI-level privacy inference results.}
    \vspace{-0.25in}
    \label{fig:device_ui_level_inference}
\end{figure}

\noindent\textbf{Device Identification (DI).}
\autoref{fig:overall_device_fingerprinting_results} presents the effectiveness of \sysname in fingerprinting $10$ mobile devices, including four VR headsets, \major{based on} the captured wireless power.
In general, \sysname achieves $97.8\%$ accuracy on the testing set of the captured wireless power traces from the $10$ devices.
In particular, it achieves $95\%$--$100\%$ in recognizing the four VR headsets, as they require transmitting more packets in network traffic when connecting to the Wi-Fi router, which also results in higher amplitudes in the captured wireless power.
Hence, when the VR headset connects to the Wi-Fi router, \sysname can leverage the Wi-Fi power to accurately detect the connection and determine the type of device without prior knowledge of the connected VR devices.
% \looseness=-1

\begin{figure*}[t]
    \centering
	\begin{subfigure}[b]{.195\linewidth}
         \centering
         \includegraphics[width=\linewidth]{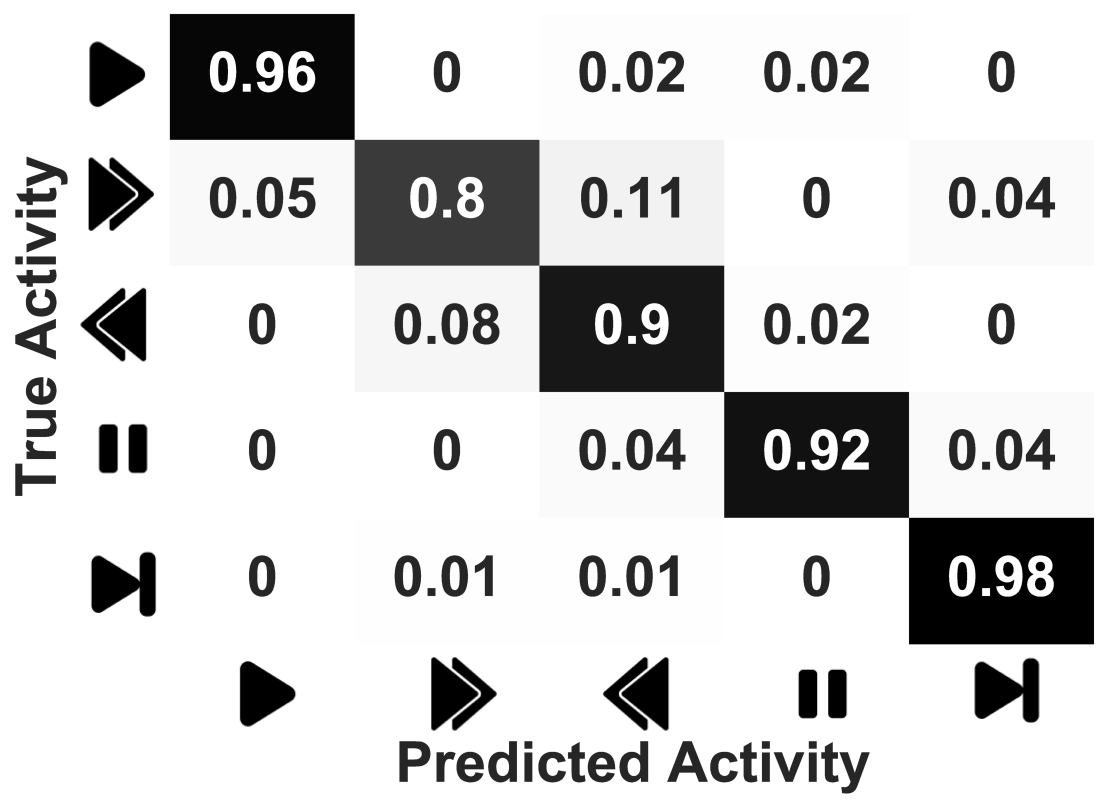}
         \vspace{-0.2in}
         \caption{\youtubevr\raisebox{-.05\height}{\includegraphics[width=24pt]{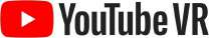}}}
         \label{fig:inapp_youtubevr_results}
    \end{subfigure}
    \centering
	\begin{subfigure}[b]{.195\linewidth}
         \centering
         \includegraphics[width=\linewidth]{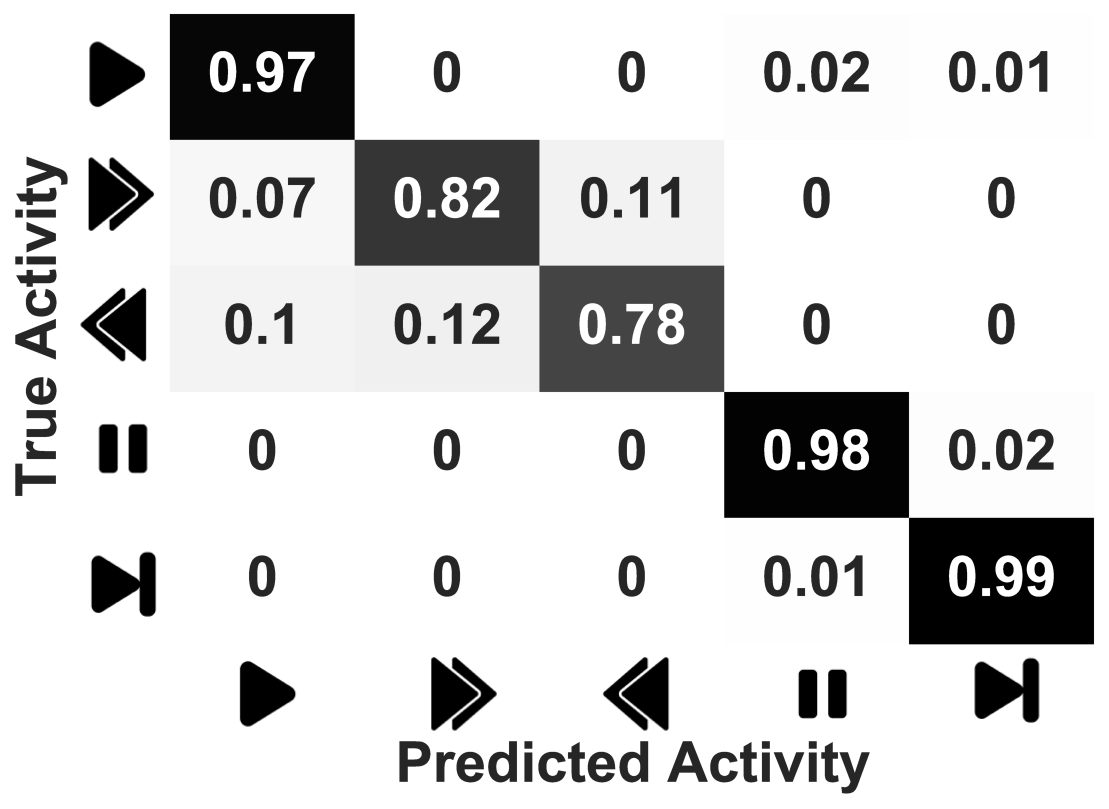}
         \vspace{-0.2in}
         \caption{\netflix\raisebox{-.15\height}{\includegraphics[width=8pt]{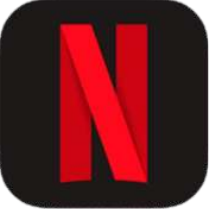}}}
         \label{fig:inapp_netflix_results}
    \end{subfigure}
    \begin{subfigure}[b]{.195\linewidth}
         \centering
         \includegraphics[width=\linewidth]{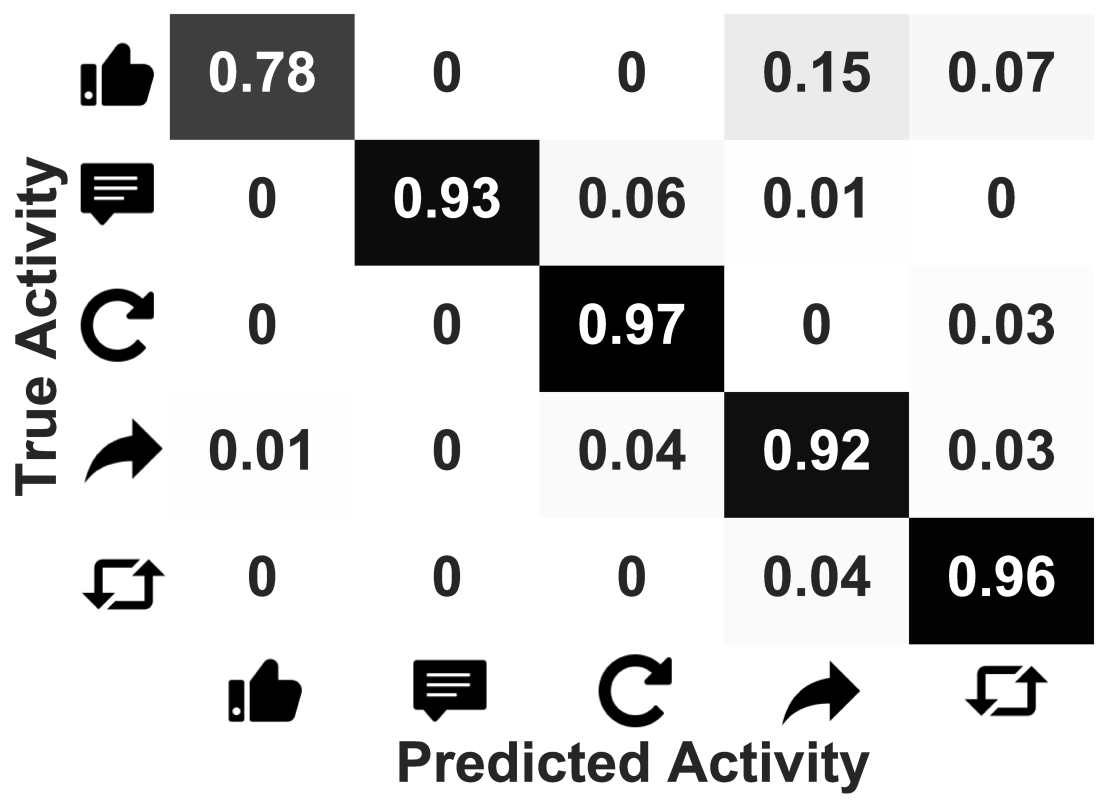}
         \vspace{-0.2in}
         \caption{\facebook\raisebox{-.15\height}{\includegraphics[width=8pt]{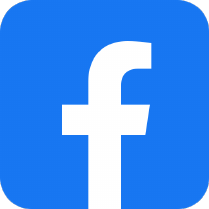}}}
         \label{fig:inapp_facebook_results}
    \end{subfigure}
    \begin{subfigure}[b]{.195\linewidth}
         \centering
         \includegraphics[width=\linewidth]{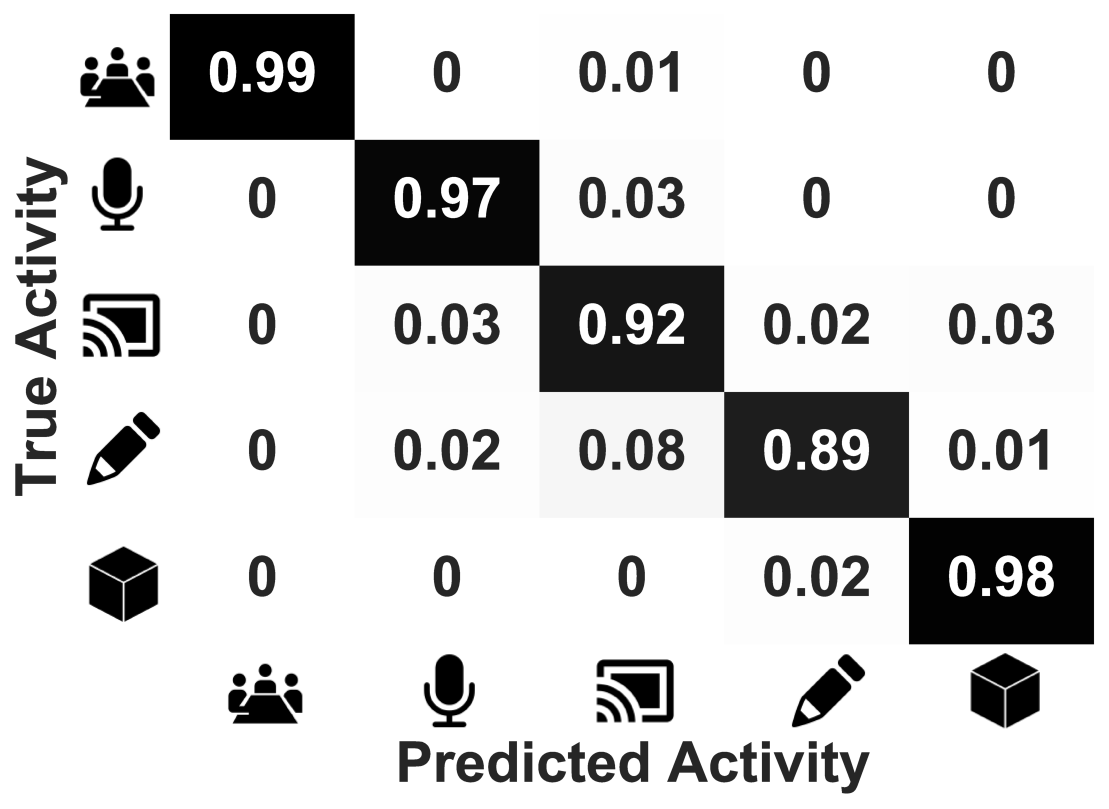}
         \vspace{-0.2in}
         \caption{\meetinvr\raisebox{-.05\height}{\includegraphics[width=24pt]{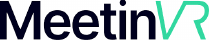}}}
         \label{fig:inapp_meetinvr_results}
    \end{subfigure}
    \begin{subfigure}[b]{.195\linewidth}
         \centering
         \includegraphics[width=\linewidth]{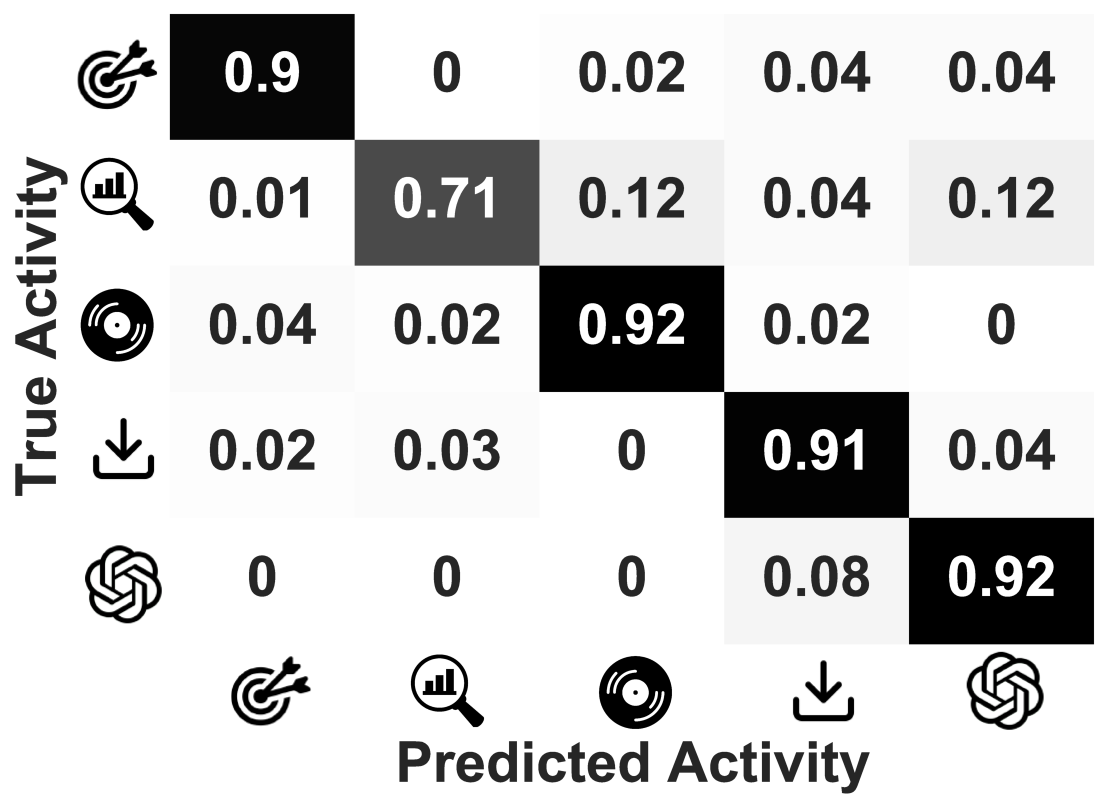}
         \vspace{-0.2in}
         \caption{\virtualspeech\raisebox{-.15\height}{\includegraphics[width=8pt]{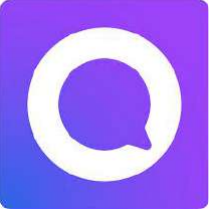}}}
         \label{fig:inapp_virtualspeech_results}
    \end{subfigure}
    \vspace{-0.1in}
    \caption{Evaluated in-app activities of \youtubevr\raisebox{0.01in}{\includegraphics[width=24pt]{icons/youtube_icon.pdf}} and \netflix\raisebox{-0.02in}{\includegraphics[width=8pt]{icons/netflix_icon.pdf}}: \raisebox{-0.02in}{\includegraphics[width=8pt]{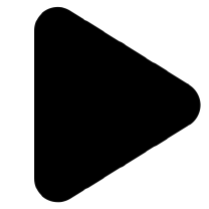}}--Play videos, \raisebox{-0.02in}{\includegraphics[width=8pt]{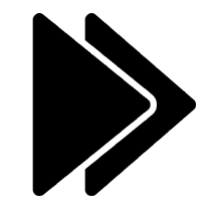}}--Fast forward, \raisebox{-0.02in}{\includegraphics[width=8pt]{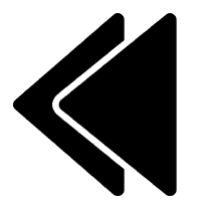}}--Backward, \raisebox{-0.02in}{\includegraphics[width=8pt]{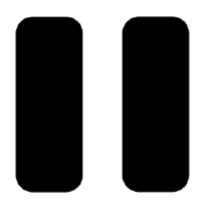}}--Pause, \raisebox{-0.02in}{\includegraphics[width=8pt]{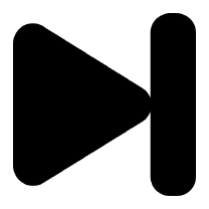}}--Switch to next video;
    \facebook\raisebox{-0.02in}{\includegraphics[width=8pt]{icons/facebook_icon.pdf}}:
    \raisebox{-0.02in}{\includegraphics[width=8pt]{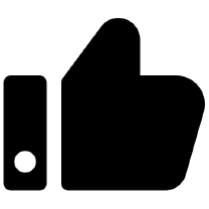}}--Thumb-up,
    \raisebox{-0.02in}{\includegraphics[width=8pt]{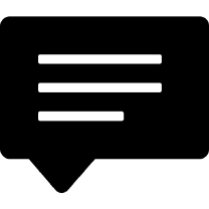}}--Comment,
    \raisebox{-0.02in}{\includegraphics[width=8pt]{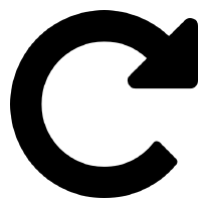}}--Refresh content,
    \raisebox{-0.02in}{\includegraphics[width=8pt]{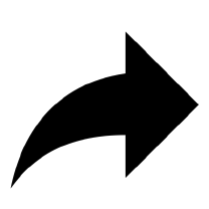}}--Share post,
    \raisebox{-0.02in}{\includegraphics[width=8pt]{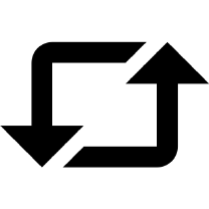}}--Repost;
    \meetinvr\raisebox{0.01in}{\includegraphics[width=24pt]{icons/meetinvr_icon.pdf}}:
    \raisebox{-0.02in}{\includegraphics[width=8pt]{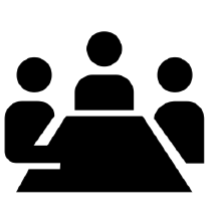}}--Join a meeting,
    \raisebox{-0.02in}{\includegraphics[width=8pt]{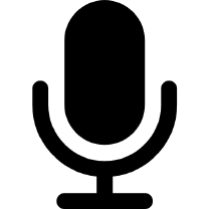}}--Speaking,
    \raisebox{-0.02in}{\includegraphics[width=8pt]{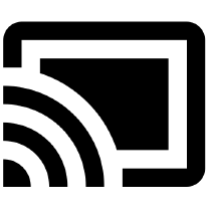}}--Casting slides,
    \raisebox{-0.02in}{\includegraphics[width=8pt]{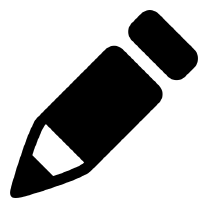}}--Scribbling,
    \raisebox{-0.02in}{\includegraphics[width=8pt]{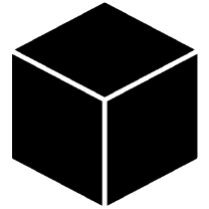}}--Build 3D models;
    \virtualspeech\raisebox{-0.02in}{\includegraphics[width=8pt]{icons/virtualspeech_icon.pdf}}:
    \raisebox{-0.02in}{\includegraphics[width=8pt]{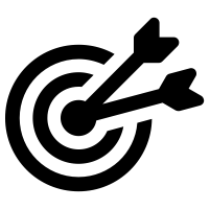}}--Practice speech,
    \raisebox{-0.02in}{\includegraphics[width=8pt]{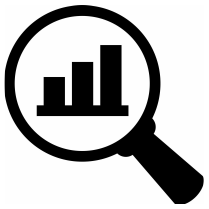}}--Analysis,
    \raisebox{-0.02in}{\includegraphics[width=8pt]{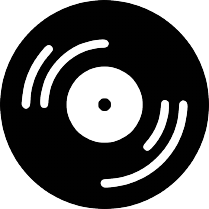}}--Record speech,
    \raisebox{-0.02in}{\includegraphics[width=8pt]{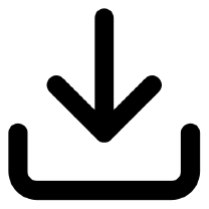}}--Download report,
    \raisebox{-0.02in}{\includegraphics[width=8pt]{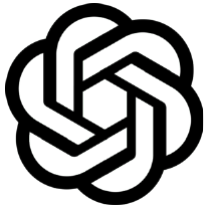}}--Practice speech with GPT.
    }
    \vspace{-0.1in}
    \label{fig:overall_inapp_activity_recognition_results}
\end{figure*}

\noindent\textbf{Website Fingerprinting (WF).} 
\autoref{fig:overall_website_fingerprinting_results} shows the empirical results of the effectiveness of website fingerprinting, where \sysname achieves an overall accuracy of $89.2\%$ in distinguishing $50$ websites.
Among the evaluated websites, we notice that websites such as \twitchtv ($100\%$), \ticktockcom ($100\%$), and \cnncom ($98\%$) are more recognizable when \major{browsed} because these websites automatically load video \major{streams} or many high-resolution images.
In contrast, \sysname presents an accuracy of only $89\%$ and $82\%$ on fingerprinting websites such as \googlecom and \baiducom, which are both widely used search engines with only a simple search bar and fewer content to render.
Thus, these websites present similar patterns in the captured wireless power, which can be easily misclassified by the reconstruction model.

\noindent\textbf{App Fingerprinting (AF).} 
\autoref{fig:overall_app_fingerprinting_results} shows \sysname achieves an accuracy of $85.1\%$ in fingerprinting $50$ VR apps from different categories. Specifically, \sysname achieves the highest performance ($100\%$) in fingerprinting VR apps from the game category (\eg, \populationone and \cardstankards) and the media and streaming category (\eg, \netflix and \plutotv), whereas it performs worse in \major{the} recognition of VR apps that adopt \major{the} default app launching interface (\eg, white static background with a simple logo), \ie, social apps such as \multiverse ($56\%$) and lifestyle apps such as \maloka ($68\%$).
\looseness=-1

% \begin{figure*}[t]
%     \centering
% 	\begin{subfigure}[b]{.3\linewidth}
%          \centering
%          \includegraphics[width=\linewidth]{figures/overall_effectiveness_user_identification_2025.pdf}
%          \vspace{-0.2in}
%          \caption{Avatar-based user identification.}
%          \label{fig:overall_user_deanonymization_results}
%     \end{subfigure}
%     \centering
% 	\begin{subfigure}[b]{.35\linewidth}
%          \centering
%          \includegraphics[width=\linewidth]{figures/overall_effectiveness_keystroke_inference_2025.pdf}
%          \vspace{-0.2in}
%          \caption{Avatar-based keystroke inference.}
%          \label{fig:overall_keystroke_inference_results}
%     \end{subfigure}
%     \centering
% 	\begin{subfigure}[b]{.3\linewidth}
%          \centering
%          \includegraphics[width=\linewidth]{figures/topk_results_2025.pdf}
%          \vspace{-0.2in}
%          \caption{Top-k AKI.}
%          \label{fig:overall_keystroke_inference_top_results}
%     \end{subfigure}
%     \vspace{-0.1in}
%     \caption{Avatar-based user privacy inference results.}
%     \vspace{-0.1in}
%     \label{fig:avatar_level_inference}
% \end{figure*}

\noindent\textbf{In-app Activity Recognition (IAR).} 
\autoref{fig:overall_inapp_activity_recognition_results} shows the effectiveness of \sysname in the recognition of in-app activity among VR apps, where we find it achieves \major{an} accuracy of $91.2\%$, $90.8\%$, $91.2\%$, $95.0\%$, and $87.2\%$ in the recognition of the five in-app activities in \youtubevr, \netflix, \facebook, \meetinvr, and \virtualspeech, respectively.
We also notice that in-app activities, \eg, forward and backward in \netflix, present similar patterns, which lead to relatively low accuracy ($80\%$ and $90\%$) \major{in distinguishing} them.
In contrast, in-app activities such as switching to the next video in video streaming apps \youtubevr ($99\%$) and \netflix ($98\%$), joining a meeting in \meetinvr ($99\%$), and practicing speech with GPT-based AI in the \virtualspeech ($92\%$), are easily recognized by \sysname with high accuracy.
Therefore, we have demonstrated the effectiveness of \sysname in recognizing in-app activities in VR apps.
\looseness=-1

\begin{figure}[t]
\centering
    \begin{subfigure}[b]{.49\linewidth}
         \centering
         \includegraphics[width=\linewidth]{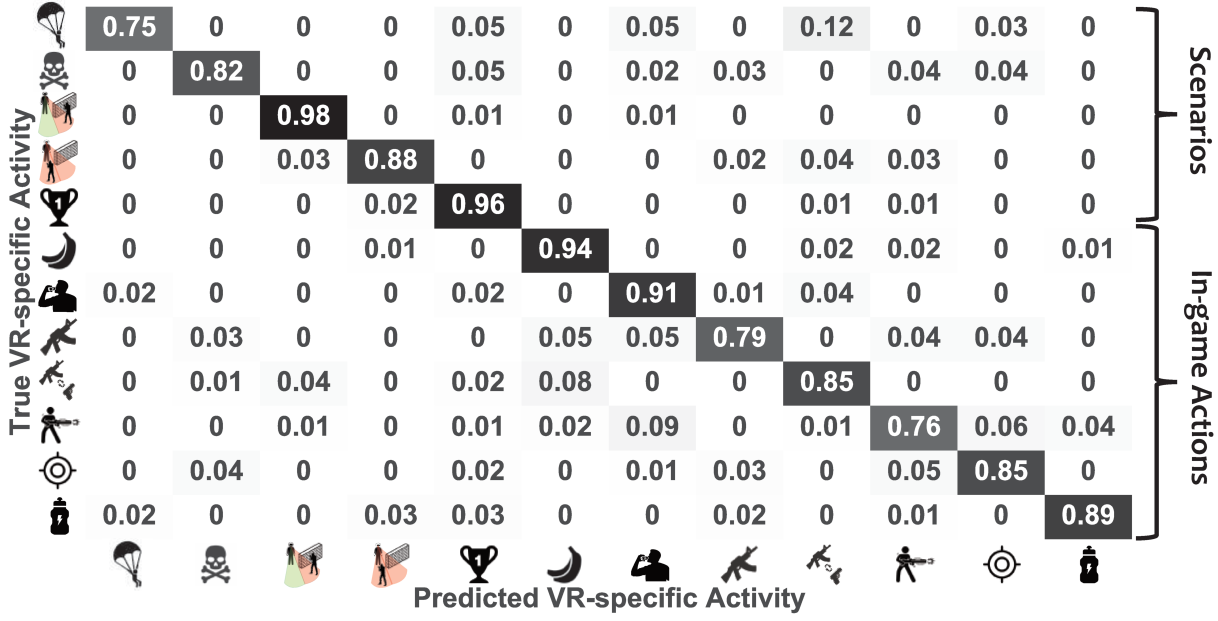}
         \vspace{-0.2in}
         \caption{In-game action detection.}
         %\caption{Overall VR-specific activity recognition results.}
         \label{fig:overall_effectiveness_vr_specific}
    \end{subfigure}
    \centering
	\begin{subfigure}[b]{.49\linewidth}
         \centering
         \includegraphics[width=\linewidth]{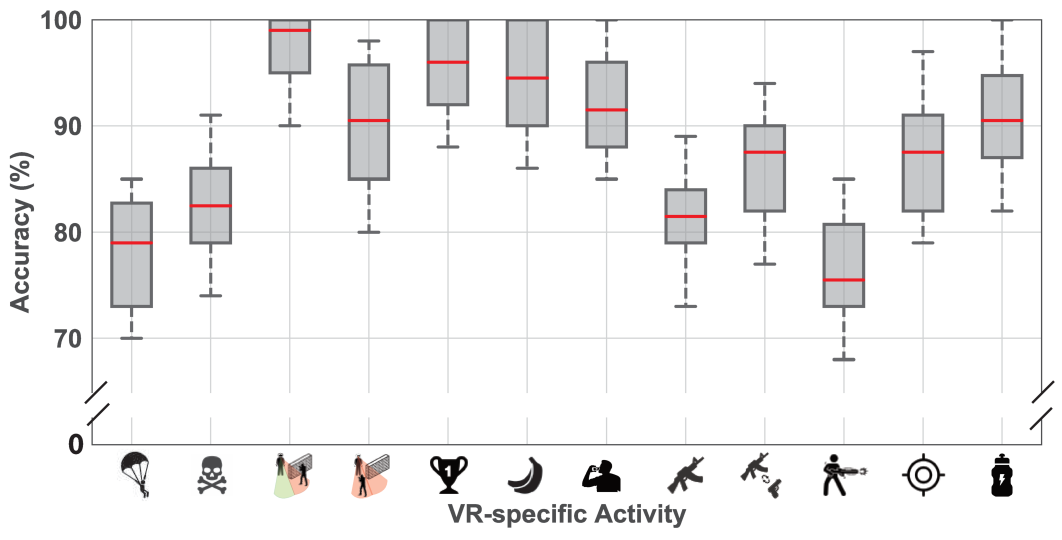}
         \vspace{-0.2in}
         \caption{Recognition accuracy.}
         %in $50$ rounds of game.}
         \label{fig:overall_effectiveness_vr_specific_activity_boxplot}
    \end{subfigure}
    \vspace{-0.15in}
    \caption{VR-specific activity recognition results in $50$ rounds of VR game \populationone. We have evaluated five common scenarios in the virtual scenes and seven in-game actions while users interact with VR. (a) Overall VR-specific activity recognition results, (b) Recognition accuracy of each VR-specific activity in $50$ rounds of the game.}
    \vspace{-0.2in}
    \label{fig:overall_vr_specific_activity_recognition_results}
\end{figure}

\noindent\textbf{VR-specific Activity Recognition (VAR).} 
\autoref{fig:overall_effectiveness_vr_specific} shows the confusion matrix of the VR-specific activity recognition results \major{for} five scenarios and seven in-game actions in \populationone, where \sysname can infer the scenarios and in-game actions with accuracy rates of $87.8\%$ and $85.6\%$, respectively.
Unlike previous in-app activity recognition on smartphones~\cite{li2022foap, ni2023eavesdropping}, we notice that \sysname is capable of recognizing unseen VR-specific activities, \ie, when the enemy hides behind an obstacle (\eg, wall) and enters the unseen NLoS frustum of the VR user. Specifically, \sysname achieves an average recognition accuracy of $95\%$ by using captured wireless power, which reflects changes in network traffic when hiding the enemy in the virtual scene. We further evaluate the effectiveness of VR-specific activity recognition in the $50$ rounds of \populationone and plot the results in \autoref{fig:overall_effectiveness_vr_specific_activity_boxplot}. As shown, \sysname achieves over $70\%$ recognition accuracy in differentiating the aforementioned $12$ VR-specific activities, and most of them present over $80\%$ accuracy.
In particular, the captured wireless power patterns in scenarios such as ``enemy hide'' and ``game over'', as well as in-game actions such as ``pick up banana'' and ``shake to charge'', are more distinctive, and \sysname can achieve over $93.0\%$ accuracy.
\looseness=-1

\begin{figure}[t]
    \centering
	\begin{subfigure}[b]{.3\linewidth}
         \centering
         \includegraphics[width=\linewidth]{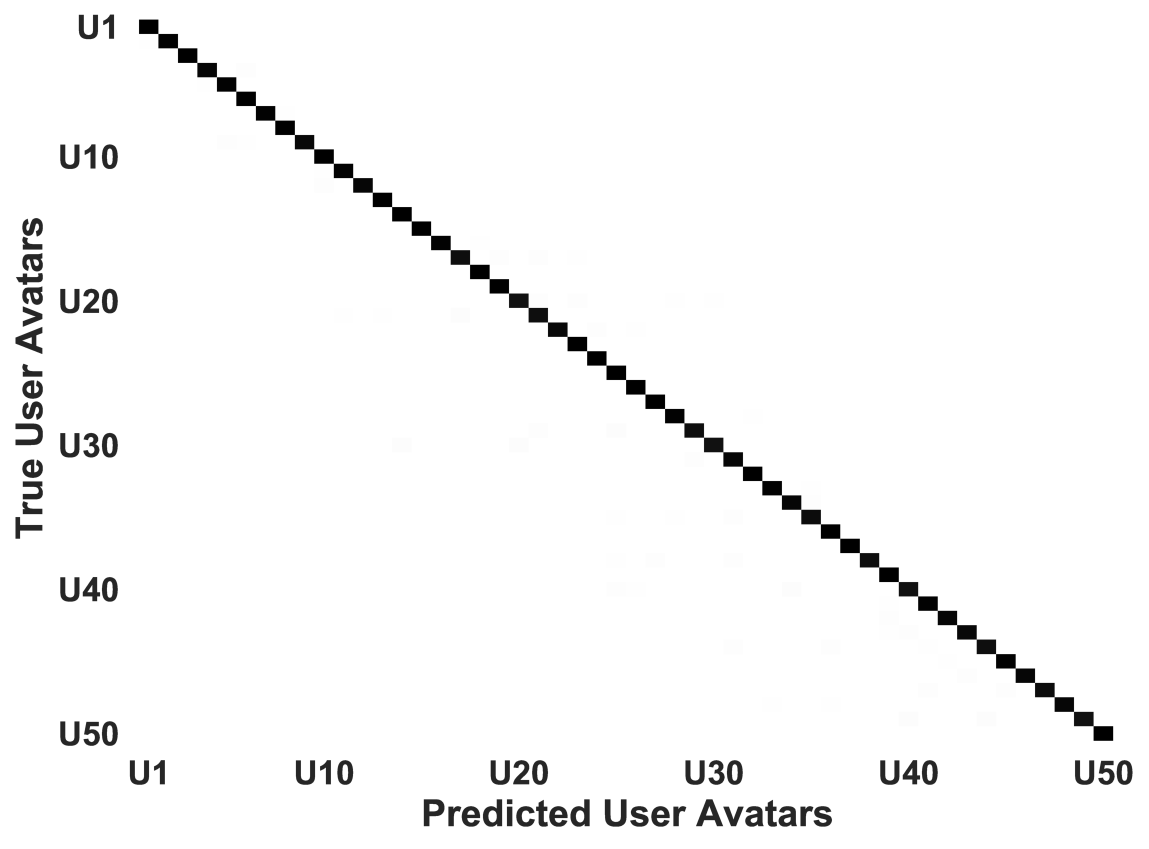}
         \vspace{-0.2in}
         \caption{AUI.}
         \label{fig:overall_user_deanonymization_results}
    \end{subfigure}
    \centering
	\begin{subfigure}[b]{.35\linewidth}
         \centering
         \includegraphics[width=\linewidth]{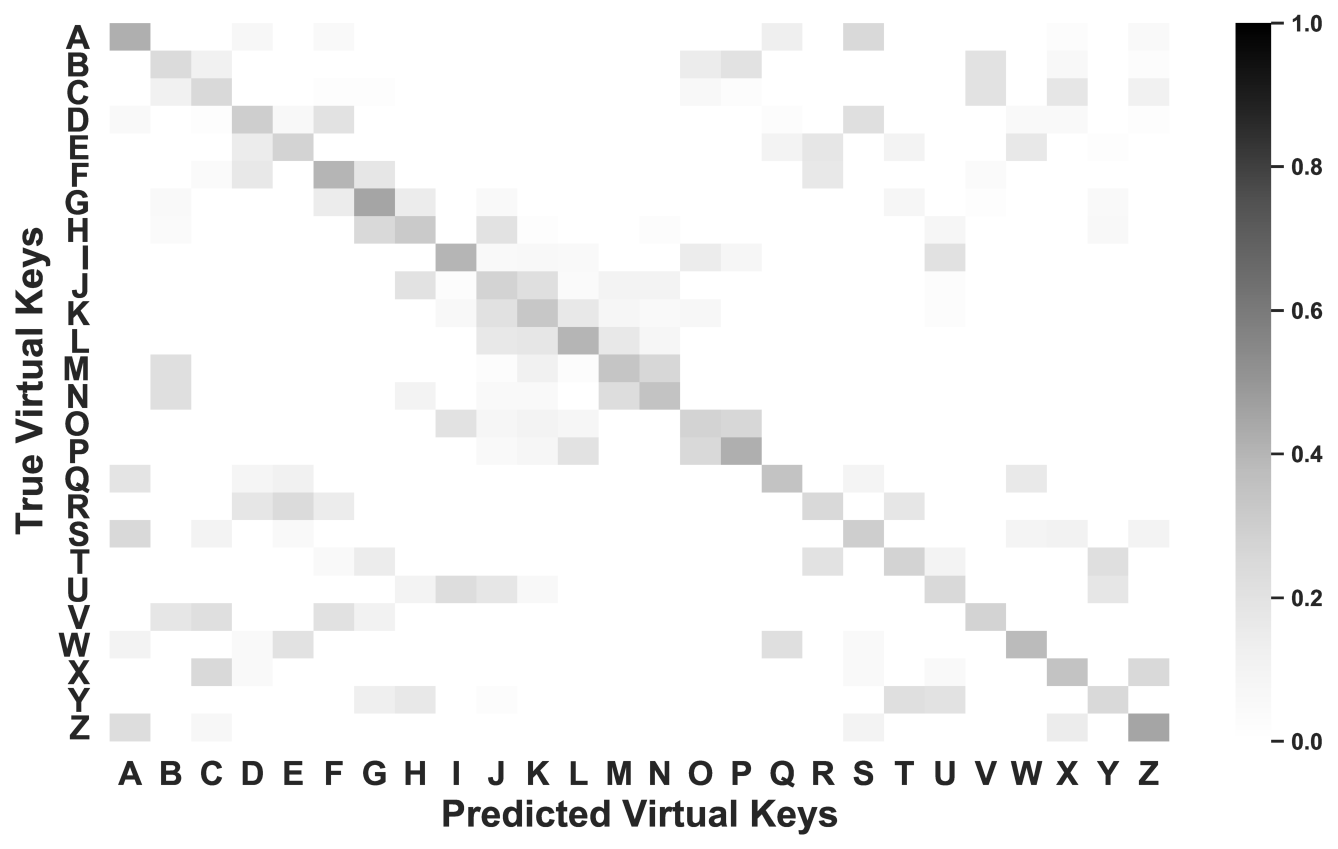}
         \vspace{-0.2in}
         \caption{AKI.}
         \label{fig:overall_keystroke_inference_results}
    \end{subfigure}
    \centering
	\begin{subfigure}[b]{.3\linewidth}
         \centering
         \includegraphics[width=\linewidth]{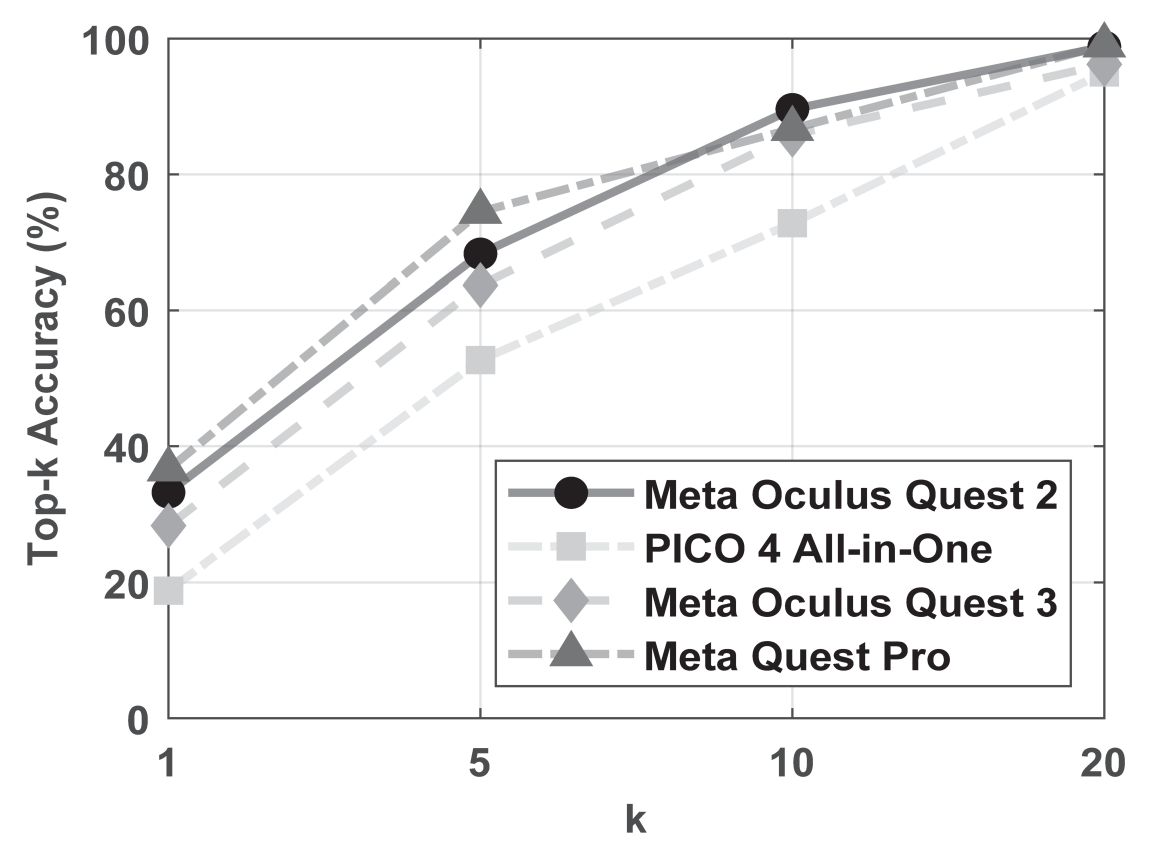}
         \vspace{-0.2in}
         \caption{Top-k AKI.}
         \label{fig:overall_keystroke_inference_top_results}
    \end{subfigure}
    \vspace{-0.15in}
    \caption{Avatar-based user privacy inference results.}
    \vspace{-0.2in}
    \label{fig:avatar_level_inference}
\end{figure}

\noindent\textbf{Avatar-based User Identification (AUI).} 
We use data samples from $\mathcal{D}_{AUD}$ to train the recognition model to assess the performance of \sysname in fingerprinting user identification when joining a shared online meeting in \meetinvr.
\autoref{fig:overall_user_deanonymization_results} shows the results of \sysname in recognizing $50$ customized avatars that represent different users, \major{and} we find that it achieves an overall accuracy of $97.9\%$.
Specifically, VR apps like \meetinvr generate the avatars by rendering the facial characteristics of VR users, which induce different traffic loads in wireless transmission and result in distinctive Wi-Fi energy patterns.
Therefore, unlike previous studies that require invasively acquiring the avatar or sensor data (\eg, \cite{meng2023anonymization, sabra2024anonymizing}) in VR headsets, \sysname can effectively de-anonymize the user identities via this non-intrusive power side channel.
\looseness=-1

\noindent\textbf{Avatar-based Keystroke Inference (AKI).} 
We leverage $\mathcal{D}_{AKI}$ to train the recognition model
to bridge the gap between captured wireless power, head movements, and keyboard zones.
 \autoref{fig:overall_keystroke_inference_results} shows the avatar-based keystroke inference results in recognizing $26$ alphabetic keys, where \sysname achieves only $35.5\%$ accuracy in inferring single keys.
\major{This} is because each zone on the virtual keyboard contains multiple keys, \eg, the up zone contains keys ``E'', ``R'', ``T'', ``Y'', ``U'', and ``I'', which increases the misclassification rate for keys in the same zone and \major{decreases} the overall performance.
Therefore, 
we have adopted the top-$k$ settings by generating candidates $k$ for each combination of keys to infer the keystrokes typed on the virtual keyboard.
Specifically, we set the value of top-$k$ as top-1, top-5, top-10, and top-20, and evaluate the keystroke inference performance of \sysname, respectively.
\autoref{fig:overall_keystroke_inference_top_results} shows the empirical results, and we can observe that \sysname achieves overall accuracy rates $68.3\%$, $89.6\%$, and $98.8\%$ in Meta Oculus Quest 2; $52.7\%$, $72.8\%$, and $95.0\%$ in PICO 4 All-in-One; $63.7\%$, $85.8\%$, and $96.2\%$ in Meta Quest 3; and $74.5\%$, $86.7\%$, and $99.0\%$ in Meta Quest Pro, under the settings of top-5, top-10, and top-20, respectively.
\looseness=-1

% \begin{figure}[t]
%     \centering
% 	\begin{subfigure}[b]{.3\linewidth}
%          \centering
%          \includegraphics[width=\linewidth]{figures/overall_effectiveness_user_identification_2025.pdf}
%          \vspace{-0.2in}
%          \caption{AUI.}
%          \label{fig:overall_user_deanonymization_results}
%     \end{subfigure}
%     \centering
% 	\begin{subfigure}[b]{.35\linewidth}
%          \centering
%          \includegraphics[width=\linewidth]{figures/overall_effectiveness_keystroke_inference_2025.pdf}
%          \vspace{-0.2in}
%          \caption{AKI.}
%          \label{fig:overall_keystroke_inference_results}
%     \end{subfigure}
%     \centering
% 	\begin{subfigure}[b]{.3\linewidth}
%          \centering
%          \includegraphics[width=\linewidth]{figures/topk_results_2025.pdf}
%          \vspace{-0.2in}
%          \caption{Top-k AKI.}
%          \label{fig:overall_keystroke_inference_top_results}
%     \end{subfigure}
%     \vspace{-0.1in}
%     \caption{Avatar-based user privacy inference results.}
%     \vspace{-0.25in}
%     \label{fig:avatar_level_inference}
% \end{figure}

% \input{tables/evaluation_results_impact_factors}

% \input{tables/more_through_wall_attack_results}

\subsection{External Impact Factors}
\label{subsec:impact_practical_factors}

\noindent\textbf{Different COTS VR Headsets.}
To evaluate whether \sysname can be launched \major{on} other VR headsets, we conducted further experiments by separately collecting data and training models from three other VR devices, \ie, PICO 4, Meta Quest 3, and Meta Quest Pro.
\autoref{tab:impact_factors} shows the evaluation results of uncovering different user interactions in the virtual scenes on the four
VR headsets, where we find that it also achieves high recognition accuracy rates, \ie, $87.5\%$ on fingerprinting websites, $82.3\%$ in fingerprinting VR apps, $88.3\%$ in recognizing different in-app activities,  $83.0\%$ in inferring VR-specific activities in \populationone, $93.3\%$ in avatar-based user identification, and $72.8\%$ in virtual keystroke inference in the PICO 4 VR headset.
Meanwhile, \sysname also shows comparative performance in eavesdropping \major{on} multi-level user privacy from Meta Quest 3 and Meta Quest Pro. 
Note that we choose not to evaluate VR headsets such as the HTC VIVE VR headset series because \textit{(i)} they require connecting to \major{a} laptop to run apps through the laptop, and \textit{(ii)} these VR devices need multiple stationary base stations to track the VR users' positions and playing zones~\cite{wu2023privacy}, which \major{are} outdated and \major{have} been abandoned by newly-released VR headsets.
\looseness=-1

\input{tables/evaluation_results_impact_factors}

\input{tables/more_through_wall_attack_results}

\noindent\textbf{Different Ambient Environments.} 
To evaluate the impact of environments with different SNRs, we follow the same procedure and conduct experiments in the four indoor scenarios (\autoref{subsec:experimental_setup}): home (\SI{35}{\deci\bel}), office (\SI{20}{\deci\bel}), corridor (\SI{40}{\deci\bel}), and cafe (\SI{15}{\deci\bel}).
\autoref{tab:impact_factors} shows the results of uncovering user interactions in different experimental environments.
Specifically, we find \major{that} the performance of \sysname in the home and corridor, where there is only one Wi-Fi router, is higher than the effectiveness in the office and cafe, where multiple Wi-Fi routers exist. This is because the Wi-Fi energy from other non-target Wi-Fi routers brings extra interference (low SNRs) in the captured wireless power, which reflects the VR user's interactions in the virtual scenes.
Despite this, \sysname still achieves average accuracy rates of $83.6\%$ and $83.2\%$ in inferring user interactions inside the VR device in the office and cafe, respectively.
\looseness=-1

\noindent\textbf{Different Attack Distances.} 
According to \autoref{subsec:rf_energy_network_traffic}, the captured wireless power attenuates as the distance between the Wi-Fi router and the wireless power sniffer increases. Hence, we evaluated the effectiveness of \sysname in uncovering user interactions by placing the custom-built wireless power sniffer at different attack distances: \SI{5.0}{\meter}, \SI{6.0}{\meter}, \SI{7.0}{\meter}, and \SI{8.0}{\meter}.
\autoref{tab:impact_factors} shows the evaluation results at different attack distances, where we find that \sysname's performance degrades with increasing \major{attack} distance due to signal attenuation %of the Wi-Fi energy 
and interference from other wireless signal sources (\eg, other Wi-Fi routers, and mobile devices).
However, \sysname still achieves \major{a} recognition accuracy of $68.3\%$ in website fingerprinting, $63.5\%$ in app fingerprinting, $78.7\%$ in \major{the} in-app activity recognition, $77.5\%$ in VR-specific activity recognition, $80.6\%$ in user identification, and $62.5\%$ in virtual keystroke inference at the distance of \SI{8.0}{\meter}.
Note that the performance at longer \major{distances} can be improved by fine-tuning the reconstruction models or leveraging high-gain antennas to enhance power-scavenging efficiency.
\looseness=-1

\noindent\textbf{Different Number of Connected Non-target Devices.} 
To further understand and evaluate the impact of different non-target devices, 
we conduct similar experiments when the target Wi-Fi router is connected to different non-target mobile devices.
In practice, we capture wireless power by separately connecting non-target devices (\ie, Apple Watch 3, HomePod mini, and iPhone 13 Pro) to the target Wi-Fi and setting these devices to work in default mode.
\autoref{tab:impact_factors}
shows \major{that} performance degrades as more non-target devices are involved in the Wi-Fi network.
In particular, the average recognition accuracy decreases approximately from $89.9\%$ to $83.1\%$ and $78.2\%$ when connected to only the Apple Watch 3 and \major{to} both the Apple Watch 3 and the HomePod mini, whereas it degrades drastically to $71.0\%$ when connecting to an extra iPhone 13 Pro.
\major{This} is because \major{the} Apple Watch 3 and HomePod mini barely require light packet transmissions to support text or voice services, whereas \major{the} iPhone 13 Pro
is normally traffic intensive, as \major{it requires} a large amount of data transmission to provide versatile functionalities.
Since we have adopted interference removal (\autoref{subsec:preprocessing_harvested_voltage}) and shape-related features (\autoref{subsec:energy_based_feature_extraction}), the fluctuations in the captured wireless power from other non-target devices incite limited distortion on the signal's shape.

% \noindent\textbf{Different Number of Background Apps.} In addition, we further explore the performance of \sysname when running different numbers of background apps simultaneously. Specifically, we conduct experiments \major{while} running one, two, and three apps (\ie, \spotify, \facebook, and \youtubevr) in the background.
% \autoref{tab:impact_factors} depicts the performance when running one, two, and three background apps, where we found that performance decreases with \major{an increasing number of} background apps.
% In particular, the overall performance of \sysname decreases from approximately $89.9\%$ (without any background apps) to $74.2\%$ (running the three background apps mentioned) \major{based on} the empirical results, because background apps introduce extra traffic loads into the wireless network and cause fluctuations in captured wireless power.
% However, \sysname still achieves an acceptable recognition accuracy (\eg, larger than $68\%$) in inferring fine-grained user behaviors, and most of the commodity VR apps (\eg, games) will automatically disable all background apps to provide immersive experiences.
% \looseness=-1

\input{tables/feature_ablation_study}

\noindent\textbf{Different Number of Background Apps.} \major{To evaluate how it handles simultaneous multi-app and background network traffic that may overlap with VR-related wireless power patterns,} we further explore the performance of \sysname when running different numbers of background apps simultaneously. Specifically, we conduct experiments \major{while} running one, two, and three apps (\ie, \spotify, \facebook, and \youtubevr) in the background \major{alongside the target VR app, simulating realistic deployment scenarios where multiple network-active apps coexist on the user's headset.}
\autoref{tab:impact_factors} depicts the performance when running one, two, and three background apps, where we found that performance decreases with \major{an increasing number of} background apps.
In particular, the overall performance of \sysname decreases from approximately
$89.9\%$ (without any background apps) to $74.2\%$ (running the three background apps mentioned) \major{based on} the empirical results, because background apps introduce extra traffic loads into the wireless network and cause fluctuations in captured wireless power. \major{Despite this, \sysname's adaptive interference removal pipeline (\autoref{subsec:preprocessing_harvested_voltage}) effectively suppresses the additional traffic noise by leveraging KL-divergence-based optimization, preserving the dominant temporal patterns of VR activities.}
However, \sysname still achieves an acceptable recognition accuracy (\eg, larger than $68\%$) in inferring fine-grained user behaviors, and most of the commodity VR apps (\eg, games) will automatically disable all background apps to provide immersive experiences.
\looseness=-1

\subsection{More Real-World Through-Wall Scenarios}
\label{subsec:through-wall_attack_scenarios}

To demonstrate the effectiveness of \sysname in more real-world through-wall attack scenarios, we conducted further experiments in multiple scenarios (\autoref{fig:experiment_setup}) with walls of different materials and thicknesses.
Specifically, we selected four common types of walls in different rooms \major{for} empirical experiments, including plastic walls in research labs (\SI{6.7}{\centi\meter} and \SI{11.3}{\centi\meter}), glass walls in workplaces (\SI{1.0}{\centi\meter} and \SI{1.2}{\centi\meter}), wooden walls in libraries (\SI{4.3}{\centi\meter} and \SI{8.6}{\centi\meter}), and concrete walls in buildings (\SI{9.5}{\centi\meter} and \SI{25.0}{\centi\meter}).
We conducted the experiments to evaluate \sysname in inferring the aforementioned multi-level user privacy from VR users when wearing \major{the} Meta Oculus Quest 2 at attack distances of \SI{5.0}{\meter} and \SI{8.0}{\meter}.

\autoref{tab:through_wall_attack_results} presents the empirical results for through-wall attack scenarios, showing that \sysname achieves
the highest performance with common obstacles (\eg, plastic walls, glass walls, and wood walls), especially under the attack distance \SI{5.0}{\meter}. The performance degrades when increasing the attack distance to \SI{8.0}{\meter} due to the superposition effects of power attenuation.
Furthermore, the results also demonstrate that \sysname can penetrate \major{thin} concrete walls (\eg, \SI{9.5}{\centi\meter}) to launch eavesdropping attacks on VR users, but performance decreases drastically when wireless power is captured through a thick concrete wall (\eg, \SI{25.0}{\centi\meter}).
Additionally, \sysname's performance drops to zero when the wireless power sniffer is obstructed by the thick concrete wall at the attack distance \SI{8.0}{\meter}, which results in severe signal attenuation that prevents wireless power from penetrating and being captured.
% However, \sysname still shows competitive performance with high practicality and effectiveness compared to previous studies (\eg, \cite{wu2023privacy, slocum2023going, zhang2023s, ling2019know, luo2022holologger, al2021vr, gopal2023hidden, nguyen2024penetration, meteriz2022keylogging, luo2024eavesdropping, su2024remote, wang2024gazeploit, mei2024mmspyvr, ni2025non, shi2021face, zhang2023facereader, li2024dangers, huang2025siren, wei2025vreaves, cayir2025speak}).
% \looseness=-1

% \input{tables/feature_ablation_study}

\begin{figure}[t]
    \centering
    \includegraphics[width=\linewidth]{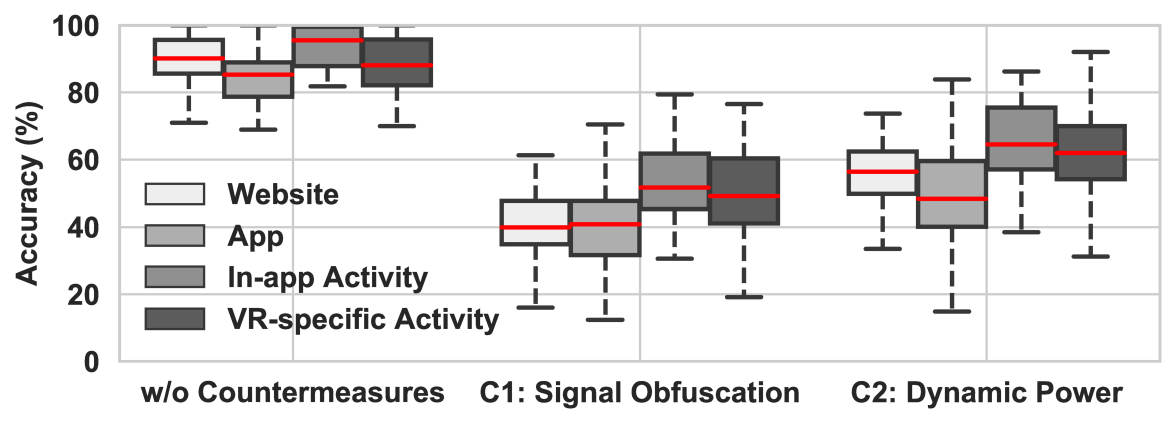}
    \vspace{-0.25in}
    \caption{Results in defending against \sysname by adopting the signal obfuscation and dynamic power countermeasures.}
    \vspace{-0.2in}
    \label{fig:results_defense}
\end{figure}

\subsection{\major{Ablation Study on Feature Selection}}
\label{subsec:ablation_study_feature_selection}

\major{To justify our choice of 15 time-domain and 2 frequency-domain features (\autoref{subsec:energy_based_feature_extraction}), we conducted an ablation study evaluating \sysname's inference accuracy when individual feature groups are removed. \autoref{tab:ablation_features} depicts the accuracy across the six privacy inference tasks under different feature configurations, which validate our optimal feature selection. First, removing the three shape-related features (\eg, shape factor, impulsive factor, crest factor) leads to a noticeable accuracy drop of $5\%$--$8\%$ across all tasks, with the largest degradation in AF and AKI, which further confirms that shape-related features play a critical role in capturing waveform contour patterns that are robust to interference from co-channel devices. 
Second, removing the two frequency-domain features (\eg, FFT mean and entropy) causes a $3\%$--$4\%$ degradation, especially in tasks involving periodic traffic patterns such as IAR, which demonstrates that frequency-domain information complements time-domain descriptors. Third, using only the 12 basic time-domain statistical features presents the results with the largest decrease (\eg, $8\%$--$10\%$), which confirms that both shape-related and frequency-domain features are necessary for robust inference. 
In addition, we also evaluated additional candidate features such as wavelet coefficients and autocorrelation-based descriptors during recursive feature elimination~\cite{yan2015feature}, but they yielded no statistically significant improvement while increasing model complexity and overfitting risk. These results justify our choice of the feature set as a balanced and effective configuration.}

%% file: tables/evaluation_results_impact_factors.tex
\begin{table*}[t]
\centering
% \resizebox{\textwidth}{!}{%
\scriptsize
\setlength{\tabcolsep}{3pt}
\renewcommand{\arraystretch}{0.9}
\caption{Evaluation results of five external factors, including four different COTS VR headsets, four different environments with external Wi-Fi interference, four different attack distances ranging from \SI{5.0}{\meter}--\SI{8.0}{\meter}, different number of connected non-target devices, and different number of running background apps. \includegraphics[width=8pt]{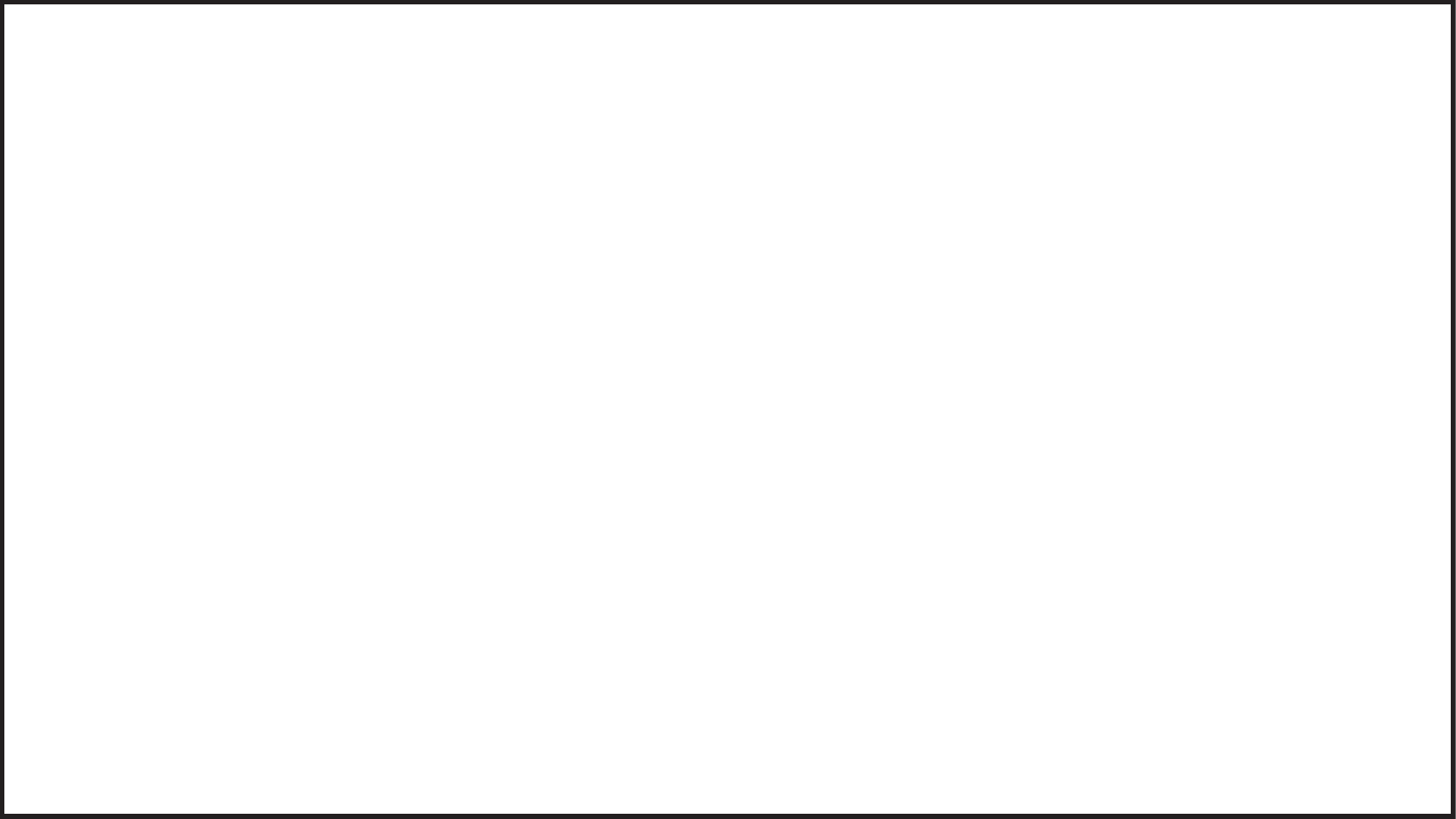} : $< 70\%$, \includegraphics[width=8pt]{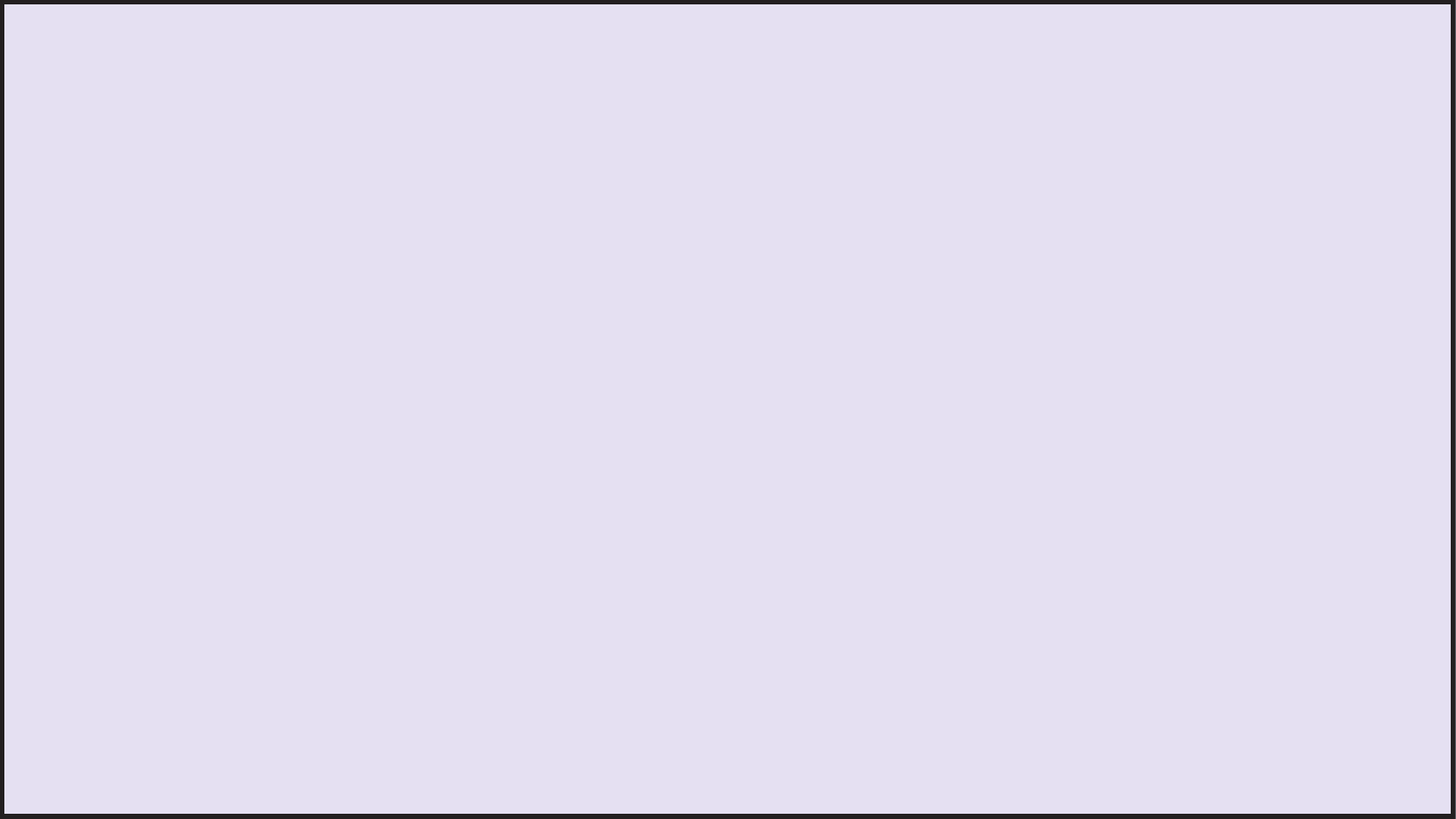} : $70\%$--$80\%$, \includegraphics[width=8pt]{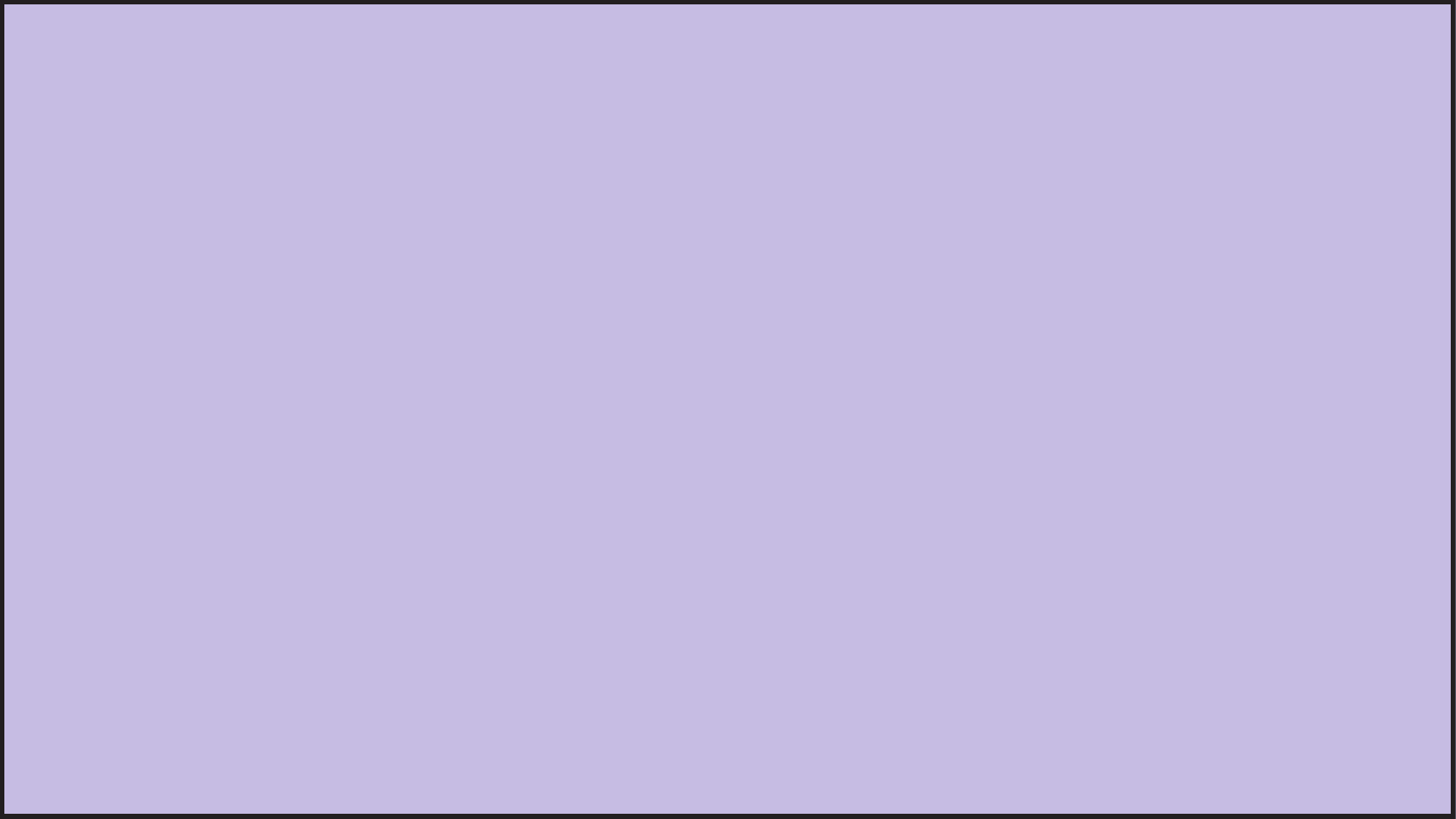} : $80\%$--$90\%$, \includegraphics[width=8pt]{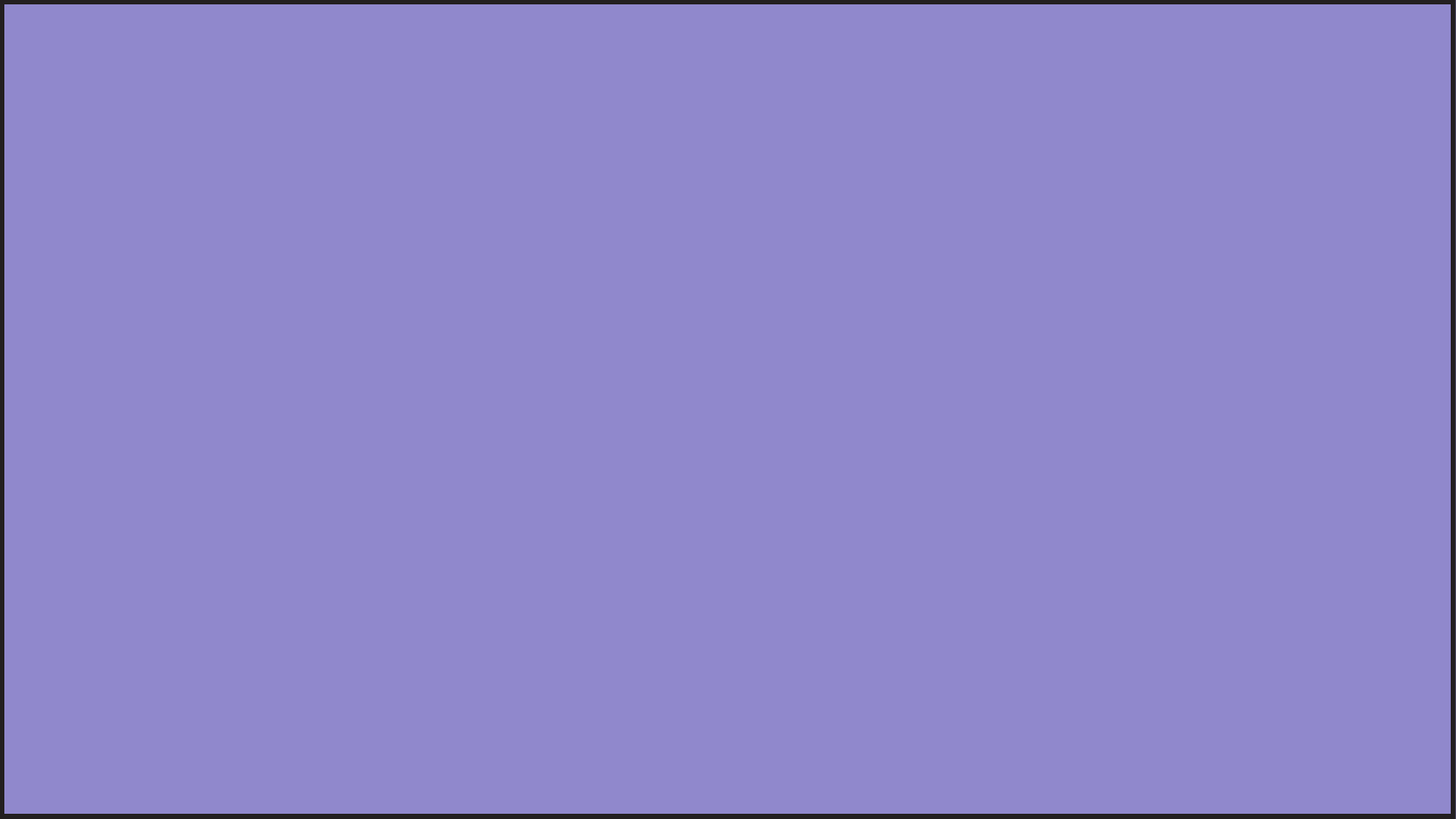} : $> 90\%$.}
\vspace{-0.1in}
\label{tab:impact_factors}
\begin{tabular}{ccccccccccccccccccccc}
\toprule
\multirow{2}{*}{\textbf{Acc. (\%)}} & \multicolumn{4}{c}{\textbf{VR Headsets}} & \multicolumn{4}{c}{\textbf{Environments}} & \multicolumn{4}{c}{\textbf{Attack Distances}}  & \multicolumn{4}{c}{\textbf{\# of Non-target Devices}} & \multicolumn{4}{c}{\textbf{\# of Background Apps}}\\ \cmidrule(lr){2-5} \cmidrule(lr){6-9} \cmidrule(lr){10-13} \cmidrule(lr){14-17} \cmidrule(lr){18-21}   
 & \textbf{Quest 2} & \textbf{PICO 4} & \textbf{Quest 3} & \textbf{Quest Pro} & \textbf{Home} & \textbf{Office} & \textbf{Corridor} & \textbf{Cafe} & \textbf{5.0 m} & \textbf{6.0 m} & \textbf{7.0 m} & \textbf{8.0 m} & \textbf{w/o} & \textbf{w/ one} & \textbf{w/ two} & \textbf{w/ three} & \textbf{w/o} & \textbf{w/ one} & \textbf{w/ two} & \textbf{w/ three}\\ \midrule\midrule
\multicolumn{1}{c|}{WF} & \multicolumn{1}{c|}{\cellcolor{mycolor3}89.2} & \multicolumn{1}{c|}{\cellcolor{mycolor3}87.5} & \multicolumn{1}{c|}{\cellcolor{mycolor3}89.6} & \multicolumn{1}{c|}{\cellcolor{mycolor3}88.5} & \multicolumn{1}{c|}{\cellcolor{mycolor3}89.2} & \multicolumn{1}{c|}{\cellcolor{mycolor3}80.6} & \multicolumn{1}{c|}{\cellcolor{mycolor4}90.7} & \multicolumn{1}{c|}{\cellcolor{mycolor3}84.3} & \multicolumn{1}{c|}{\cellcolor{mycolor3}89.2} & \multicolumn{1}{c|}{\cellcolor{mycolor3}87.3} & \multicolumn{1}{c|}{\cellcolor{mycolor2}78.7} & \multicolumn{1}{c|}{\cellcolor{mycolor1}68.3} & \multicolumn{1}{c|}{\cellcolor{mycolor3}89.2} & \multicolumn{1}{c|}{\cellcolor{mycolor3}81.6} & \multicolumn{1}{c|}{\cellcolor{mycolor2}76.3} & \multicolumn{1}{c|}{\cellcolor{mycolor1}69.6} & \multicolumn{1}{c|}{\cellcolor{mycolor3}89.2} & \multicolumn{1}{c|}{\cellcolor{mycolor3}83.9} & \multicolumn{1}{c|}{\cellcolor{mycolor3}81.1} & \multicolumn{1}{c}{\cellcolor{mycolor2}75.5} \\ \hline

\multicolumn{1}{c|}{AF} & \multicolumn{1}{c|}{\cellcolor{mycolor3}85.1} & \multicolumn{1}{c|}{\cellcolor{mycolor3}82.3} & \multicolumn{1}{c|}{\cellcolor{mycolor3}86.7} & \multicolumn{1}{c|}{\cellcolor{mycolor3}86.0} & \multicolumn{1}{c|}{\cellcolor{mycolor3}85.1} & \multicolumn{1}{c|}{\cellcolor{mycolor2}79.7} & \multicolumn{1}{c|}{\cellcolor{mycolor3}89.6} & \multicolumn{1}{c|}{\cellcolor{mycolor2}74.7} & \multicolumn{1}{c|}{\cellcolor{mycolor3}85.1} & \multicolumn{1}{c|}{\cellcolor{mycolor3}82.0} & \multicolumn{1}{c|}{\cellcolor{mycolor2}76.7} & \multicolumn{1}{c|}{\cellcolor{mycolor1}63.5} & \multicolumn{1}{c|}{\cellcolor{mycolor3}85.1} & \multicolumn{1}{c|}{\cellcolor{mycolor2}74.5} & \multicolumn{1}{c|}{\cellcolor{mycolor1}65.5} & \multicolumn{1}{c|}{\cellcolor{mycolor1}61.3} & \multicolumn{1}{c|}{\cellcolor{mycolor3}85.1} & \multicolumn{1}{c|}{\cellcolor{mycolor3}82.3} & \multicolumn{1}{c|}{\cellcolor{mycolor2}74.7} & \multicolumn{1}{c}{\cellcolor{mycolor2}70.5} \\ \hline

\multicolumn{1}{c|}{IAR} & \multicolumn{1}{c|}{\cellcolor{mycolor4}91.1} & \multicolumn{1}{c|}{\cellcolor{mycolor3}88.3} & \multicolumn{1}{c|}{\cellcolor{mycolor4}93.3} & \multicolumn{1}{c|}{\cellcolor{mycolor4}90.5} & \multicolumn{1}{c|}{\cellcolor{mycolor4}91.1} & \multicolumn{1}{c|}{\cellcolor{mycolor3}84.3} & \multicolumn{1}{c|}{\cellcolor{mycolor4}94.5} & \multicolumn{1}{c|}{\cellcolor{mycolor3}82.3} & \multicolumn{1}{c|}{\cellcolor{mycolor4}91.1} & \multicolumn{1}{c|}{\cellcolor{mycolor4}90.6} & \multicolumn{1}{c|}{\cellcolor{mycolor3}86.5} & \multicolumn{1}{c|}{\cellcolor{mycolor2}78.7} & \multicolumn{1}{c|}{\cellcolor{mycolor4}91.1} & \multicolumn{1}{c|}{\cellcolor{mycolor3}85.5} & \multicolumn{1}{c|}{\cellcolor{mycolor2}79.7} & \multicolumn{1}{c|}{\cellcolor{mycolor2}73.5} & \multicolumn{1}{c|}{\cellcolor{mycolor4}91.1} & \multicolumn{1}{c|}{\cellcolor{mycolor3}89.0} & \multicolumn{1}{c|}{\cellcolor{mycolor3}84.5} & \multicolumn{1}{c}{\cellcolor{mycolor2}78.6} \\ \hline

\multicolumn{1}{c|}{VAR} & \multicolumn{1}{c|}{\cellcolor{mycolor3}86.5} & \multicolumn{1}{c|}{\cellcolor{mycolor3}83.0} & \multicolumn{1}{c|}{\cellcolor{mycolor3}87.7} & \multicolumn{1}{c|}{\cellcolor{mycolor3}84.6} & \multicolumn{1}{c|}{\cellcolor{mycolor3}86.5} & \multicolumn{1}{c|}{\cellcolor{mycolor3}85.5} & \multicolumn{1}{c|}{\cellcolor{mycolor3}88.3} & \multicolumn{1}{c|}{\cellcolor{mycolor2}79.7} & \multicolumn{1}{c|}{\cellcolor{mycolor3}86.5} & \multicolumn{1}{c|}{\cellcolor{mycolor3}87.3} & \multicolumn{1}{c|}{\cellcolor{mycolor3}82.5} & \multicolumn{1}{c|}{\cellcolor{mycolor2}77.5} & \multicolumn{1}{c|}{\cellcolor{mycolor3}86.5} & \multicolumn{1}{c|}{\cellcolor{mycolor3}80.3} & \multicolumn{1}{c|}{\cellcolor{mycolor3}81.6} & \multicolumn{1}{c|}{\cellcolor{mycolor2}72.3} & \multicolumn{1}{c|}{\cellcolor{mycolor3}86.5} & \multicolumn{1}{c|}{\cellcolor{mycolor3}82.0} & \multicolumn{1}{c|}{\cellcolor{mycolor2}76.3} & \multicolumn{1}{c}{\cellcolor{mycolor1}68.9} \\ \hline

\multicolumn{1}{c|}{AUI} & \multicolumn{1}{c|}{\cellcolor{mycolor4}97.9} & \multicolumn{1}{c|}{\cellcolor{mycolor4}93.3} & \multicolumn{1}{c|}{\cellcolor{mycolor4}98.2} & \multicolumn{1}{c|}{\cellcolor{mycolor4}95.6} & \multicolumn{1}{c|}{\cellcolor{mycolor4}97.9} & \multicolumn{1}{c|}{\cellcolor{mycolor4}94.5} & \multicolumn{1}{c|}{\cellcolor{mycolor4}98.0} & \multicolumn{1}{c|}{\cellcolor{mycolor4}95.6} & \multicolumn{1}{c|}{\cellcolor{mycolor4}97.9} & \multicolumn{1}{c|}{\cellcolor{mycolor3}88.3} & \multicolumn{1}{c|}{\cellcolor{mycolor3}85.5} & \multicolumn{1}{c|}{\cellcolor{mycolor3}80.6} & \multicolumn{1}{c|}{\cellcolor{mycolor4}97.9} & \multicolumn{1}{c|}{\cellcolor{mycolor4}93.3} & \multicolumn{1}{c|}{\cellcolor{mycolor3}89.6} & \multicolumn{1}{c|}{\cellcolor{mycolor3}84.5} & \multicolumn{1}{c|}{\cellcolor{mycolor4}97.9} & \multicolumn{1}{c|}{\cellcolor{mycolor4}96.0} & \multicolumn{1}{c|}{\cellcolor{mycolor3}88.7} & \multicolumn{1}{c}{\cellcolor{mycolor3}81.2} \\ \hline

\multicolumn{1}{c|}{AKI (Top-10)} & \multicolumn{1}{c|}{\cellcolor{mycolor3}89.6} & \multicolumn{1}{c|}{\cellcolor{mycolor2}72.8} & \multicolumn{1}{c|}{\cellcolor{mycolor3}85.8} & \multicolumn{1}{c|}{\cellcolor{mycolor3}86.7} & \multicolumn{1}{c|}{\cellcolor{mycolor3}89.6} & \multicolumn{1}{c|}{\cellcolor{mycolor2}77.0} & \multicolumn{1}{c|}{\cellcolor{mycolor4}90.1} & \multicolumn{1}{c|}{\cellcolor{mycolor3}82.3} & \multicolumn{1}{c|}{\cellcolor{mycolor3}89.6} & \multicolumn{1}{c|}{\cellcolor{mycolor3}81.6} & \multicolumn{1}{c|}{\cellcolor{mycolor2}71.7} & \multicolumn{1}{c|}{\cellcolor{mycolor1}62.5} & \multicolumn{1}{c|}{\cellcolor{mycolor3}89.6} & \multicolumn{1}{c|}{\cellcolor{mycolor3}83.5} & \multicolumn{1}{c|}{\cellcolor{mycolor2}76.6} & \multicolumn{1}{c|}{\cellcolor{mycolor1}65.0} & \multicolumn{1}{c|}{\cellcolor{mycolor3}89.6} & \multicolumn{1}{c|}{\cellcolor{mycolor3}84.8} & \multicolumn{1}{c|}{\cellcolor{mycolor3}82.0} & \multicolumn{1}{c}{\cellcolor{mycolor2}70.7} \\ \bottomrule
\end{tabular}%
% }
\vspace{-0.15in}
\end{table*}

%% file: tables/more_through_wall_attack_results.tex
% \usepackage{multirow}
\begin{table}[t]
\centering
% \resizebox{\textwidth}{!}{%
\scriptsize
\setlength{\tabcolsep}{3pt}
\renewcommand{\arraystretch}{0.95}
\caption{Evaluation results of through-wall attacks with walls that have different types and thicknesses at different attack distances. \raisebox{-0.04in}{\includegraphics[width=10pt]{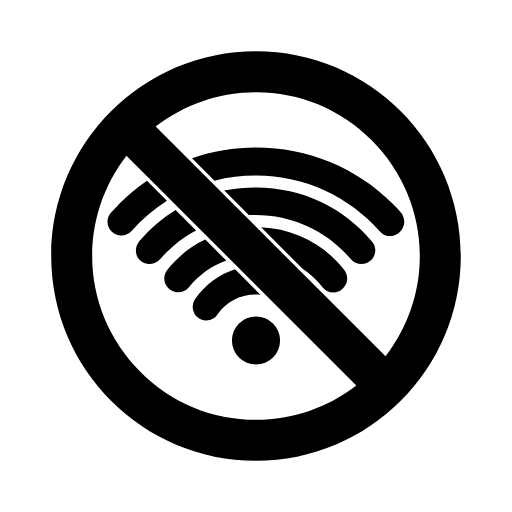}}: No wireless powers can be captured by the Wi-Fi sniffer through a thick concrete wall (\SI{25.0}{\centi\meter}) at \SI{8.0}{\meter}.}
\vspace{-0.1in}
\label{tab:through_wall_attack_results}
\begin{tabular}{ccccccccc}
\toprule
\textbf{Type}                             & \textbf{Thickness}        & \textbf{Dist.}         & \textbf{WF (\%)}         & \textbf{AF (\%)}         & \textbf{IAR (\%)}        & \textbf{VAR (\%)}        & \textbf{AUI (\%)}         & \textbf{AKI (\%)} \\ \midrule\midrule
\multicolumn{1}{c|}{\multirow{4}{*}{Plastic}}  & \multicolumn{1}{c|}{\SI{6.7}{\centi\meter}} & \multicolumn{1}{c|}{\SI{5.0}{\meter}} & \multicolumn{1}{c|}{\cellcolor{mycolor3}89.2} & \multicolumn{1}{c|}{\cellcolor{mycolor3}85.1} & \multicolumn{1}{c|}{\cellcolor{mycolor4}91.1} & \multicolumn{1}{c|}{\cellcolor{mycolor3}86.5} & \multicolumn{1}{c|}{\cellcolor{mycolor4}97.9} & \cellcolor{mycolor3}89.6               \\ \cline{2-9} 

\multicolumn{1}{c|}{}                          & \multicolumn{1}{c|}{\SI{6.7}{\centi\meter}} & \multicolumn{1}{c|}{\SI{8.0}{\meter}} & \multicolumn{1}{c|}{\cellcolor{mycolor1}68.3} & \multicolumn{1}{c|}{\cellcolor{mycolor1}63.5} & \multicolumn{1}{c|}{\cellcolor{mycolor2}78.7} & \multicolumn{1}{c|}{\cellcolor{mycolor2}77.5} & \multicolumn{1}{c|}{\cellcolor{mycolor3}80.6} & \cellcolor{mycolor1}62.5               \\ \cline{2-9} 

\multicolumn{1}{c|}{}                          & \multicolumn{1}{c|}{\SI{11.3}{\centi\meter}} & \multicolumn{1}{c|}{\SI{5.0}{\meter}} & \multicolumn{1}{c|}{\cellcolor{mycolor3}84.8} & \multicolumn{1}{c|}{\cellcolor{mycolor2}79.3} & \multicolumn{1}{c|}{\cellcolor{mycolor3}89.6} & \multicolumn{1}{c|}{\cellcolor{mycolor3}83.5} & \multicolumn{1}{c|}{\cellcolor{mycolor4}95.0} & \cellcolor{mycolor3}87.3               \\ \cline{2-9} 

\multicolumn{1}{c|}{}                          & \multicolumn{1}{c|}{\SI{11.3}{\centi\meter}} & \multicolumn{1}{c|}{\SI{8.0}{\meter}} & \multicolumn{1}{c|}{\cellcolor{mycolor1}61.9} & \multicolumn{1}{c|}{\cellcolor{mycolor1}57.7} & \multicolumn{1}{c|}{\cellcolor{mycolor2}72.3} & \multicolumn{1}{c|}{\cellcolor{mycolor2}71.7} & \multicolumn{1}{c|}{\cellcolor{mycolor2}74.7} & \cellcolor{mycolor1}52.5               \\ \hline

\multicolumn{1}{c|}{\multirow{4}{*}{Glass}}  & \multicolumn{1}{c|}{\SI{1.0}{\centi\meter}} & \multicolumn{1}{c|}{\SI{5.0}{\meter}} & \multicolumn{1}{c|}{\cellcolor{mycolor4}92.7} & \multicolumn{1}{c|}{\cellcolor{mycolor3}88.6} & \multicolumn{1}{c|}{\cellcolor{mycolor4}94.5} & \multicolumn{1}{c|}{\cellcolor{mycolor4}90.6} & \multicolumn{1}{c|}{\cellcolor{mycolor4}98.0} & \cellcolor{mycolor4}93.3               \\ \cline{2-9} 

\multicolumn{1}{c|}{}                          & \multicolumn{1}{c|}{\SI{1.0}{\centi\meter}} & \multicolumn{1}{c|}{\SI{8.0}{\meter}} & \multicolumn{1}{c|}{\cellcolor{mycolor2}79.3} & \multicolumn{1}{c|}{\cellcolor{mycolor2}75.5} & \multicolumn{1}{c|}{\cellcolor{mycolor3}81.6} & \multicolumn{1}{c|}{\cellcolor{mycolor2}78.7} & \multicolumn{1}{c|}{\cellcolor{mycolor4}90.7} & \cellcolor{mycolor3}80.6               \\ \cline{2-9} 

\multicolumn{1}{c|}{}                          & \multicolumn{1}{c|}{\SI{1.2}{\centi\meter}} & \multicolumn{1}{c|}{\SI{5.0}{\meter}} & \multicolumn{1}{c|}{\cellcolor{mycolor4}91.9} & \multicolumn{1}{c|}{\cellcolor{mycolor3}87.3} & \multicolumn{1}{c|}{\cellcolor{mycolor4}92.7} & \multicolumn{1}{c|}{\cellcolor{mycolor3}88.6} & \multicolumn{1}{c|}{\cellcolor{mycolor4}97.3} & \cellcolor{mycolor4}92.0               \\ \cline{2-9} 
\multicolumn{1}{c|}{}                          & \multicolumn{1}{c|}{\SI{1.2}{\centi\meter}} & \multicolumn{1}{c|}{\SI{8.0}{\meter}} & \multicolumn{1}{c|}{\cellcolor{mycolor2}75.5} & \multicolumn{1}{c|}{\cellcolor{mycolor2}71.7} & \multicolumn{1}{c|}{\cellcolor{mycolor2}77.9} & \multicolumn{1}{c|}{\cellcolor{mycolor2}74.7} & \multicolumn{1}{c|}{\cellcolor{mycolor3}85.8} & \cellcolor{mycolor2}74.5               \\ \hline

\multicolumn{1}{c|}{\multirow{4}{*}{Wood}}     & \multicolumn{1}{c|}{\SI{4.3}{\centi\meter}} & \multicolumn{1}{c|}{\SI{5.0}{\meter}} & \multicolumn{1}{c|}{\cellcolor{mycolor3}85.6} & \multicolumn{1}{c|}{\cellcolor{mycolor2}79.7} & \multicolumn{1}{c|}{\cellcolor{mycolor3}87.3} & \multicolumn{1}{c|}{\cellcolor{mycolor3}81.6} & \multicolumn{1}{c|}{\cellcolor{mycolor4}90.3} & \cellcolor{mycolor3}80.3               \\ \cline{2-9} 

\multicolumn{1}{c|}{}                          & \multicolumn{1}{c|}{\SI{4.3}{\centi\meter}} & \multicolumn{1}{c|}{\SI{8.0}{\meter}} & \multicolumn{1}{c|}{\cellcolor{mycolor1}62.7} & \multicolumn{1}{c|}{\cellcolor{mycolor1}58.9} & \multicolumn{1}{c|}{\cellcolor{mycolor1}64.5} & \multicolumn{1}{c|}{\cellcolor{mycolor1}59.0} & \multicolumn{1}{c|}{\cellcolor{mycolor2}71.7} & \cellcolor{mycolor1}55.3               \\ \cline{2-9} 

\multicolumn{1}{c|}{}                          & \multicolumn{1}{c|}{\SI{8.6}{\centi\meter}} & \multicolumn{1}{c|}{\SI{5.0}{\meter}} & \multicolumn{1}{c|}{\cellcolor{mycolor2}79.3} & \multicolumn{1}{c|}{\cellcolor{mycolor2}73.7} & \multicolumn{1}{c|}{\cellcolor{mycolor3}80.6} & \multicolumn{1}{c|}{\cellcolor{mycolor2}72.0} & \multicolumn{1}{c|}{\cellcolor{mycolor3}82.3} & \cellcolor{mycolor2}70.9               \\ \cline{2-9} 

\multicolumn{1}{c|}{}                          & \multicolumn{1}{c|}{\SI{8.6}{\centi\meter}} & \multicolumn{1}{c|}{\SI{8.0}{\meter}} & \multicolumn{1}{c|}{\cellcolor{mycolor1}53.8} & \multicolumn{1}{c|}{\cellcolor{mycolor1}51.3} & \multicolumn{1}{c|}{\cellcolor{mycolor1}53.5} & \multicolumn{1}{c|}{\cellcolor{mycolor1}46.7} & \multicolumn{1}{c|}{\cellcolor{mycolor1}60.8} & \cellcolor{mycolor1}44.8               \\ \hline

\multicolumn{1}{c|}{\multirow{4}{*}{Concrete}} & \multicolumn{1}{c|}{\SI{9.5}{\centi\meter}} & \multicolumn{1}{c|}{\SI{5.0}{\meter}} & \multicolumn{1}{c|}{\cellcolor{mycolor2}78.7} & \multicolumn{1}{c|}{\cellcolor{mycolor2}72.5} & \multicolumn{1}{c|}{\cellcolor{mycolor3}80.6} & \multicolumn{1}{c|}{\cellcolor{mycolor2}73.6} & \multicolumn{1}{c|}{\cellcolor{mycolor3}84.8} & \cellcolor{mycolor2}71.7               \\ \cline{2-9} 
\multicolumn{1}{c|}{}                          & \multicolumn{1}{c|}{\SI{9.5}{\centi\meter}} & \multicolumn{1}{c|}{\SI{8.0}{\meter}} & \multicolumn{1}{c|}{\cellcolor{mycolor1}56.7} & \multicolumn{1}{c|}{\cellcolor{mycolor1}48.5} & \multicolumn{1}{c|}{\cellcolor{mycolor1}62.5} & \multicolumn{1}{c|}{\cellcolor{mycolor1}50.3} & \multicolumn{1}{c|}{\cellcolor{mycolor1}63.2} & \cellcolor{mycolor1}45.5               \\ \cline{2-9} 
\multicolumn{1}{c|}{}                          & \multicolumn{1}{c|}{\SI{25.0}{\centi\meter}} & \multicolumn{1}{c|}{\SI{5.0}{\meter}} & \multicolumn{1}{c|}{\cellcolor{mycolor1}45.5} & \multicolumn{1}{c|}{\cellcolor{mycolor1}33.3} & \multicolumn{1}{c|}{\cellcolor{mycolor1}46.7} & \multicolumn{1}{c|}{\cellcolor{mycolor1}30.9} & \multicolumn{1}{c|}{\cellcolor{mycolor1}52.5} & \cellcolor{mycolor1}31.7               \\ \cline{2-9} 
\multicolumn{1}{c|}{}                          & \multicolumn{1}{c|}{\SI{25.0}{\centi\meter}} & \multicolumn{1}{c|}{\SI{8.0}{\meter}} & \multicolumn{1}{c|}{\raisebox{-0.04in}{\includegraphics[width=10pt]{icons/nosignal_icon.png}}} & \multicolumn{1}{c|}{\raisebox{-0.04in}{\includegraphics[width=10pt]{icons/nosignal_icon.png}}} & \multicolumn{1}{c|}{\raisebox{-0.04in}{\includegraphics[width=10pt]{icons/nosignal_icon.png}}} & \multicolumn{1}{c|}{\raisebox{-0.04in}{\includegraphics[width=10pt]{icons/nosignal_icon.png}}} & \multicolumn{1}{c|}{\raisebox{-0.04in}{\includegraphics[width=10pt]{icons/nosignal_icon.png}}} & \raisebox{-0.04in}{\includegraphics[width=10pt]{icons/nosignal_icon.png}}               \\ \bottomrule
\end{tabular}
\vspace{-0.2in}
\end{table}

%% file: tables/feature_ablation_study.tex
\begin{table}[t]
\centering
\scriptsize
\setlength{\tabcolsep}{2pt}
\renewcommand{\arraystretch}{1.0}
\caption{\major{Ablation study on feature selection. Red values: Accuracy degradation relative to the full configuration.}}
\vspace{-0.1in}
\label{tab:ablation_features}
\begin{tabular}{ccccccc}
\toprule
\textbf{Configuration} & \textbf{WF (\%)} & \textbf{AF (\%)} & \textbf{IAR (\%)} & \textbf{VAR (\%)} & \textbf{AUI (\%)} & \textbf{AKI (\%)} \\ \midrule\midrule
\multicolumn{1}{c|}{Full (15T+2F)}
& \multicolumn{1}{c|}{\cellcolor{mycolor3}89.2}
& \multicolumn{1}{c|}{\cellcolor{mycolor3}85.1}
& \multicolumn{1}{c|}{\cellcolor{mycolor4}91.1}
& \multicolumn{1}{c|}{\cellcolor{mycolor3}86.5}
& \multicolumn{1}{c|}{\cellcolor{mycolor4}97.9}
& \cellcolor{mycolor3}89.6 \\ \hline
\multicolumn{1}{c|}{w/o Shape (12T+2F)}
& \multicolumn{1}{c|}{\cellcolor{mycolor3}83.5 (\textcolor{red}{$-$5.7})}
& \multicolumn{1}{c|}{\cellcolor{mycolor2}78.6 (\textcolor{red}{$-$6.5})}
& \multicolumn{1}{c|}{\cellcolor{mycolor3}85.8 (\textcolor{red}{$-$5.3})}
& \multicolumn{1}{c|}{\cellcolor{mycolor3}80.9 (\textcolor{red}{$-$5.6})}
& \multicolumn{1}{c|}{\cellcolor{mycolor4}92.6 (\textcolor{red}{$-$5.3})}
& \cellcolor{mycolor3}83.4 (\textcolor{red}{$-$6.2}) \\ \hline
\multicolumn{1}{c|}{w/o Frequency (15T)}
& \multicolumn{1}{c|}{\cellcolor{mycolor3}85.4 (\textcolor{red}{$-$3.8})}
& \multicolumn{1}{c|}{\cellcolor{mycolor3}81.7 (\textcolor{red}{$-$3.4})}
& \multicolumn{1}{c|}{\cellcolor{mycolor3}87.6 (\textcolor{red}{$-$3.5})}
& \multicolumn{1}{c|}{\cellcolor{mycolor3}83.2 (\textcolor{red}{$-$3.3})}
& \multicolumn{1}{c|}{\cellcolor{mycolor4}94.5 (\textcolor{red}{$-$3.4})}
& \cellcolor{mycolor3}86.0 (\textcolor{red}{$-$3.6}) \\ \hline
\multicolumn{1}{c|}{Only Time (12T)}
& \multicolumn{1}{c|}{\cellcolor{mycolor3}80.7 (\textcolor{red}{$-$8.5})}
& \multicolumn{1}{c|}{\cellcolor{mycolor2}75.3 (\textcolor{red}{$-$9.8})}
& \multicolumn{1}{c|}{\cellcolor{mycolor3}83.5 (\textcolor{red}{$-$7.6})}
& \multicolumn{1}{c|}{\cellcolor{mycolor2}77.8 (\textcolor{red}{$-$8.7})}
& \multicolumn{1}{c|}{\cellcolor{mycolor3}89.7 (\textcolor{red}{$-$8.2})}
& \cellcolor{mycolor3}80.5 (\textcolor{red}{$-$9.1}) \\ \bottomrule
\end{tabular}
\vspace{-0.15in}
\end{table}

%% file: sections_major/06_discussion.tex
\section{Discussion}
\label{sec:discussion}

\subsection{Countermeasures}
\label{subsec:countermeasures}

% Below, we introduce two countermeasures (\eg, $C_{1}$ and $C_{2}$) to defend against the proposed eavesdropping attacks from \sysname.

% \begin{figure}[t]
%     \centering
%     \includegraphics[width=\linewidth]{figures/countermeasures_results_2025.pdf}
%     \vspace{-0.2in}
%     \caption{Results in defending against \sysname by adopting the signal obfuscation and dynamic power countermeasures.}
%     \vspace{-0.2in}
%     \label{fig:results_defense}
% \end{figure}

\noindent\textbf{$C_{1}$: Signal Obfuscation.} 
One defense method is to implement signal obfuscation in wireless network traffic. This involves allowing the Wi-Fi router to randomly transmit redundant data packets~\cite{li2022foap}, creating additional Wi-Fi energy in traffic transmission, and interfering with signals collected by \sysname. Moreover, placing other Wi-Fi sources (\eg, routers, mobile devices) in close proximity can generate additional RF signals to obfuscate captured wireless power.
To explore this defense solution, we placed another Wi-Fi router (Huawei Pro 2) near the target router at the same distance while randomly generating network transmissions.
\autoref{fig:results_defense} shows \major{that} the accuracy rates decrease by \major{an average of} $46.8\%$, $49.5\%$, $33.4\%$, and $36.8\%$ in the recognition of websites, apps, in-app activities, and VR-specific activities, respectively.
Although these signal obfuscation methods could degrade the performance of \sysname, transmitting redundant packets can result in increased power consumption, and nearby Wi-Fi sources could potentially interfere with each other~\cite{sarbu2018non}.
\looseness=-1

\noindent\textbf{$C_{2}$: Dynamic Transmission Power.} 
As illustrated in \autoref{eq:friis_transmission_equation}, the captured wireless power depends on the power sent from the transmitter (\eg, Wi-Fi router). 
If the Wi-Fi router employs a dynamic transmission power strategy~\cite{sanchez2010dynamic}, \ie, bursting transmission in the low-power model while transmitting small packets in high-power mode, the captured wireless power will change accordingly and \major{may not} reflect the real network traffic of the VR devices' activities. 
Hence, we enable the dynamic transmission mode in the Wi-Fi router (Huawei Pro 2) and repeat the same experiments, where the results in \autoref{fig:results_defense} show that the performance decreases by $33.6\%$, $35.7\%$, $25.8\%$, and $25.2\%$ in inferring the privacy of the aforementioned user.
Hence, \sysname could be deceived by the modified patterns in the transmitted Wi-Fi energy. 
However, the dynamic transmission power scheme cannot guarantee the quality of the link since it is utilized to increase spectral efficiency and reduce interference by transmitting at the minimum power without overload.
\looseness=-1

% \begin{figure}[t]
%     \centering
%     \includegraphics[width=\linewidth]{figures/countermeasures_results_2025.pdf}
%     \vspace{-0.25in}
%     \caption{Results in defending against \sysname by adopting the signal obfuscation and dynamic power countermeasures.}
%     \vspace{-0.2in}
%     \label{fig:results_defense}
% \end{figure}

\subsection{Limitations and Future Works}
\label{subsec:limitations}

While the initial results in this paper are promising, several limitations still exist in our current research.
First, following the research line of relevant studies~\cite{al2021vr, mei2024mmspyvr, ni2023eavesdropping}, \sysname requires the attacker to \major{physically} place a small wireless power sniffer in a space adjacent to \major{the victim within \SI{5.0}{\meter}--\SI{8.0}{\meter} of} the target Wi-Fi router to capture through-wall wireless power leakage, which could raise concerns about its practicality.
Nevertheless, prior VR attacks (\eg, \cite{gopal2023hidden, nguyen2024penetration, khalili2024virtual, luo2024eavesdropping, meteriz2022keylogging}) require placing multiple cameras or smartphones near the VR user to record videos and audio, which may not be effective in environments with low light conditions or other noisy interference.
Thus, \sysname provides an orthogonal attack surface for eavesdropping on multi-level AR/VR privacy in a long-range and through-wall manner.
Moreover, we propose \sysname as a proof-of-concept study to validate the vulnerabilities from wireless power leakage. We implemented the models with heuristic settings and parameter configurations while evaluating their performance on datasets from limited websites, apps, and activities, raising concerns about their generalization ability.
Hence, \sysname can be easily extended to other attack scenarios or platforms by collecting \major{a} few data samples to finetune the open-world inference models.

In addition, we have not evaluated \sysname's performance on more VR/AR headsets such as Apple Vision Pro, and users can see the ambient environments in mixed reality scenes while visionOS is a closed OS that does not allow unauthorized installation, leading to the failure of the aforementioned studies.
Nevertheless, it still requires a network connection (\eg, Wi-Fi, or cellular) for communication and interactions, causing bursts of network traffic and changes in the transmitted wireless power.
\major{Additionally, \sysname's effectiveness is inherently affected by the deployment environment. As shown in \autoref{subsec:rf_energy_network_traffic} and \autoref{tab:through_wall_attack_results}, attack effectiveness degrades with increasing wall thickness, denser obstacles, longer attack distances, and the presence of co-channel wireless devices generating heavy traffic. While \sysname remains effective across common settings such as residential apartments, offices, and cafes, performance may further drop in extreme conditions such as multi-wall or multi-floor scenarios, dense Wi-Fi environments with many simultaneously transmitting devices, or rooms shielded with metal-lined or thick reinforced concrete walls. We plan to explore signal enhancement techniques (\eg, higher-gain antennas, multi-sniffer fusion) to extend the effective operating envelope in such challenging environments, and to address the above limitations in our future work.}
\looseness=-1

%% file: sections_major/07_related_work.tex
\section{Related Works}
\label{sec:related_works}
% \vspace{-0.05in}

\input{tables/comparison_ni}

\noindent\textbf{Attacks on VR Devices.}
Most of the recent studies exploit pre-installed malicious software or apps to acquire headset traces from their motion sensors to identify users' head and hand gestures~\cite{choi2023revisiting}, snoop virtual keyboard input~\cite{wu2023privacy, zhang2023s, slocum2023going, luo2022holologger, ling2019know, lee2025eyes}, reconstruct speech and vital signals~\cite{shi2021face, cayir2025speak, zhang2023passive, zhang2023facereader, ye2025bpsniff} (\eg, heartbeat rates, and blood pressure), recognize running apps~\cite{zhang2023s}, and obtain spatial mesh information~\cite{zhang2023s} and localizations~\cite{farrukh2023locin}.
Moreover, research has shown that adversaries can use hand trackers or cameras to monitor users' head/hand movements to infer 360$^{\circ}$ videos~\cite{nguyen2024penetration} and virtual keystrokes~\cite{gopal2023hidden, meteriz2022keylogging, yildiran2022airtype, meteriz2021sia}.
In addition, recent studies have demonstrated the feasibility of inferring keystrokes through acoustic signals~\cite{luo2024eavesdropping} or infrared emanations~\cite{ni2025non} from controllers, unencrypted network packets in some multi-user VR apps~\cite{su2024remote}, and avatar's gaze information in a joint virtual environment~\cite{wang2024gazeploit}.
\major{In addition, a recent study, \textsc{mmSpyVR}~\cite{mei2024mmspyvr}, exploits 60\,GHz mmWave radar to actively probe VR users' body movements through obstacles and infer five coarse-grained activities and keystrokes. However, it requires the user to play VR within the radar's field of view, emits detectable mmWave signals, and degrades to 67.7\% and 52.6\% accuracy at \SI{8.0}{\meter}~\cite{mei2024mmspyvr}.
Compared with existing works, \sysname introduces an orthogonal approach that addresses these limitations 
% by passively exploiting a power side channel in wireless communication between the VR headset and the Wi-Fi access point, 
by imposing no positional constraint on the user and unveiling fine-grained multi-level privacy, particularly identifying explicit user interactions, unseen VR-specific in-game activities, and avatar-based user identifications and virtual keystrokes.}
\looseness=-1

\noindent\textbf{Attacks via Power Side Channels.}
Leveraging the power side channel to reveal privacy leakages in mobile devices usually involves two primary tactics: intrusively accessing target mobile devices to obtain battery profiles~\cite{chen2017powerful, gao2024deeptheft} or acquiring power traces from charging devices (\eg, USB cables~\cite{ni2023xporter}, wireless chargers~\cite{ni2023uncovering}).
For example, \textsf{LineTalker}~\cite{li2024dangers} shows that the adversary can intrusively access the charging current of VR to eavesdrop on UI-level activities (\eg, websites or apps).
Moreover, \textsf{DeepTheft}~\cite{gao2024deeptheft} leverages the CPU energy consumption profile to reconstruct the architecture of on-device deep neural networks.

\major{A recent work, \textsf{AppListener}~\cite{ni2023eavesdropping}, pioneered RF energy harvesting as a side channel to fingerprint mobile apps by passively capturing wireless power from a connected Wi-Fi router.}
\major{Although \sysname shares the same underlying principle, it extends the attack vector along multiple dimensions, as summarized in \autoref{tab:comparison_with_applistener}. First, \textsf{AppListener} captures only \SI{2.4}{\giga\hertz} signals with a single low-gain antenna and lacks interference removal, restricting its range to under \SI{1.5}{\meter} in non-blocked proximity, whereas \sysname uses dual high-gain antennas on both \SI{2.4}{\giga\hertz} and \SI{5}{\giga\hertz} with an adaptive interference removal pipeline, achieving reliable acquisition at \SI{5.0}{\meter}--\SI{8.0}{\meter} through different walls. Second, \textsf{AppListener} performs coarse-grained app fingerprinting on smartphones, while \sysname targets VR headsets and achieves three levels of seven privacy inference tasks, including unseen in-game events, avatar-based re-identification, and virtual keystroke inference. Third, \textsf{AppListener} relies on closed-world classification, whereas \sysname adopts an open-world embedding model that handles previously unseen activities. Overall, \sysname offers an orthogonal and complementary attack surface that builds upon \textsf{AppListener} and takes a step forward in understanding wireless power leakage on AR/VR platforms.}
\looseness=-1

%% file: tables/comparison_ni.tex
\begin{table}[t]
\centering
\scriptsize
\setlength{\tabcolsep}{3pt}
\renewcommand{\arraystretch}{1.0}
\caption{\major{Comparison between \sysname and \textnormal{\textsf{AppListener}}~\cite{ni2023eavesdropping}.}}
\vspace{-0.1in}
\label{tab:comparison_with_applistener}
\begin{tabular}{cccc}
\toprule
\textbf{Metrics} & \textbf{Sub-areas} & \textsf{AppListener~\cite{ni2023eavesdropping}} & \textbf{\sysname (Ours)} \\ \midrule\midrule

\multicolumn{1}{c|}{\multirow{3}{*}{Hardware}}
& \multicolumn{1}{c|}{Frequency Band}     & \multicolumn{1}{c|}{2.4\,GHz only}        & 2.4\,GHz/5\,GHz \\ \cline{2-4}
\multicolumn{1}{c|}{}
& \multicolumn{1}{c|}{Antenna Gain}       & \multicolumn{1}{c|}{2.3\,dBi (single)}    & 36\,dBi (dual) \\ \cline{2-4}
\multicolumn{1}{c|}{}
& \multicolumn{1}{c|}{Interference Removal} & \multicolumn{1}{c|}{\myHalfcirc}             & \myYescirc \\ \hline

\multicolumn{1}{c|}{\multirow{3}{*}{Range}}
& \multicolumn{1}{c|}{Distance}      & \multicolumn{1}{c|}{$<$ 1.5\,m}             & 5\,m--8\,m \\ \cline{2-4}
\multicolumn{1}{c|}{}
& \multicolumn{1}{c|}{Target}      & \multicolumn{1}{c|}{Smartphones}             & VR headsets \\ \cline{2-4}
\multicolumn{1}{c|}{}
& \multicolumn{1}{c|}{Through-Wall}       & \multicolumn{1}{c|}{Only for thin walls} & Different thin/thick walls \\ \hline

\multicolumn{1}{c|}{\multirow{4}{*}{Privacy}}
& \multicolumn{1}{c|}{Device Identification}     & \multicolumn{1}{c|}{\myNocirc}               & \myYescirc \\ \cline{2-4}
\multicolumn{1}{c|}{}
& \multicolumn{1}{c|}{Website/App Fingerprinting} & \multicolumn{1}{c|}{\myHalfcirc}    & \myYescirc \\ \cline{2-4}
\multicolumn{1}{c|}{}
& \multicolumn{1}{c|}{Unseen In-app Activity}   & \multicolumn{1}{c|}{\myNocirc}               & \myYescirc \\ \cline{2-4}
\multicolumn{1}{c|}{}
& \multicolumn{1}{c|}{Avatar-based Re-ID \& Keystrokes} & \multicolumn{1}{c|}{\myNocirc}        & \myYescirc \\ \hline

\multicolumn{1}{c|}{Setting}
& \multicolumn{1}{c|}{Inference}            & \multicolumn{1}{c|}{Closed-world}         & Open-world\\ \bottomrule
\end{tabular}
\vspace{-0.2in}
\end{table}

%% file: sections_major/08_conclusion.tex
\section{Conclusion}
\label{sec:conclusion}

In this paper, we disclose a practical through-wall eavesdropping attack to reveal user privacy in AR/VR \major{devices} by leveraging the captured wireless power in the transmission of the Wi-Fi network.
To validate its feasibility, we design and implement \sysname, an attack framework that utilizes a custom-built wireless power sniffer to non-intrusively capture Wi-Fi energy in the wireless transmission
and then infer multi-level user interactions, unseen VR-specific activities,
as well as user identification and virtual keystrokes relevant to users' avatars.
The empirical results show that it achieves high accuracy in recognizing various user behaviors inside COTS VR headsets and presents high resilience to various external impact factors.
Furthermore, \sysname outperforms previous studies by \major{demonstrating} the ability to eavesdrop through different types of walls at longer distances.
% In the end, we \major{propose} two effective countermeasures to defend against the disclosed \major{wireless power-based traffic analysis} attacks.
% We hope our findings can raise public awareness of the privacy leakage from the network characteristics of AR/VR and spur research on detecting vulnerabilities and developing new defense approaches.
\looseness=-1

\section*{Acknowledgment}

We sincerely appreciate our shepherd and all reviewers for their constructive feedback and comments.
This work was fully supported by the Hong Kong Research Grants Council (RGC) under Grants CityU 11222226, 21219223, 11219524, R1012-21, RFS2122-1S04, C1029-22G, C6015-23G, and CRS\_HKUST601/24.
The work was also supported by InnoHK initiative, The Government of the HKSAR, and Laboratory for AI-Powered Financial Technologies, as well as Guangdong and Hong Kong Universities '1+1+1' Joint Research Collaboration Scheme.
% was supported in part by Hong Kong Research Grants Council (RGC) under Grants 11219524, 11219025, RFS2122-1S04, C1029-22G, C6015-23G, and  CRS_HKUST601/24
%
% This work was also substantially supported by the InnoHK initiative, The Government of the HKSAR, and the Laboratory for AI-Powered Financial Technologies (AIFT).
Any opinions, findings, and conclusions in this paper are those of the authors and are not necessarily those of the supported organizations. 